\documentclass[aps,prb,reprint,twocolumn,superscriptaddress,amsmath,amssymb,10pt]{revtex4-2}
\usepackage{graphicx}
\usepackage{epstopdf}
\usepackage{xcolor}
\usepackage{comment}
\usepackage[colorlinks, linkcolor=red,citecolor=blue,urlcolor=blue]{hyperref}
\usepackage[all]{hypcap} 
\begin{document}
\title{
Interplay of ferroelectricity and interlayer superconductivity in van der Waals bilayers
}

\begin{abstract}
We study the distinctive features of the interplay between the interlayer superconductivity and ferroelectricity in van der Waals heterostructures. Corresponding analysis is carried out within the framework of the quasiclassical Eilenberger equations for a tunnel coupled bilayer with inhomogeneous relative shift of the conduction bands between the layers, which describes the net charge transfer in sliding ferroelectrics. It is shown that the critical temperature of the interlayer superconductivity can be significantly enhanced for superconducting nuclei localized in the vicinity of ferroelectric domain walls. We demonstrate that the increase in the tunneling amplitude leads to the decrease (increase) in the difference between the critical temperatures for localized and homogeneous superconducting states for the spin-singlet (spin-triplet) interlayer superconductivity. We also perform an extensive analysis of the effects of the in-plane magnetic field on the interlayer superconductivity. It is shown that the orbital effect can result in the suppression of the spin-singlet interlayer superconductivity and to the enhancement of the spin-triplet one. We find that possible manifestations of the paramagnetic effect include the suppression of the interlayer superconductivity by rather weak Zeeman fields, the two-fold anisotropy of the critical magnetic field for the spin-triplet states as well as the appearance of the reentrant superconducting phases. It is shown that the joint influence of the orbital and paramagnetic mechanisms on the spin-triplet interlayer superconductivity can even lead to a nonmonotonic behavior of the superconducting critical temperature as a function of the external magnetic field. The obtained results are discussed in the context of recent experimental data on van der Waals structures with coexisting superconductivity and sliding ferroelectricity. 
\end{abstract}

\author{D. S. Annenkov}
\affiliation{L. D. Landau Institute for Theoretical Physics, Chernogolovka 142432, Russia}
\affiliation{Moscow Institute of Physics and Technology (National Research University), Dolgoprudnyi, Moscow region, 141701 Russia}
\author{A. A. Kopasov}
\affiliation{National University of Science and Technology ``MISIS'', Moscow 119049, Russia}
\author{A. S. Mel'nikov}
\affiliation{Moscow Institute of Physics and Technology (National Research University), Dolgoprudnyi, Moscow region, 141701 Russia}
\affiliation{Institute for Physics of Microstructures, Russian Academy of Sciences, 603950 Nizhny Novgorod, GSP-105, Russia}

\maketitle

\section{Introduction}

The physics of van der Waals heterostructures attracts a significant attention of both experimentalists and theoreticians~\cite{GeimN2013,DasARMR2015,LiuNRM2016,NovoselovS2016,RobinsonACSN2016,SongNN2018} due to the fact that these systems offer a unique possibility to study the interplay between different phases of matter including superconductivity~\cite{StaleyPRB2009}, ferromagnetism~\cite{DengN2018,GongN2017,HuangN2017,GibertiniNN2019} and ferroelectricity~\cite{WangNM2023,ZhangNRM2023,LiAM2024}. The electronic band structure of such materials and the properties of the emergent phases are known to be highly sensitive to the layer stacking, which can be engineered during the sample fabrication through a relative layer twist~\cite{CarrPRB2017,HennighausenES2021}. In addition, changes in the layer stacking can be triggered by the external electric field or stress leading to the so-called interlayer sliding~\cite{FeiN2018,SternS2021,YasudaS2021}. For several stackings characterized by the intrinsic charge imbalance between the layers (in the presence of a spontaneous electric polarization directed perpendicular to the plane of the layers), the application of the external electric field can switch the layer stacking and the direction of the electric dipole moment. This phenomenon, known as the sliding ferroelectricity, was experimentally observed in a number of systems including bilayer and trilayer WTe$_2$~\cite{FeiN2018} and AB/BA stacking h-BN~\cite{SternS2021,YasudaS2021}.

Recent experimental results for bilayer T$_{\rm d}$-MoTe$_2$, which exhibits both ferroelectricity and superconductivity, demonstrate that the sliding ferroelectricity can provide the mechanism for electrostatic control over the superconducting state~\cite{JindalN2023}. In particular, it was shown that the critical temperature of the superconducting transition $T_c$ strongly depends on the transverse electric field, which governs the switching between the two states with homogeneous electric polarization (up and down). Rather strong electric fields were found to be detrimental to superconductivity, whereas the increase in $T_c$ was observed upon the decrease in the field magnitude. Such behavior reflects the competition between ferroelectricity and superconductivity and suggests that the superconductivity nucleation in sliding ferroelectrics can be more favorable in the vicinity of ferroelectric domain walls. Another important experimental result for the similar system is related to the effects of the in-plane magnetic field~\cite{LiPRL2024}. It was demonstrated that the in-plane critical magnetic field destroying superconductivity is two-fold anisotropic and both the maximum and minimum critical magnetic fields exceed the Pauli limiting field for sufficiently low temperatures. Such features point to the importance of the spin-dependent effects in studies of the peculiarities of the superconducting state in such systems. Note that currently the experimental behavior of the in-plane critical magnetic field is attributed to the effects of the tilted Ising spin-orbit coupling~\cite{CuiNC2019,RhodesNL2021}.

The underlying microscopic mechanism of superconductivity in van der Waals ferroelectrics is currently under a theoretical investigation, and the complete description of this phenomenon is yet to be established. For instance, recent theoretical study shows that the domain wall fluctuations in sliding ferroelectrics can provide rather strong intralayer electronic attraction~\cite{ChaudharyPRL2024}. At the same time the above-mentioned experimental observations motivated us to study an alternative scenario of the Cooper pair formation in sliding ferroelectrics, namely, \textit{the interlayer pairing}. Note that this idea was first suggested in a seminal paper of M.~H.~Cohen and D.~H.~Douglass, Jr.~\cite{CohenPRL1967}, later by K.~B.~Efetov and A.~I.~Larkin~\cite{EfetovJETP1975}, and then further developed for layered superconductors~\cite{EfetovJETP1975,TesanovicPRB1987,BulaevskiiPRB1990,KlemmLiu,LiuKlemm,KettemannPRB1992} and van der Waals heterostructures~\cite{HosseiniPRL2012,HosseiniPRB2012,LiuPRL2017,AlidoustPRB2019,KopasovPRB2024}.

\begin{figure}[htpb]
\centering
\includegraphics[scale = 0.19]{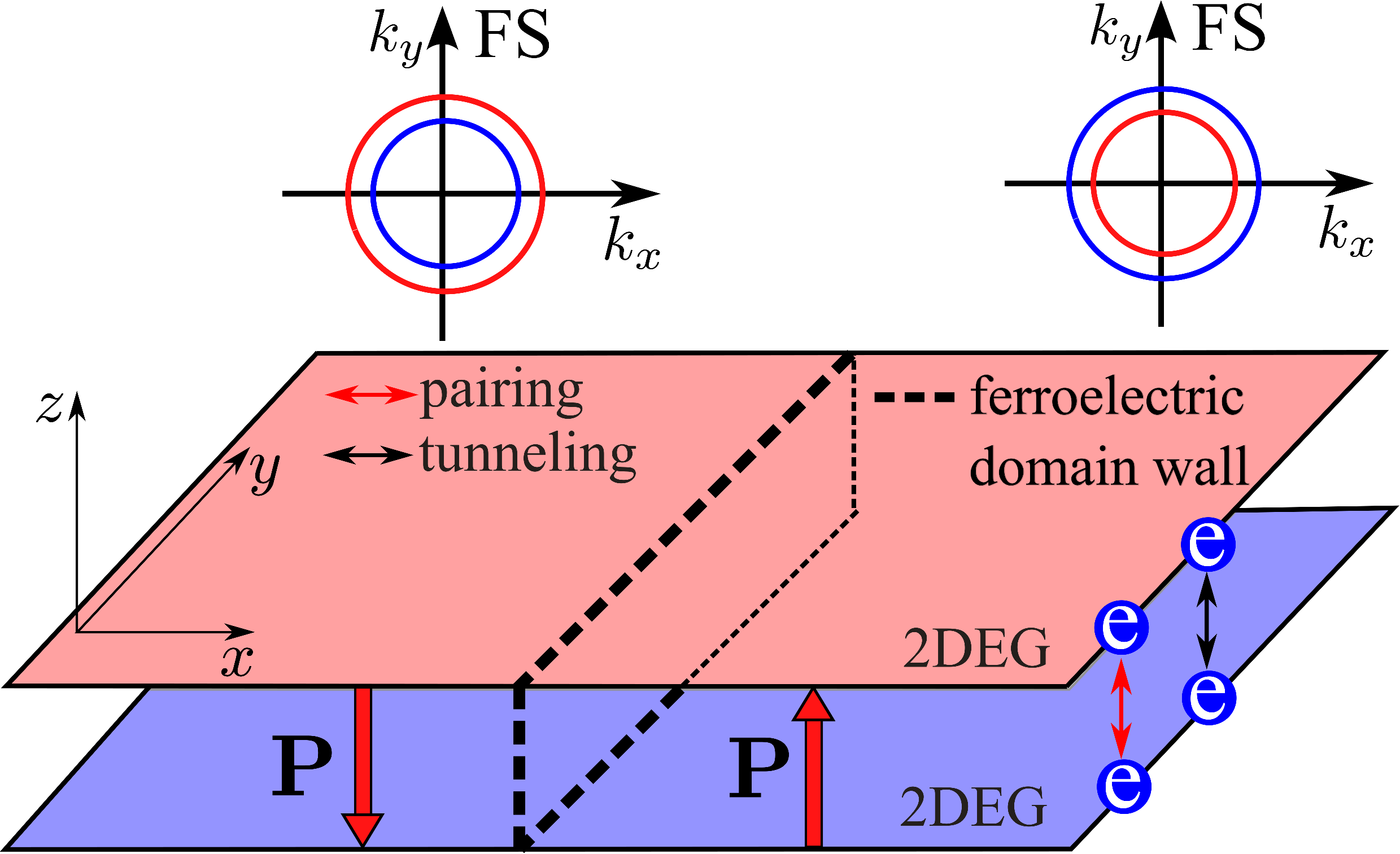}
\caption{Schematic picture of the considered tunnel coupled bilayer system with the interlayer electronic attraction. Dashed lines show the position of the ferroelectric domain wall, which separates the region with homogeneous polarization $\mathbf{P}$ directed perpendicular to the plane of the layers. Upper plots: schematic pictures of the Fermi surfaces (FS) far from the domain wall.}
\label{Fig:system_schematic}
\end{figure}

The motivation for the study of the interlayer pairing in relation to the sliding ferroelectricity can be simply clarified if one notices that the charge imbalance between the layers (equivalent to the electric polarization) can be described by a relative shift of the conduction energy bands in different layers [see Fig.~\ref{Fig:system_schematic}]. For the Cooper pairs formed by electrons from different layers, the effect of the relative band shift is known to be similar to the one of the Zeeman field on the conventional spin-singlet superconductivity~\cite{ChandrasekharAPL1962,ClogstonPRL1962,SaintJames}. The increase in the band shift (increase in the electric dipole moment) should cause the decrease in $T_c$ of the interlayer superconductivity (see, e.g., Ref.~\cite{KopasovPRB2024}). So, the ferroelectricity naturally competes with the interlayer superconductivity and one can expect that possible interlayer superconductivity can be enhanced in the vicinity of ferroelectric domain walls (in spatial regions with vanishing relative band shift). Regarding the spin-dependent effects, it is important to note that the Pauli principle, which usually hampers the formation of the spin-triplet Cooper pairs in systems with the local effective attraction, does no more impose severe restrictions on the spins of electrons in the interlayer pair~\cite{EfetovJETP1975}. The depairing effect of the in-plane magnetic field is, in turn, sensitive to the spin structure of Cooper pairs~\cite{MineevSamokhinBook} and can possess the two-fold anisotropy for the spin-triplet superconducting state. Despite a lot of theoretical works devoted to the study of the interlayer superconductivity, quantitative theoretical studies of the above mentioned phenomena is still lacking. The main goal of our work is to fill these gaps and provide an extensive analysis of the effects associated with the interplay between superconductivity and ferroelectricity.

In this manuscript we study the distinctive features of the interplay between the ferroelectricity and interlayer superconductivity in van der Waals bilayers [see Fig.~\ref{Fig:system_schematic}]. Our analysis is based on the formalism of the Eilenberger equations, which we derive for a model tunnel coupled bilayer with a finite relative shift of the conducting bands and an arbitrary spin structure of the interlayer gap function (both the spin-singlet and spin-triplet cases are considered). It is shown that the critical temperature for the superconductivity nucleation in the vicinity of the ferroelectric domain wall (in spatial region with vanishing relative band shift) can significantly exceed the critical temperature of the homogeneous superconducting state. Using analytical and numerical solutions of the linearized self-consistency equation, we find that the increase in the tunnel coupling leads to the decrease (increase) in the difference between critical temperatures of localized and homogeneous states for the spin-singlet (spin-triplet) interlayer superconductivity. As a next step, we carry out the extensive analysis of the effects of the in-plane magnetic field on the homogeneous interlayer superconducting states. Both the orbital and paramagnetic mechanisms have been taken into account. We find that the main features of the orbital effect on the interlayer superconductivity can be explained by an effective renormalization of the tunneling amplitude. It is demonstrated that the spin-singlet interlayer superconductivity is suppressed by the orbital effect and the magnitude of such suppression depends on the tunneling amplitude and the relative band shift. On the contrary, we demonstrate that the spin-triplet superconductivity can be enhanced by the orbital effect. For both spin-singlet and spin-triplet interlayer superconductivity we find that paramagnetic effect can result not only in the suppression of superconductivity with the increasing field but also in the emergence of reentrant superconducting phases, which appear due to a compensation of the relative band shift by the Zeeman shift~\cite{BuzdinPRL2005,MontielPRB2011} (see also the analysis of phase diagrams of the van der Waals superconductor - ferromagnet structures in Refs.~\cite{BokaiPRB2024,IanovskaiaPRB2024}). We demonstrate that the joint influence of the orbital and paramagnetic effect can even lead to a nonmonotonic behavior of the critical temperature of the spin-triplet interlayer superconductivity as a function of the external magnetic field. 
   
Note that the model of interlayer pairing considered in our work is purely phenomenological and we do not analyze here the microscopic mechanism of the electronic attraction underlying the formation of interlayer Cooper pairs. Generally such mechanism can be associated, for instance, with the interaction of electrons with the phonon modes involving coupled atomic displacements in neighboring layers~\cite{CohenPRL1967}, renormalization of the interlayer Coulomb interaction~\cite{SboychakovPE2025}, and the exchange of magnons in structures involving additional antiferromagnetic layers~\cite{ThingstadPRB2025}. Irrespective to the particular attraction mechanism we study here the consequences of interlayer pairing itself and consider its relevance for the experimental data in van der Waals bilayer systems.

This manuscript is organized as follows. In Sec.~\ref{section_model_equations} we present the model and basic equations. In Sec.~\ref{section_superconductivity_nucleation} we briefly discuss the main features of the superconductivity nucleation for the spin-singlet and spin-triplet interlayer superconductivity, which are useful for understanding of our main results. In Sec.~\ref{localized_superconductivity_section} we carry out the analysis of the localized nucleation of interlayer superconductivity on a ferroelectric domain wall.  In Sec.~\ref{effects_of_the_inplane_field_section} we study the effects of the in-plane magnetic field on spin-singlet and spin-triplet interlayer superconducting states. In Sec.~\ref{discussion_section} we discuss our results in context of recent experiments on van der Waals bilayers with coexisting superconductivity and ferroelectricity. Finally, the results are summarized in Sec.~\ref{conclusion_section}.

\section{Model and basic equations}\label{section_model_equations}

Hereafter, we consider a two-layer model similar to the one used in Ref.~\cite{KopasovPRB2024} with tunnel coupling and the interlayer electronic attraction [see Fig.~\ref{Fig:system_schematic}]. For simplicity, we neglect here the intralayer Cooper pairs. 
It is important to
note that the considered model possesses some similarities with the model of a two-band superconductor with a cross-band pairing
(or simply crosspairing)~\cite{VargasPRB2020,MideiPRB2023}. The main difference between them is the tunable band hybridization in the two-layer model described by a finite hopping parameter. In a real experimental situation, the tunability of the layer coupling can be achieved by changing the thickness of the intermediate insulating layer or by pressure~\cite{JayaramanRMP1983,MaNL2021,NayakACSN2015,PeiACSN2022,PeiMD2022,JiaoRPP2023}. We expect that in the presence of intralayer pairing, two order parameters (interlayer and intralayer) will compete in analogy to the multiband superconductors with competing intra-band and cross-band pairings (see, e.g., Ref.~\cite{VargasPRB2020}).

First, we present the Hamiltonian of the system and the set of the Gor'kov equations. Then we provide the quasiclassical Eilenberger equations, which form the basis of our theoretical analysis. For brevity, the main part of the discussion of the formalism is carried out disregarding the paramagnetic effect of the external magnetic field. The generalization of the derived Eilenberger equations to the case of a finite Zeeman field is straightforward and is given in the end of this section.

The Hamiltonian of the bilayer structure can be written in the following form:
\begin{equation}\label{system_Hamiltonian}
H=H_{1}+H_{2}+H_{t}+H_{\text{int}},
\end{equation}
where
\begin{equation}
H_{j}=\int d^{2}\mathbf{r} \ \psi_{j\sigma}^{\dagger}\left(\mathbf{x}\right)\hat{\xi}_{j}\psi_{j\sigma}\left(\mathbf{x}\right) \ ,
\end{equation}
describe isolated 2D layers labeled by the subscript $j = 1,2$, $\psi_{j\sigma}^{\dagger}\left(\mathbf{x}\right)\,\left(\psi_{j\sigma}\left(\mathbf{x}\right)\right)$ are fermionic creation (annihilation) operators in the layer $j$
in the Matsubara representation \cite{Kopnin_book, AGD_book}, $\mathbf{x}=\left(\mathbf{r},\tau\right)$ where
$\tau$ is the imaginary time variable in the Matsubara technique, $\sigma=\uparrow,\downarrow$ denotes
spin degree of freedom (summation over repeated spin indices is implied), $\hat{\xi}_{j}=\mathbf{P}_{j}^{2}/2m - \mu + U_{j}\left(\mathbf{r}\right)$, $\mathbf{P}_{j}=-i\nabla_{\mathbf{r}}-e\mathbf{A}_{j}/c$
is the kinetic momentum operator, $\mathbf{A}$ is the vector potential, $e$ denotes the electron charge, $c$ is the speed of light, $m$ is the effective mass, and $\mu$ is the chemical potential. The potential
$U_{j}\left(\mathbf{r}\right)$ is introduced to take into account (possibly inhomogeneous) relative band shift of the conducting bands induced by the ferroelectric polarization between layers. The last two terms in Eq.~(\ref{system_Hamiltonian}) describe the momentum-conserving tunneling
\begin{equation}
H_{t}=\int d^{2}\mathbf{r}\left[t\psi_{1\sigma}^{\dagger}\left(\mathbf{x}\right)\psi_{2\sigma}\left(\mathbf{x}\right)+t^{*}\psi_{2\sigma}^{\dagger}\left(\mathbf{x}\right)\psi_{1\sigma}\left(\mathbf{x}\right)\right] \
\end{equation}
and the interlayer electron-electron interaction
\begin{equation}
H_{\text{int}}=\frac{U_0}{2}\int d^{2}\mathbf{r} \ \psi_{1\sigma}^{\dagger}\left(\mathbf{x}\right)\psi_{2\sigma'}^{\dagger}\left(\mathbf{x}\right)\psi_{2\sigma'}\left(\mathbf{x}\right)\psi_{1\sigma}\left(\mathbf{x}\right).
\end{equation}
Here $t$ is the tunneling amplitude and an attractive electron-electron interaction and attractive spin-independent interaction with $U_0 = -|U_0|$ is assumed. Treating the interaction within the mean-field approximation and introducing the decoupling fields
\begin{equation}
\left[\hat{\Delta}_{\rm int}(\mathbf{r})\right]_{\sigma\sigma'} = -\frac{U_0}{2}\left\langle\psi_{1\sigma}(\mathbf{x})\psi_{2\sigma'}(\mathbf{x})\right\rangle \ ,
\end{equation}
we obtain the following form of the effective interaction (see also Ref.~\cite{KopasovPRB2024}):
\begin{eqnarray}
H_{\rm eff} = \int d^2\mathbf{r}\biggl[\frac{2}{|U_0|}{\rm Tr}\left(\hat{\Delta}_{\rm int}\hat{\Delta}_{\rm int}^{\dagger}\right) \\
\nonumber
 + (\hat{\Delta}_{\rm int})_{\sigma\sigma'}\psi_{1\sigma}^{\dagger}(\mathbf{x})\psi^{\dagger}_{2\sigma'}(\mathbf{x}) 
+ (\hat{\Delta}_{\rm int}^*)_{\sigma\sigma'}(\mathbf{x})\psi_{2\sigma'}\psi_{1\sigma}(\mathbf{x})
\biggl].
\end{eqnarray}

Our analysis is based on the Matsubara Green function formalism. The Green's function of the system represents $8\times8$ matrix in the generalized layer - particle-hole (Nambu) - spin space
\begin{equation}\label{Gorkov_GF_definition}
\underline{G}\left(\mathbf{x}_{1},\mathbf{x}_{2}\right)=\left\langle T_{\tau}\psi\left(\mathbf{x}_{1}\right)\psi^{\dagger}\left(\mathbf{x}_{2}\right)\right\rangle \ ,
\end{equation}
where angular brackets stand for thermodynamic average, $T_{\tau}$ is
the time ordering operator, and $\psi=\left(\psi_{1\uparrow},\,\psi_{1\downarrow},\,\psi_{1\uparrow}^{\dagger},\,\psi_{1\downarrow}^{\dagger},\,\psi_{2\uparrow},\,\psi_{2\downarrow},\,\psi_{2\uparrow}^{\dagger},\,\psi_{2\downarrow}^{\dagger}\right)^{T}$. The Green's
function~(\ref{Gorkov_GF_definition}) has the following structure in the layer and particle - hole
spaces
\begin{equation}
\underline{G}=\left(\begin{array}{cc}
\check{G}_{11} & \check{G}_{12}\\
\check{G}_{21} & \check{G}_{22}
\end{array}\right),\,\check{G}_{ij}=\left(\begin{array}{cc}
\hat{G}_{ij} & \hat{F}_{ij}\\
\hat{\overline{F}}_{ij} & \hat{\overline{G}}_{ij}
\end{array}\right) \ ,
\end{equation}
respectively. In the following we also introduce three sets of the Pauli matrices $\underline{\eta}_i$, $\check{\tau}_i$, and $\hat{\sigma}_i$ ($i = x,y,z$) corresponding to the layer, particle-hole (Nambu), and spin space, respectively. More specifically, in the following text 8$\times$8 matrices acting in the generalized layer-Nambu-spin space are denoted by underscores, 4$\times$4 matrices acting in the Nambu-spin space are denoted by inverse hats, and 2$\times$2 matrices acting only in the spin space are denoted by hats. Using the standard procedure~\cite{Svidzinski_book, Kopnin_book}, one obtains the following system of the Gor'kov equations
in the Matsubara frequency - coordinate representation (see also Ref.~\cite{KopasovPRB2024} and textbooks \cite{Kopnin_book, AGD_book})
\begin{widetext} 
\begin{equation}\label{Gorkov_equations}
\left(\begin{array}{cc}
\begin{array}{cc}
-i\omega_{n}+\hat{\xi}_{1}(\mathbf{r}) & 0\\
0 & -i\omega_{n}-\hat{{\xi}}^*_{1}(\mathbf{r})
\end{array} & \check{t}(\mathbf{r})\\
\check{t}^{\dagger}(\mathbf{r}) & \begin{array}{cc}
-i\omega_{n}+\hat{\xi}_{2}(\mathbf{r}) & 0\\
0 & -i\omega_{n}-\hat{{\xi}}^*_{2}(\mathbf{r})
\end{array}
\end{array}\right)\left(\begin{array}{cc}
\check{G}_{11}(\mathbf{r},\mathbf{r}') & \check{G}_{12}(\mathbf{r},\mathbf{r}')\\
\check{G}_{21}(\mathbf{r},\mathbf{r}') & \check{G}_{22}(\mathbf{r},\mathbf{r}')
\end{array}\right)=\delta\left(\mathbf{r}-\mathbf{r}'\right) \ .
\end{equation}
\end{widetext}
Here $\omega_n = 2\pi T(n+1/2)$ are Matsubara frequencies, $T$ is temperature, $n$ is an integer, asterisk symbol stands for complex conjugation, and, hereafter, we omit the frequency arguments of the Green's functions for brevity. The interlayer coupling matrix
\begin{equation}
\check{t}(\mathbf{r})=\left(\begin{array}{cc}
t & \hat{\Delta}_{\text{int}}(\mathbf{r})\\
-\hat{\Delta}_{\text{int}}^{*}(\mathbf{r}) & -t^{*}
\end{array}\right),
\end{equation}
contains the interlayer gap function $\hat{\Delta}_{\rm int}$, which satisfies
the self-consistency equation \cite{Kopnin_book, AGD_book, MineevSamokhinBook}:
\begin{equation}\label{self_cons_Gorkov}
\hat{\Delta}_{\text{int}}\left(\mathbf{r}\right)=-\frac{U_0}{2}T\sum_{\omega_{n}}\hat{F}_{12}\left(\mathbf{r},\mathbf{r}\right) \ ,
\end{equation}
where the anomalous Green's function is taken at coinciding arguments $\textbf{r}=\textbf{r}'$. In the present work we consider both the spin-singlet and spin-triplet interlayer pairing described by the following forms of the gap function
\begin{subequations}
    \begin{align}
        \hat{\Delta}_{\rm int}(\mathbf{r}) = \Delta_{\rm int}(\mathbf{r})(i\hat{\sigma}_y) \ , \\
        \hat{\Delta}_{\rm int}(\mathbf{r}) = \Delta_{\rm int}(\mathbf{r})\mathbf{d}\hat{\boldsymbol{\sigma}}(i\hat{\sigma}_y) \ ,
    \end{align}
\end{subequations}
respectively. Here $\hat{\boldsymbol{\sigma}}$ is the vector of Pauli matrices in the spin space and the $\textbf{d}$ vector defines the spin structure of the spin-triplet Cooper pairs~\cite{MineevSamokhinBook}. In the following consideration we restrict ourselves to superconducting phases, which do not break time-reversal symmetry.

In the present work we consider the case when the tunneling amplitude, typical relative band shift as well as the interlayer gap function are much less than the Fermi energy in both layers, which allows one to use the standard quasiclassical approach~\cite{Svidzinski_book,Kopnin_book} for the description of the superconducting state. Thus, in the considered case the Gor'kov equations~(\ref{Gorkov_equations}) can be transformed into the transport-like Eilenberger equations for the quasiclassical Green's function $\underline{g}$ (more details on the transition to the quasiclassical theory can be found in Appendix~\ref{Eilenberger_derivation}), defined
as 
\begin{equation}\label{quasiclassical_GF_definition}
\underline{g}\left(\mathbf{R},\mathbf{n}\right)=-\int\frac{d\xi}{i\pi}\check{\tau}_z\underline{G}\left(\mathbf{R},\mathbf{n},\xi \right) \ .
\end{equation}
In the above expression the integration is performed over the quasiparticle energy near the Fermi surface,
$\mathbf{R} = (\mathbf{r} + \mathbf{r}')/2$, the unit vector $\mathbf{n}$ parameterizes the momentum direction $\mathbf{p}$ near the Fermi surface $\mathbf{p} = \mathbf{n}(p_F + \xi/v_F)$, $\xi = p^2/2m-\mu$, $p_F$ is the Fermi momentum, $v_F = p_F/m$ is the Fermi velocity, and $\underline{G}\left(\mathbf{R},\mathbf{n},\xi\right)$
is the Green's function in the mixed representation (Wigner-transformed function). 
The resulting Eilenberger equations together with the self-consistency
equation can be written as (see Appendix~\ref{Eilenberger_derivation} for details of the derivation and cf. Ref.~\cite{BobkovPRB2025} for the case of intralayer pairing):
\begin{widetext}
\begin{subequations}\label{eilenberger_self_cons}
\begin{align}\label{eq:eilenberger_main}
-iv_{F}\mathbf{n}{\nabla}_{\mathbf{R}}\underline{g}-i\omega_{n}\left[\check{\tau}_{z},\underline{g}\right]+\left[\frac{1}{2}\left(\check{t}\check{\tau}_{z}\underline{\eta_{+}}+\check{t}^{\dagger}\check{\tau}_{z}\underline{\eta_{-}}\right),\underline{g}\right]+\left(\begin{array}{cc}
0 & U_{12}\left(\mathbf{R}\right)\check{g}_{12}\\
-U_{12}\left(\mathbf{R}\right)\check{g}_{21} & 0
\end{array}\right)+ \\
\nonumber
+\left(\begin{array}{cc}
-\frac{e}{c}v_{F}\mathbf{n}\mathbf{A}_{1}\left[\check{\tau}_{z},\check{g}_{11}\right] & -\frac{e}{c}v_{F}\mathbf{n}\left(\mathbf{A}_{1}\check{\tau}_{z}\check{g}_{12}-\mathbf{A}_{2}\check{g}_{12}\check{\tau}_{z}\right)\\
-\frac{e}{c}v_{F}\mathbf{n}\left(\mathbf{A}_{2}\check{\tau}_{z}\check{g}_{21} - \mathbf{A}_{1}\check{g}_{21}\check{\tau}_{z}\right) & -\frac{e}{c}v_{F}\mathbf{n}\mathbf{A}_{2}\left[\check{\tau}_{z},\check{g}_{22}\right]
\end{array}\right)=0 \ ,\\
\hat{\Delta}_{\text{int}}\left(\mathbf{R}\right)= -i\pi \lambda T\sum_{\omega_{n}}\int\frac{d\mathbf{n}}{2\pi}\hat{f}_{12}\left(\mathbf{R},\mathbf{n}\right) \ .\label{eq:self_cons_main}
\end{align}
\end{subequations}
\end{widetext}
Here $\underline{\eta}_{\pm}=\underline{\eta}_{x}\pm i\underline{\eta}_{y}$, $U_{12}\left(\mathbf{R}\right) = U_{1}\left(\mathbf{R}\right)-U_{2}\left(\mathbf{R}\right)$, $\lambda = |U_{0}|\nu_{F}/2$ is the dimensionless pairing constant, and
$\nu_F = m/2\pi$ is the density of states at the Fermi level per spin projection. To address the paramagnetic effects of the external magnetic field $\mathbf{H}$, one should make the following replacement $-i\omega_n\check{\tau}_z \to (-i \omega_n\check{\tau}_z + \check{M})$ in Eq.~(\ref{eq:eilenberger_main}), where 
\begin{equation}
\check{M} = \begin{pmatrix}\mathbf{h}\hat{\boldsymbol{\sigma}}&0\\0&\mathbf{h}\hat{\boldsymbol{\sigma}}^*\end{pmatrix} \ ,
\end{equation}
and $\mathbf{h} = g\mu_B\mathbf{H}/2$, $g$ is the Land\'{e} $g$-factor, $\mu_B$ is the Bohr magneton. The Eilenberger equations~(\ref{eq:eilenberger_main}) should be supplemented with the normalization condition $(\underline{\eta}_z\underline{g})^2 = \underline{1}$ (see Appendix~\ref{Eilenberger_derivation} for the corresponding discussion). Eqs.~(\ref{eq:eilenberger_main}) and (\ref{eq:self_cons_main}) form the basis of our theoretical analysis of the interlayer pairing
effects.

\section{Homogeneous superconducting state}\label{section_superconductivity_nucleation}

A lot of insights into the physics of the interlayer superconductivity effects can be obtained from the analysis of the distinctive features of the superconductivity nucleation for a homogeneous state, see also \cite{KopasovPRB2024}. Here we briefly discuss the effects of the electron tunneling and homogeneous relative band shift on the critical temperature of the spin-singlet and spin-triplet interlayer superconductivity.

\begin{figure*}[htpb]
\centering
\includegraphics[scale = 0.9]{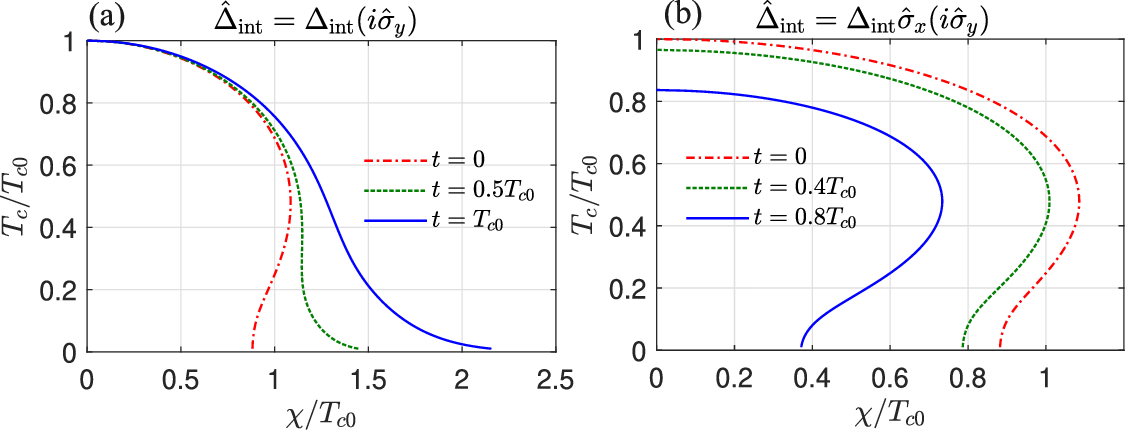}
\caption{Typical dependencies of the critical temperature of the interlayer superconductivity $T_c$ for homogeneous spin-singlet (a) and spin-triplet (b) states vs. the relative band shift $U_1 - U_2 = -2\chi$. Here $T_{c0}$ is the critical temperature for $t = 0$ and $\chi = 0$.}
\label{Fig:Tc_homogeneous_illustration}
\end{figure*}

Considering, for instance, a homogeneous spin-singlet interlayer pairing $\hat{\Delta}_{\rm int} = \Delta_{\rm int}(i\hat{\sigma}_y)$, homogeneous relative band shift $U_1 - U_2 = -2\chi$, and zero external magnetic field, the equation governing the superconducting critical temperature can be written in the following form:
\begin{eqnarray}\label{Tc_homogeneous_equation_spin_singlet}
\log\left(\frac{T}{T_{c0}}\right) = \\
\nonumber
2\pi T {\rm Re} \sum_{\omega_n > 0}\left[\frac{t^2}{\varkappa^2\omega_n} + \frac{\chi^2}{\varkappa^2}\frac{1}{(\omega_n - i\varkappa)}-\frac{1}{\omega_n}\right] \ ,
\end{eqnarray}
where $\varkappa = \sqrt{t^2 + \chi^2}$ and $T_{c0}$ denotes the critical temperature for $\chi = 0$ and $t = 0$. The above linearized self-consistency equation allows an analytical solution in the vicinity of $T_{c0}$. Within this parameter range the solution of Eq.~(\ref{Tc_homogeneous_equation_spin_singlet}) can be expanded over small parameters $\chi/T_{c0}$, $t/T_{c0}$, $\varkappa/T_{c0}\ll 1$. The resulting expansion can be cast to the form

\begin{subequations}
\begin{align}
\label{Tc_homogeneous_result_spin_singlet}
[T_c(\chi,t) - T_{c0}]/T_{c0} \approx c_1\tilde{\varkappa}^2 
 + c_2\tilde{\varkappa}^4 + c_3\tilde{\varkappa}^6 \ ,\\
c_1 = \chi^2\psi_2/2\varkappa^2 \ ,\\
 c_2 = c_1^2/2 + \chi^2\left(-c_1\psi_2 - \psi_4/4!\right)/\varkappa^2 \ ,\\
c_3 = -c_1^3/3 + c_1c_2 + \chi^2(3c_1^2 - 2c_2)\psi_2/2\varkappa^2 \\
\nonumber
 +\chi^2\left(c_1\psi_4/3! + \psi_6/6!\right)/\varkappa^2 \ .
\end{align}
\end{subequations}
Here $\tilde{\varkappa} = \varkappa/2\pi T_{c0}$, $\psi_n = \psi_n(1/2)$ and $\psi_n(x) = d^n\psi(x)/dx^n$ is the $n$-th derivative of the digamma function $\psi(x)$~\cite{Abramowitz_book}. Typical dependencies of $T_c$ vs. the relative band shift $\chi$ for several tunnel amplitudes $t/T_{c0} = 0$, 0.5, and 1 are presented in Fig.~\ref{Fig:Tc_homogeneous_illustration}(a). These plots clearly show that ferroelectricity competes with the interlayer superconductivity: the relative band shift $\chi$ suppresses the spin-singlet interlayer superconductivity. The increase in the tunneling amplitude $t$ stabilizes the superconducting state at larger band splittings.

The distinctive features of the superconductivity nucleation for the spin-triplet superconducting state are qualitatively different from the ones discussed above. In the absence of the external magnetic field, the superconducting critical temperature does not depend on the particular spin structure of the triplet Cooper pairs, for definiteness we choose $\hat{\Delta}_{\rm int} = \Delta_{\rm int}\hat{\sigma}_x(i\hat{\sigma}_y)$. Taking a homogeneous relative band shift and zero external magnetic field in the Eqs.~(\ref{eilenberger_self_cons}), one finds the following equation for the critical temperature of the interlayer spin-triplet superconducting state:
\begin{equation}\label{Tc_homogeneous_equation_spin_triplet}
\ln\left(\frac{T}{T_{c0}}\right) = 2\pi T\sum_{\omega_n>0}\left[\frac{\omega_n}{\omega_n^2 + \chi^2 + |t|^2}-\frac{1}{\omega_n}\right] \ .
\end{equation}
In the vicinity of $T_{c0}$ the solution for the critical temperature of the interlayer spin-triplet superconducting state can be presented in the form of expansion
\begin{subequations}
\begin{align}
\label{Tc_homogeneous_result_spin_triplet}
[T_c(\chi,t) - T_{c0}]/T_{c0} \approx c_1\tilde{\varkappa}^2 + c_2\tilde{\varkappa}^4 + c_3\tilde{\varkappa}^6 \ ,\\
c_1 = \psi_2/2 \ ,\\
c_2 = c_1^2/2 -c_1\psi_2 - \psi_4/4! \ ,\\
c_3 = -c_1^3/3 + c_1c_2 + (3c_1^2 - 2c_2)\psi_2/2\\
\nonumber
 + c_1\psi_4/3! + \psi_6/6! \ .
\end{align}
\end{subequations}
Typical behavior of the critical temperature for the spin-triplet state~(\ref{Tc_homogeneous_equation_spin_triplet}) is shown in Fig.~\ref{Fig:Tc_homogeneous_illustration}(b). Direct comparison with the case of the spin-singlet pairing reveals that the spin structure of the interlayer superconductivity does not affect the behavior of $T_c(\chi)$ for decoupled layers ($t = 0$). The difference between $T_c(\chi,t)$ behavior for spin-singlet and spin-triplet states manifests itself only for a finite tunnel coupling. The plots in Fig.~\ref{Fig:Tc_homogeneous_illustration}(b) clearly demonstrates that the tunneling as well as the relative band shift suppresses the spin-triplet interlayer superconductivity. Such drastic difference in the influence of tunneling comparing with the spin-singlet case is the consequence of the Pauli principle.
The tunneling results in hybridization of electronic states in different layers and formation of two states with a certain energy splitting, which grows with increasing tunneling amplitude. The Pauli exclusion principle forbids the formation of the spin-triplet Cooper pairs, which have the same orbital structure. The spin-triplet Cooper pairs forming from electrons, which possess different orbital structure still can exist. The superconducting critical temperature for the latter pairs should obviously be suppressed by the increasing tunneling due to the splitting of corresponding energy levels of different orbitals mentioned above (somewhat similar to the $T_c$ suppression by the band shift).

Note that some of the $T_c\left(\chi\right)$ plots in Fig.~\ref{Fig:Tc_homogeneous_illustration} are multivalued, which reflects the possibility of the change in the order of the phase transition. The detailed analysis of this issue requires comparison between free energies of the normal state and the superconducting states with zero and finite momentum and is behind the scope of this work. Nevertheless, qualitative structure of the phase diagram can be understood from the analogy of the considered problem with the case of the spin-split superconductor~\cite{SaintJames}. So, if we restrict ourselves to the states with zero momentum then the typical situation is that the two branches of the phase diagram in Fig.~\ref{Fig:Tc_homogeneous_illustration}(b) (with $\partial\chi/\partial T >0$ and $\partial\chi/\partial T <0$) should cross at the tricritical point $T_*(\chi)$: the superconducting phase transition should be of the second (first) order for $T_c>T_*$($T_c<T_*$), and the branch with $\partial \chi/\partial T > 0$ corresponds to the line of instability of the nonsuperconducting phase for the first order phase transition (indicates the temperature dependences of the over-cooling relative band shift).

\section{Superconducting states localized at ferroelectric domain walls}\label{localized_superconductivity_section}

In this section we apply the derived formalism of Eilenberger equations
to a bilayer structure with an isolated ferroelectric
domain wall between layers and without the external magnetic field. The ferroelectric
polarization can be considered as a relative shift of the conduction
bands of different layers, so in its presence one can expect suppression
of interlayer superconducting correlations, but in the region of domain
walls the effective polarization is partially averaged out on a scale of superconducting nucleus, which can make localized
superconducting states more favorable. It reminds the situation in ferromagnetic superconductors or
superconductor/ferromagnet bilayers, where localized superconducting
nuclei are formed in the vicinity of ferromagnetic domain walls~\cite{BuzdinZhETP1984,HouzetPRB2006}. It is worth noting, that in the
absence of tunneling between layers $\left(t=0\right)$ such analogy
is exact under the corresponding replacement of layer indices with
spin indices.

\begin{figure*}[htpb]
\centering
\includegraphics[scale = 0.72]{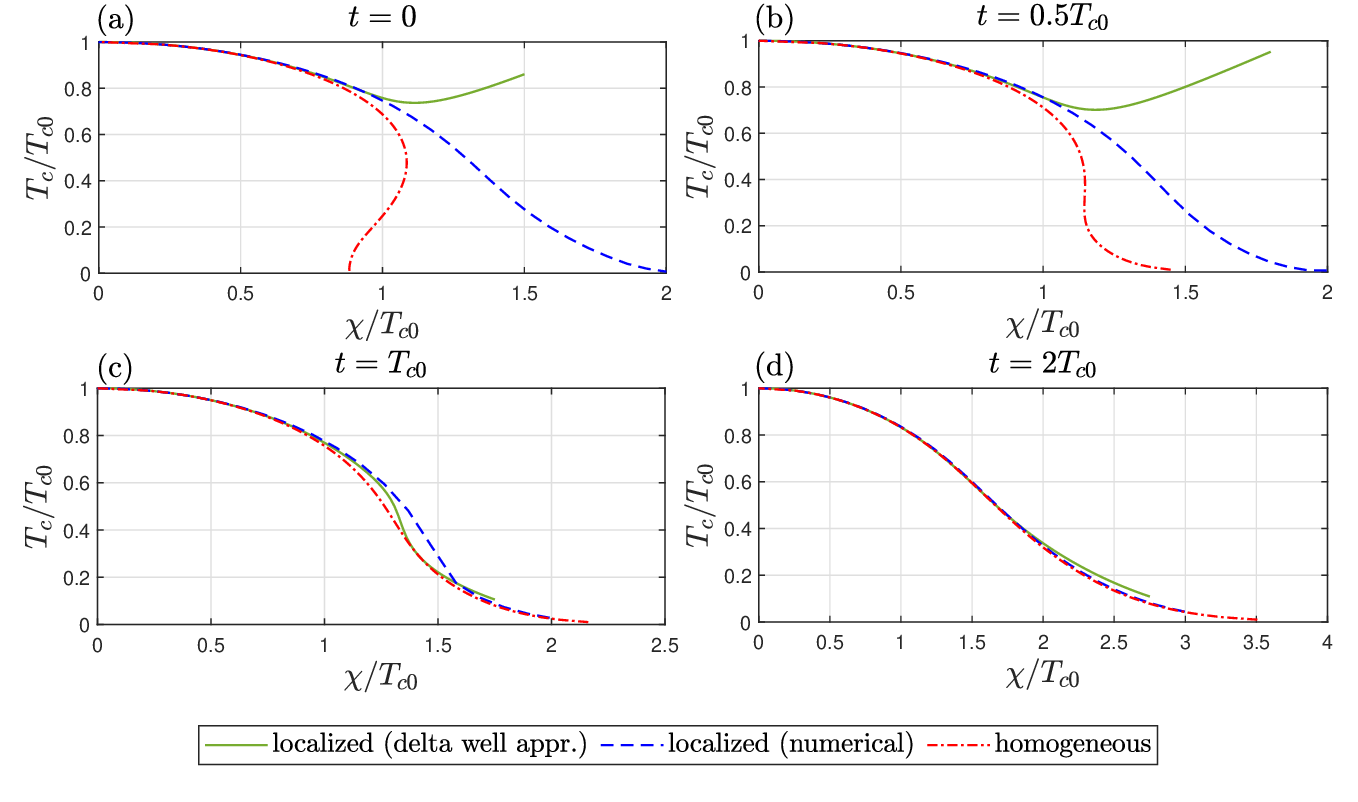}
\caption{Phase diagrams for different tunneling amplitudes in the case of a spin-singlet order parameter. Green solid and blue dashed lines correspond to domain wall superconductivity, but were obtained in different ways: by direct numerical solution of the Eqs.~(\ref{self_cons_domain_wall_main_text}) or by numerical solution of Eq.~(\ref{Tc_localized_equation}), obtained in delta well approximation, respectively. Red dashed-dotted lines are critical lines of the transition into uniform superconducting state. Panels (a), (b), (c) and (d) correspond to $t=0$, $t = 0.5T_{c0}$, $t = T_{c0}$ and $t = 2T_{c0}$ respectively.}
\label{Fig:dw_singlet_plots}
\end{figure*}

The domain walls in van der Waals structures can be identified with dislocations \cite{EnaldievPRL2020, EngelkePRB2023, Enaldiev2DMAT2024}. Theoretical investigations \cite{Enaldiev2DMAT2024} suggest that the widths of the ferroelectric domain walls in transition metal dichalcogenides are of the atomic scale. This allows us to assume the domain wall to be thin compared to the superconducting coherence length and account effects of the domain wall in Eqs.~(\ref{eq:eilenberger_main}) through the stepwise spatial profile of the relative band shift 
\begin{equation}
U_{12}\left(\mathbf{r}\right)=2\chi \ \text{sgn}(x) \ ,
\end{equation}
where $x$ is the coordinate in the direction perpendicular to the
domain wall. Hereafter, we change the notation $\mathbf{R}\to\mathbf{r}$ in Eq.~(\ref{eq:eilenberger_main}) for convenience. To find the critical temperature, we linearize Eqs.~(\ref{eq:eilenberger_main}) with respect to the superconducting gap.
We assume that the introduced parameters $\chi$ and $t$ are small, compared to the Fermi energy $\left(\chi/E_F,\,t/E_F\ll1\right)$, thus allowing us to neglect the corrections to the normal-state Green's function and also to solve the Eilenberger equations on straight-line quasiparticle trajectories, disregarding the normal reflection of quasiparticles from the interlayer potential inhomogeneity.
So, linearizing the Eq.~(\ref{eq:eilenberger_main}) using the known~\cite{Kopnin_book} normal-state expressions  $g_{11}=g_{22}=\overline{g}_{11}=\overline{g}_{22}=-\text{sgn}(\omega_{n})$, we get the system defining the interlayer anomalous Green's function $\hat{f}_{12}$ 
\begin{widetext}
\begin{subequations}\label{eilenberger_domain_wall_with_spin}
\begin{align}
\left[i\mathbf{v}\nabla_{\mathbf{r}}+2i\omega_{n}-2\chi\text{sgn}(x)\right]\hat{f}_{12} 
-t\left(\hat{f}_{22}-\hat{f}_{11}\right)+2\hat{\Delta}_{\text{int}}\left(\mathbf{r}\right)\text{sgn}(\omega_{n}) = 0 \ ,\\
\left[i\mathbf{v}\nabla_{\mathbf{r}}+2i\omega_{n}+2\chi\text{sgn}(x)\right]\hat{f}_{21} 
+t\left(\hat{f}_{22}-\hat{f}_{11}\right)+2\hat{\Delta}_{\text{int}}\left(\mathbf{r}\right)\text{sgn}(\omega_{n}) = 0 \ ,\\
\left[i\mathbf{v}\nabla_{\mathbf{r}}+2i\omega_{n}\right]\hat{f}_{11}+t\left(\hat{f}_{12}-\hat{f}_{21}\right)=0 \ ,\\
\left[i\mathbf{v}\nabla_{\mathbf{r}}+2i\omega_{n}\right]\hat{f}_{22}-t\left(\hat{f}_{12}-\hat{f}_{21}\right)=0 \ .
\end{align}
\end{subequations}
\end{widetext}
Here we used the fact that in the absence of magnetic field time-reversal symmetry imposes $t=t^{*}$. 

In the following subsections we solve Eqs.~(\ref{eilenberger_domain_wall_with_spin}) for singlet and triplet spin structure of the superconducting gap to obtain the critical temperature of the localized solution formation. We study the influence of the tunneling $t$ and the relative band shift $\chi$ on it in comparison with the critical temperature of the transition into uniform phase, discussed in Sec.~\ref{section_superconductivity_nucleation}.

\subsection{Domain-wall superconductivity in the spin-singlet case}

The spin-singlet interlayer superconducting gap has the structure $\hat{\Delta}_{\rm int}=\Delta_{\rm int}\left(\mathbf{r}\right)\left(i\hat{\sigma}_y\right)$ with
\begin{equation}
    \Delta_{\rm int}\left(\mathbf{r}\right) = \Delta_{\rm int}\left(x\right)e^{iq_yy},
\end{equation}
where $q_{y}$ describes the possible modulation of a localized state along the domain
wall in a way similar to the Fulde–Ferrell–Larkin–Ovchinnikov (FFLO) state. Let us start from the $q_y=0$ case. Substituting $\hat{f}_{ij}=f_{ij}\left(i\hat{\sigma}_y\right)$ and introducing the angle $\varphi$ between the trajectory and the $x-$axis
we rewrite Eqs.~(\ref{eilenberger_domain_wall_with_spin}) as ($v_x=v_{F}\cos\varphi$)
\begin{subequations}\label{eilenberger_domain_wall_without_spin_spin_singlet_case}
\begin{align}
\left[iv_x\frac{d}{dx}+2i\omega_{n}\right]f_{11}+t\left(f_{12}-f_{21}\right)=0 \ ,\\
\left[iv_x\frac{d}{dx}+2i\omega_{n}\right]f_{22}-t\left(f_{12}-f_{21}\right)=0 \ ,\\
\left[iv_x\frac{d}{dx}+2i\omega_{n}-2\chi\text{sgn}(x)\right]f_{12} \\
\nonumber
 - t\left(f_{22}-f_{11}\right)+2\Delta_{\rm int}\left(x\right)\text{sgn}(\omega_{n})=0 \ ,\\
\left[iv_x\frac{d}{dx}+2i\omega_{n}+2\chi\text{sgn}(x)\right]f_{21}\\
\nonumber
+t\left(f_{22}-f_{11}\right)+2\Delta_{\rm int}\left(x\right)\text{sgn}(\omega_{n})=0 \ .
\end{align} 
\end{subequations}
To derive the equation governing the superconducting critical temperature of a localized state, we, first, write down the solutions of the above equations within $x>0$ and $x<0$ regions and then match them continuously at the domain wall (at $x = 0$). As a next step, we substitute the resulting solutions into the self-consistency equation~(\ref{eq:self_cons_main}). Performing these calculations, which are rather lengthy and presented in details in Appendix~\ref{kernel_derivation_appendix}, we obtain the following form of the
self-consistency equation written for the Fourier-transformed order parameter 
\begin{subequations} \label{self_cons_domain_wall_main_text}
    \begin{align}        \Delta_{\text{int}}\left(k\right)=\int_{-\infty}^{+\infty}\frac{dk'}{2\pi}K\left(k,k'\right)\Delta_{\text{int}}\left(k'\right) \ , \label{selfcons_singlet_form} \\ 
        K(k,k')=2\pi\delta\left(k-k'\right)K_h\left(k\right)+K_{\rm inh}\left(k,k'\right). \label{selfcons_kernel_structure}
    \end{align}
\end{subequations}
Explicit expressions for the kernel $K(k,k')$, it's homogeneous $K_h(k)$ and inhomogeneous $K_{\rm inh}\left(k,k'\right)$ parts are given in the Appendix~\ref{kernel_derivation_appendix}

Below we derive analytical expressions for the deviations of the superconducting critical temperature for a localized state $T_{cw}(\chi,t)$ from the one corresponding to the homogeneous case $T_c(\chi,t)$. 
We note that the first part $K_h(k,k') \propto \delta(k-k')$, which we refer to as the homogeneous part of the kernel, describes the superconductivity nucleation for a homogeneous relative band shift $\chi$. So, disregarding the inhomogeneous part of the kernel $[K(k,k') - K_h(k,k')]$ and taking $\Delta_{\rm int}(k) = \Delta_{\rm int}(2\pi)\delta(k)$, one gets the equation governing the critical temperature of a homogeneous state $T_c(\chi, t)$ in the form of Eq.~(\ref{Tc_homogeneous_equation_spin_singlet}).

To solve the inhomogeneous problem~(\ref{self_cons_domain_wall_main_text}), we use an analytical approach similar to the one used in Ref.~\cite{HouzetPRB2006} for the analysis of localized superconductivity nucleation at the domain walls in superconductor-ferromagnet hybrids. This approach is based on the separation of characteristic spatial scales, which is inherent to the considered problem in the near vicinity of the superconducting critical temperature $T_{c0}$. Indeed, within this temperature range the inhomogeneous part of the kernel in the real-space representation $[K(x,x')-K_h(x-x')]$ varies on the scale $v_F/T$ whereas the spatial scale of the interlayer order parameter $\xi \left(T\right)\propto \sqrt{T_{c0}/\left(T_{c0}-T\right)}\to +\infty$ diverges in the critical region for $T\rightarrow T_{c0}$. In other words, in the vicinity of $T_{c0}$ we can put $k=k' = 0$ in the inhomogeneous part of the kernel in Eqs.~(\ref{self_cons_domain_wall_main_text}), so that its effect is approximated by an effective Dirac delta potential. Keeping this in mind, and also expanding the homogeneous part of the kernel in Eq.~(\ref{self_cons_domain_wall_main_text}) up to the terms $\propto k^2$, the critical temperature of the localized superconducting nucleus is governed by the following linearized Ginzburg -- Landau-type equation  
\begin{equation}\label{delta_potential_approximation_equation}
\frac{k^{2}}{2M}\Delta\left(k\right)-\frac{\varkappa^{2}}{4\pi T}\alpha\int\Delta\left(k'\right)\frac{dk'}{2\pi}=E\Delta\left(k\right) \ .
\end{equation}
Here the temperature-dependent coefficients $M$, $\alpha$, and $E$ are given by the expressions (see Appendix~\ref{Ginzburg_Landau_appendix} for details of the derivation)
\begin{widetext}
\begin{subequations}\label{temperature_dependent_coefficients}
\begin{align}
\frac{1}{2M}=-\frac{v_{F}^{2}}{16}\left(\frac{\chi^{2}}{16\pi^{3}T^{3}}\text{Re}\left[\psi_2\left(\frac{1}{2}-i\frac{\varkappa}{2\pi T}\right)\right]-\frac{7\zeta\left(3\right)t^{2}}{8\pi^{3}T^{3}}\right) \ ,\\
E=-\frac{\varkappa^{2}}{4\pi T}\ln\left(\frac{T}{T_{c0}}\right)+\frac{\chi^{2}}{4\pi T}\text{Re}\left[\psi\left(\frac{1}{2}\right)-\psi\left(\frac{1}{2}-i\frac{\varkappa}{2\pi T}\right)\right] \ ,\\
\alpha=-\frac{v_{F}}{4T}\frac{\chi^{2}}{T^{2}}\frac{\left[\left(\frac{\varkappa}{T}\right)^{3}+\left(\frac{\varkappa}{T}\right)\left(\frac{t}{T}\right)^{2}\left(2+\cosh\left(\frac{\varkappa}{T}\right)\right)-\left(\left(\frac{\varkappa}{T}\right)^{2}+3\left(\frac{t}{T}\right)^{2}\right)\sinh\left(\frac{\varkappa}{T}\right)\right]}{\left(\frac{\varkappa}{T}\right)^{5}\cosh^{2}\left(\frac{\varkappa}{2T}\right)} \ ,
\end{align}
\end{subequations}
\end{widetext}
and the critical temperature $T_{cw}$ is obtained from the equation $E_0 = E(T_{cw})$, where $E_0$ denotes the minimal eigenvalue of the problem~(\ref{delta_potential_approximation_equation}). Here $\zeta(x)$ is the Riemann zeta function~\cite{Abramowitz_book}. It is easy to see that in the considered case both $M$ and $\alpha$ are positive, so that the additional potential $\propto \alpha$ in Eq.~(\ref{delta_potential_approximation_equation}) has the form of the Dirac delta potential well and the equation, defining the $T_{cw}\left(\chi,t\right)$ can be also presented in the following form
\begin{equation}\label{Tc_localized_equation}
E(T_{cw})=-\frac{M\varkappa^{4}\left[\alpha(T_{cw})\right]^{2}}{32\pi^2T_{cw}^{2}} \ .
\end{equation}
Analytical expressions for the deviation of $T_{cw}\left(\chi,t\right)$ from $T_c\left(\chi,t\right)$ (the latter is given by Eq.~(\ref{Tc_homogeneous_result_spin_singlet})) can be obtained in the form of the expansion over small parameters $\chi/T_{c0}$, $t/T_{c0}$, $\varkappa/T_{c0}\ll 1$. We find as a result (see Appendix~\ref{Ginzburg_Landau_appendix} for details of the derivation)
\begin{eqnarray}\label{Tc_inhomogeneous_main_result}
\frac{T_{cw}\left(\chi,t\right)-T_{c}\left(\chi,t\right)}{T_{c0}}\approx 0.016\left(\frac{\chi}{T_{c0}}\right)^4 \\
\nonumber
 - 0.01\left(\frac{t}{T_{c0}}\right)^{2}\left(\frac{\chi}{T_{c0}}\right)^4 + 0.02\left(\frac{\chi}{T_{c0}}\right)^{6} \ .
\end{eqnarray}
The above Eq.~(\ref{Tc_inhomogeneous_main_result}) is the main analytical result of this subsection. This expression shows that the superconducting states localized at the ferroelectric domain wall, indeed, can have higher critical temperature in comparison with a homogeneous solution. In the near vicinity of $T_{c0}$, the corresponding temperature shift $\propto (\chi/T_{c0})^4$ and doesn't depend on the tunneling amplitude. The tunneling manifests itself rather far from $T_{c0}$ and leads to the decrease in $T_{cw}-T_c$ (suppresses the enhancement of the critical temperature for a localized state). The reason for the $T_c$ matching is the suppression of the band offset by the tunneling. Indeed, it is the sign change in the band offset, which causes the increase of the critical temperature at the domain wall. Naturally, the decrease of this jump in the band offset caused by increasing tunneling suppresses the effect of the domain wall superconductivity.

It is important to note that the applicability of the delta potential approximation (Eqs.~(\ref{temperature_dependent_coefficients}) and (\ref{Tc_localized_equation})) rather far from $T_{c0}$ is questionable and requires further justification. For this purpose, we carry out numerical solution of the self-consistency equation~(\ref{self_cons_domain_wall_main_text}). Adding and subtracting the normal-state contribution to the self-consistency equation~(\ref{self_cons_domain_wall_main_text}), we can eliminate the dimensionless coupling constant $\lambda$ in favor of the critical temperature $T_{c0}$ and rewrite the equation in the following form
\begin{equation}
\Delta_{\rm int}(k)\log\left(\frac{T}{T_{c0}}\right) = \int_{-\infty}^{+\infty}dk' \ \tilde{K}(k,k')\Delta_{\rm int}(k') \ ,
\end{equation}
with $\tilde{K}(k,k') = K(k,k')/2\pi \lambda - 2\pi \delta(k-k') T\sum_{\omega_n > 0}1/\omega_n$, and $K(k,k')$ is defined \textcolor{red}{in} Eq.~(\ref{self_cons_domain_wall_main_text}). The above equation is then discretized on a grid in the momentum space $k \to k_n = n\Delta k$, $k' \to k'_m = m\Delta k'$ ($n,m$ are integer numbers), so that $\tilde{K}(k,k') \to \tilde{K}(k_n,k'_m) = \tilde{K}_{n,m}$, and $\Delta_{\rm int}(k) \to \Delta_{\rm int}(k_n)$. The critical temperature is obtained from the condition of vanishing determinant of the linear system
\begin{equation}
\sum_m\left[\tilde{K}_{n,m} - \delta_{n,m}\log\left(\frac{T}{T_{c0}}\right)\right]\Delta_{\rm int}(k_m) = 0 \ .
\end{equation}

\begin{figure*}[htpb]
\centering
\includegraphics[scale = 0.75]{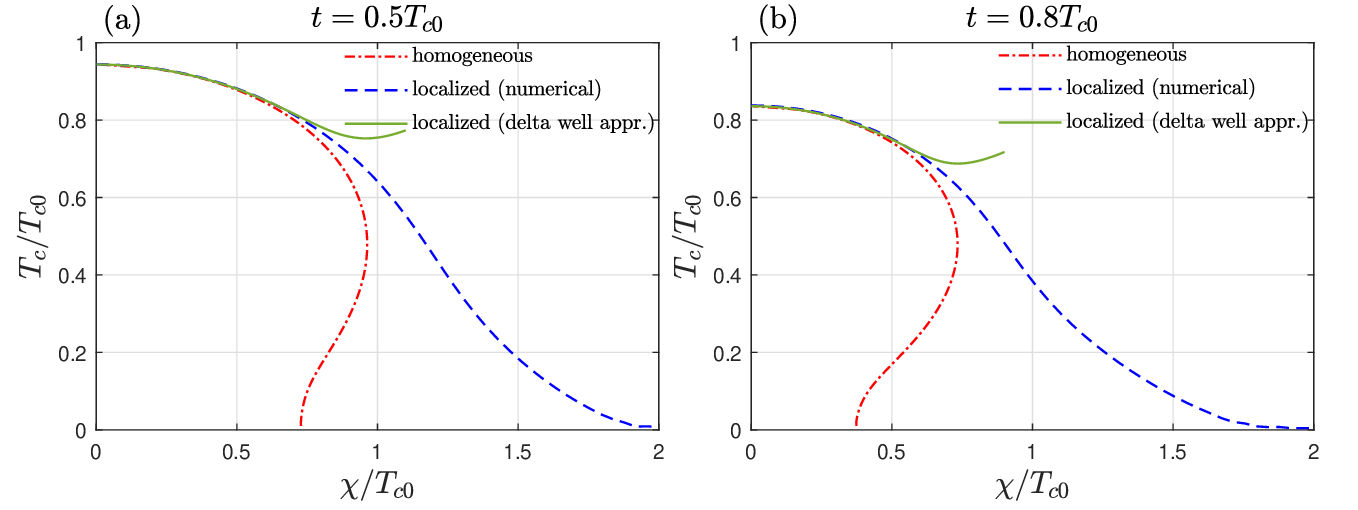}
\caption{Phase diagrams for different tunneling amplitudes in the case of a spin-triplet order parameter. Green solid and blue dashed lines correspond to domain wall superconductivity, but were obtained in different ways: by direct numerical solution of the Eq.~(\ref{self_cons_domain_wall_triplet_main_text}) or by numerical solution of Eq.~(\ref{Tc_localized_triplet_equation}), obtained in delta well approximation. Red dash-dotted lines are critical lines of the transition into uniform superconducting state. Panels (a), (b) correspond to $t = 0.5T_{c0}$ and $t = 0.8T_{c0}$, respectively.}
\label{Fig:dw_triplet_plots}
\end{figure*}

The direct comparison between the exact numerical solution, the numerical results obtained using the delta potential approximation, as well as the analytical results are provided in Fig.~\ref{Fig:dw_triplet_plots}. The resulting phase diagrams show that for small tunneling amplitudes the critical line for the localized state is significantly increased compared to the critical line for the transition into uniform superconducting state, while for larger values of $t$ domain wall superconductivity is suppressed. Comparison of green solid and blue dashed-dotted curves in Fig.~\ref{Fig:dw_singlet_plots} shows that the approximation of the inhomogeneous part of the kernel by an effective Dirac delta potential is applicable in a rather wide range of values of the parameters $\chi$ and $t$.

All previous analysis referred to the case of the absence of modulation of the order parameter along the domain wall $(q_y = 0)$. In general, it is necessary to solve the problem for a finite value of the longitudinal momentum $q_y$ and choose the value corresponding to the maximum of $T_{cw}(\chi, t)$. To account a finite value of the longitudinal momentum $q_y$, it is necessary to replace $kv_x$, $k'v_x$ by $kv_x+q_yv_y$ and $k'v_x+q_yv_y$, respectively, in the kernel $K(k,k')$ of the self-consistency equation~(\ref{self_cons_domain_wall_main_text}). We used the kernel modified for finite $q_y$ values in our numerical calculations and verified that the maximum of the critical temperature for a domain wall solution formation corresponds to the $q_y=0$ case. Note that for the problem of superconductivity nucleation at the domain wall in ferromagnetic superconductors, which is analogous to the considered one  in the limit $t=0$, the nonvanishing longitudinal momentum $q_y$ can appear in the three-dimensional case at zero temperature~\cite{BuzdinZhETP1984}.

\subsection{Domain-wall superconductivity in the spin-triplet state}

The analogue of Eqs.~(\ref{eilenberger_domain_wall_without_spin_spin_singlet_case}) for the triplet spin structure of the interlayer order parameter, obeying the relation $\hat{\Delta}_{\rm int} = \hat{\Delta}_{\rm int}^T$, has the form
\begin{subequations}\label{eilenberger_domain_wall_without_spin_spin_triplet_case}
\begin{align}
\left[iv_x\frac{d}{dx}+2i\omega_{n}\right]f_{11}+t\left(f_{12}-f_{21}\right)=0 \ ,\\
\left[iv_x\frac{d}{dx}+2i\omega_{n}\right]f_{22}-t\left(f_{12}-f_{21}\right)=0 \ ,\\
\left[iv_x\frac{d}{dx}+2i\omega_{n}-2\chi\text{sgn}(x)\right]f_{12} \\
\nonumber
-t\left(f_{22}-f_{11}\right)+2\Delta_{\rm int}\left(x\right)\text{sgn}(\omega_{n})=0 \ ,\\
\left[iv_x\frac{d}{dx}+2i\omega_{n}+2\chi\text{sgn}(x)\right]f_{21} \\
\nonumber
+t\left(f_{22}-f_{11}\right)-2\Delta_{\rm int}\left(x\right)\text{sgn}(\omega_{n})=0 \ .
\end{align} 
\end{subequations}

Performing similar calculations, the details of which are given in Appendix \ref{kernel_derivation_triplet_appendix}, we obtain the self-consistency equation written for the Fourier-transformed order parameter in the same form as in Eq.~(\ref{self_cons_domain_wall_main_text}), but with different kernel. Explicit expression for the kernel of the self-consistency equation in spin-triplet case is provided in Appendix~\ref{kernel_derivation_triplet_appendix}.
One can also check that in the $t=0$ limit, kernels in singlet and triplet cases coincide with each other.
The obtained kernel leads to the following linearized Ginzburg--Landau-type equation (see Appendix~\ref{GL_triplet_appendix} for details of the derivation)
\begin{equation}\label{GL_spin_triplet_main_text}
\frac{k^{2}}{2M}\Delta_{{\rm int}}\left(k\right)-\frac{\alpha}{4\pi T}\int\Delta_{{\rm int}}\left(k'\right)\frac{dk'}{2\pi}=E\Delta_{{\rm int}}\left(k\right),
\end{equation}
with temperature dependent coefficients defined as follows
\begin{subequations}\label{GL_coefficients_triplet}
    \begin{align}
    \frac{1}{2M}=-\frac{v_{F}^{2}}{16}\cdot\frac{1}{16\pi^{3}T^{3}}\text{Re}\left[\psi_2\left(\frac{1}{2}-i\frac{\varkappa}{2\pi T}\right)\right], \\
    \alpha=\frac{v_{F}}{4T}\frac{\left(\frac{\chi}{T}\right)^{2}\left(\sinh\left(\frac{\varkappa}{T}\right)-\frac{\varkappa}{T}\right)}{\left(\frac{\varkappa}{T}\right)^{3}\cosh^{2}\left(\frac{\varkappa}{2T}\right)}, \\
    E=-\frac{1}{4\pi T}\ln\left(\frac{T}{T_{c0}}\right)\\
    \nonumber
    +\frac{1}{4\pi T}\text{Re}\left[\psi\left(\frac{1}{2}\right)-\psi\left(\frac{1}{2}-i\frac{\varkappa}{2\pi T}\right)\right].
    \end{align}
\end{subequations}
Solving the resulting equation, we get the equation governing the critical temperature of the localized state 
\begin{equation}\label{Tc_localized_triplet_equation}
E\left(T_{cw}\right)=-\frac{M\left[\alpha\left(T_{cw}\right)\right]^{2}}{32\pi^{2}T_{cw}^{2}}.
\end{equation}
which leads to the following shift of the critical temperature for a localized superconducting state (details of calculation can be found in Appendix~\ref{GL_triplet_appendix})
\begin{eqnarray}\label{Tc_DW_expansion_spin_triplet}
        \frac{T_{cw}\left(\chi,t\right)-T_{c}\left(\chi,t\right)}{T_{c0}} \approx 0.016\left(\frac{\chi}{T_{c0}}\right)^4 \\ \nonumber +0.02\left(\frac{t}{T_{c0}}\right)^2\left(\frac{\chi}{T_{c0}}\right)^4
        +0.02\left(\frac{\chi}{T_{c0}}\right)^6.
\end{eqnarray}
This expansion demonstrates that in spin-triplet case critical temperature of domain wall solution formation can exceed the critical temperature of transition into uniform superconducting state. We also note, that in contrast with spin-singlet case, increase in the tunneling amplitude leads to the enhancement of $T_{cw}\left(\chi,t\right)-T_{c}\left(\chi,t\right)$.
Results of the exact numerical solution of the self-consistency equation, as well as the results obtained, using delta-potential approximation are given in Fig.~\ref{Fig:dw_triplet_plots}. These plots demonstrate that critical temperature of the domain wall solution formation is significantly enhanced, compared to the critical temperature of transition into the uniform state. The tunneling suppresses localized states weaker than homogeneous states leading, thus, to an increase in $T_{cw}\left(\chi,t\right)-T_c\left(\chi,t\right)$, which is consistent with the analytical result in the Eq.~(\ref{Tc_DW_expansion_spin_triplet}).

\section{Effects of the in-plane magnetic field}\label{effects_of_the_inplane_field_section}

\begin{figure*}[htpb]
\centering
\includegraphics[scale = 0.75]{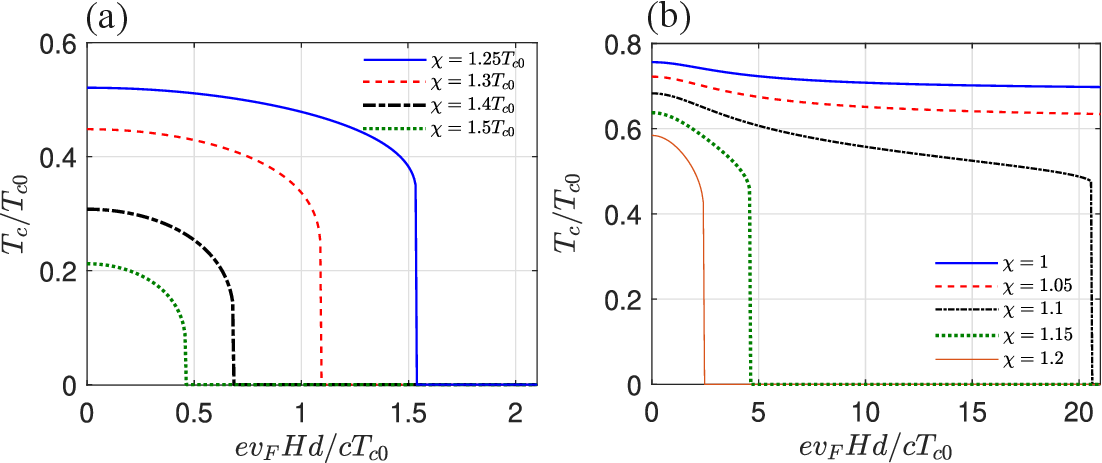}
\caption{Typical dependencies of the critical temperature $T_c$ on the external magnetic field $H$ for the spin-singlet case (only the orbital mechanism of the superconductivity suppression is taken into account). The plots are obtained on the basis of Eq.~(\ref{self_cons_orbital_main_text}) for $t = T_{c0}$, and different band offsets: (a) $\chi/T_{c0} = 1.25$, 1.3, 1.4, 1.5, (b) $\chi/T_{c0} = 1$, 1.05, 1.1, 1.15, 1.2.}
\label{Fig:orbital_effect_illustration}
\end{figure*}

We proceed with the analysis of the effects of the in-plane magnetic field $\mathbf{H} = H(\mathbf{e}_x\cos\theta  + \mathbf{e}_y\sin\theta )$ on the interlayer superconductivity. For simplicity, the influence of the orbital and paramagnetic mechanisms are studied for a homogeneous interlayer superconducting state, so that the difference between the intralayer potentials $U_{12} = 2\chi$ is constant. The vector potential $\mathbf{A}$ is taken to be orthogonal to $\mathbf{H}$ and depends only on the coordinate perpendicular to the plane of the layer $z$. So, we choose the vector potential within each layer in the following form: $\mathbf{A}_1 = -\mathbf{A}_2 = (Hd/2)(\mathbf{e}_x\sin\theta - \mathbf{e}_y\cos\theta)$, where $d$ is the interlayer distance. We address the joint effect of the orbital and paramagnetic mechanisms for an arbitrary spin structure of the interlayer order parameter by solving the linearized Eilenberger equation
\begin{widetext}
\begin{eqnarray}\label{Eilenberger_magnetic_field_general}
2i\omega_n \begin{pmatrix}\hat{f}_{11}&\hat{f}_{12}\\ \hat{f}_{21}&\hat{f}_{22}\end{pmatrix} - 2\chi\begin{pmatrix}0&\hat{f}_{12}\\-\hat{f}_{21}&0\end{pmatrix}-\frac{2e}{c}(\mathbf{v}\mathbf{A}_1)\begin{pmatrix}\hat{f}_{11}&0\\0&-\hat{f}_{22}\end{pmatrix} + \begin{pmatrix}t(\hat{f}_{12}-\hat{f}_{21})&t^*\hat{f}_{11}-t\hat{f}_{22}\\ -t^*\hat{f}_{11}+t\hat{f}_{22}&t^*(\hat{f}_{21}-\hat{f}_{12})\end{pmatrix} \\
\nonumber
+\begin{pmatrix}\hat{f}_{11}\mathbf{h}\hat{\boldsymbol{\sigma}}^*-\mathbf{h}\hat{\boldsymbol{\sigma}}\hat{f}_{11}&\hat{f}_{12}\mathbf{h}\hat{\boldsymbol{\sigma}}^*-\mathbf{h}\hat{\boldsymbol{\sigma}}\hat{f}_{12}\\ \hat{f}_{21}\mathbf{h}\hat{\boldsymbol{\sigma}}^*-\mathbf{h}\hat{\boldsymbol{\sigma}}\hat{f}_{21}&\hat{f}_{22}\mathbf{h}\hat{\boldsymbol{\sigma}}^*-\mathbf{h}\hat{\boldsymbol{\sigma}}\hat{f}_{22}\end{pmatrix} + 2{\rm sgn}(\omega_n)\begin{pmatrix}0&\hat{\Delta}_{\rm int}\\-\hat{\Delta}_{\rm int}^{T}&0\end{pmatrix} = 0 \ ,
\end{eqnarray}
\end{widetext}
together with the self-consistency equation~(\ref{eq:self_cons_main}), from which we determine the critical temperature of the superconducting transition $T_c$. In numerical calculations we take the Doppler shift $q = e\mathbf{v}_F\mathbf{A}_1/c$ in $h$ units. For this purpose we introduce the dimensionless quantity $\beta = ev_Fd/2g\mu_Bc \sim k_F d/g$. Note that it's value depends on the electronic band structure of the layers as well as on the system geometry ($d$ is a
bilayer thickness). The limit $\beta\ll 1$ corresponds to the case when the orbital effect of the in-plane magnetic field is rather weak in comparison with the paramagnetic one. Note that we neglect here possible dependence of the Fermi velocity $v_F$ and the $g$-factor on the direction of the quasiparticle trajectory.

Let us briefly discuss 
the orbital and paramagnetic mechanisms.  
We find that on a qualitative level the effect of the orbital mechanism is reduced to a renormalization (suppression) of the quasiparticle tunneling.
This fact can be understood qualitatively: the momentum-conserving tunneling
is hampered by the shift of the energy bands of different layers by
$2e\mathbf{A}_1/c$ in the momentum space. This leads to qualitatively different influence of the orbital mechanism on the spin-singlet and spin-triplet interlayer pairs. The tunneling can partially compensate the depairing effect of the relative band shift for the spin-singlet interlayer pairs [see, e.g., Fig.~\ref{Fig:Tc_homogeneous_illustration}(a)] and one can naturally expect that the orbital effect should lead to the suppression of the spin-singlet interlayer superconductivity.
On the other hand, the tunneling suppresses the spin-triplet superconducting state [see, e.g., Fig.~~\ref{Fig:Tc_homogeneous_illustration}(b)], thus, in this case the orbital mechanism can provide an \textit{increase} in the critical temperature for spin-triplet inerlayer superconductivity. Regarding the paramagnetic effect, one can note the following. First, rather weak Zeeman fields suppress the interlayer superconductivity and such suppression is isotropic for spin-singlet state and can be highly anisotropic for spin-triplet pairs. Second, further increase in the Zeeman field can lead to the appearance of reentrant superconducting phases. This effect takes place for both spin-singlet and spin-triplet pairs and can be explained by a compensation of the relative band shift by the Zeeman field~\cite{BuzdinPRL2005, MontielPRB2011}.

\begin{figure*}[htpb]
\centering
\includegraphics[scale = 0.85]{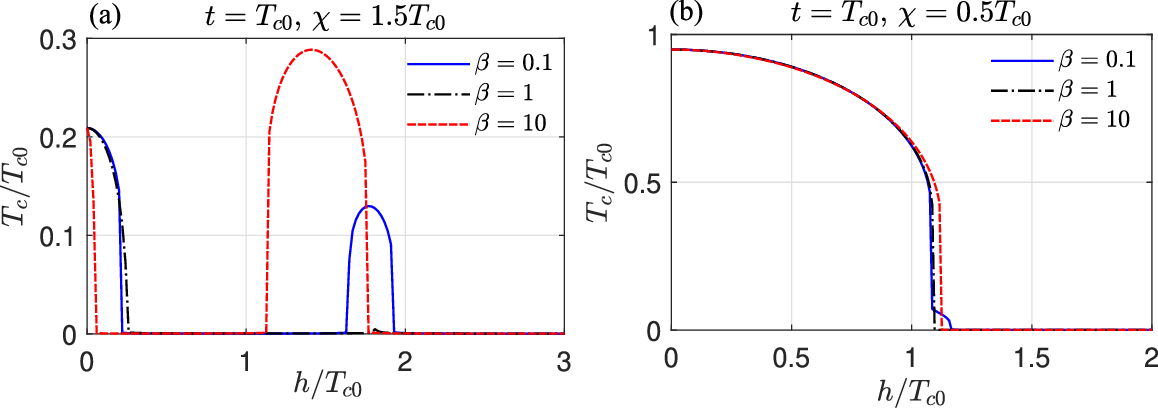}
\caption{Typical dependencies of the critical temperature $T_c$ for the interlayer spin-singlet superconducting state on the external magnetic field (both orbital and paramagnetic effects of the in-plane external magnetic field are taken into account). The plots are obtained by direct numerical solution of the self-consistency equation (\ref{selfcons_singlet_main})  
for $t = T_{c0}$, $\chi/T_{c0} = 1.5$ (a), and $t = T_{c0}$, $\chi/T_{c0} = 0.5$ (b).}
\label{Fig:joint_effect_spin_singlet}
\end{figure*} 

\subsection{Magnetic field effects for the spin-singlet case}

We start with the spin-singlet case and put $\hat{\Delta}_{\rm int} = \Delta (i\hat{\sigma}_y)$ in the linearized Eilenberger equations~(\ref{Eilenberger_magnetic_field_general}). 
It is straightforward to show that the solution of the resulting system has the following structure $\hat{f}_{ij} = \left[f_{ij}^{(0)} + \mathbf{n}_h\hat{\boldsymbol{\sigma}}f_{ij}^{(1)}\right](i\hat{\sigma}_y)$, where $\mathbf{n}_h$ is a unit vector directed along the direction of the Zeeman field $\mathbf{h}$. As a next step, we substitute the above-mentioned expressions for the anomalous Green function into the Eilenberger equations~(\ref{Eilenberger_magnetic_field_general}), solve the resulting algebraic system of equations, and then substitute the obtained solution into the self-consistency equation~(\ref{eq:self_cons_main}), 
which in the discussed case takes form
\begin{equation}\label{selfcons_singlet_main}
    \Delta_{\rm int}=-i\pi\lambda T\sum_{\omega_n}\int\frac{d\textbf{n}}{2\pi}f_{12}^{(0)} \ .
\end{equation} 
Note that it is convenient to recast the frequency summation to the one over the positive Matsubara frequencies. As a result, we find the following equation for the superconducting critical temperature. It is rather lengthy, so it's explicit form can be found in the Appendix~\ref{Eilenberger_equations_spin_triplet_magnetic_field}.

Focusing on the orbital effect, we consider $h = 0$ limit, which leads us to the following equation governing the behavior of the superconducting critical temperature:
\begin{widetext}
\begin{eqnarray}\label{self_cons_orbital_main_text}
\ln\left(\frac{T_{c0}}{T}\right)=2\pi T\sum_{\omega_{n}>0}\frac{\chi^{2}}{\left(\chi^{2}+\omega_{n}^{2}\right)}
\left\{\frac{1}{\omega_{n}}-\frac{\left|t\right|^{2}}{\sqrt{\left(\omega_{n}^{2}+\left|t\right|^{2}+\chi^{2}\right)\left[\left(\omega_{n}^{2}+\tilde{q}^{2}\right)\left(\omega_{n}^{2}+\chi^{2}\right)+\left|t\right|^{2}\omega_{n}^{2}\right]}}\right\} \ ,
\end{eqnarray}
\end{widetext}
 where $\tilde{q} = ev_FdH/2c$. 
We note that for the orbital effect $T_c$, of course, does not depend on the magnetic field direction within the plane of the layers (under the assumption of an isotropic Fermi velocity). The above equation implies that the orbital effect of the magnetic field manifests itself only for a bilayer with a finite tunnel coupling and nonzero relative band shift. One can see that within the limit $t\to 0$ the vector potential simply drops out of the linearized self-consistency equation. On the other hand, in the limit $\chi \to 0$ the right-hand-side of Eq.~(\ref{self_cons_orbital_main_text}) vanishes and one gets $T_c = T_{c0}$. Typical behavior of the superconducting critical temperature on the external magnetic field $T_c(H)$ due to the orbital mechanism is shown in Fig.~\ref{Fig:orbital_effect_illustration} for $t = T_{c0}$. The plots in Fig.~\ref{Fig:orbital_effect_illustration}(a) are obtained for rather strong relative band shifts ($\chi/T_{c0} = 1.25$, 1.3, 1.4, and 1.5) in comparison with the tunneling amplitude $t/T_{c0} = 1$ and reveal that the interlayer pairing is fully suppressed by the orbital mechanism for magnetic fields $H \sim cT_c(H=0)/ev_Fd$. On the other hand, for relative band shifts, which are closer to the tunneling amplitude [see Fig.~\ref{Fig:orbital_effect_illustration}(b)], one can clearly see that interlayer pairing becomes much more insensitive to the orbital mechanism upon the decrease in the band shift.

\begin{figure}[htpb]
\centering
\includegraphics[scale = 1]{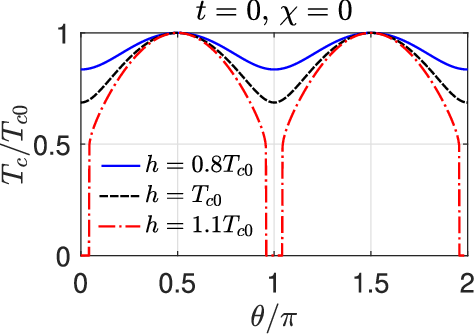}
\caption{Typical dependence of the critical temperature of the spin-triplet interlayer superconductivity described by the gap function $\hat{\Delta}_{\rm int} = \Delta_{\rm int}\hat{\sigma}_x(i\hat{\sigma}_y)$ on the angle of the in-plane Zeeman field $\theta$ for $t = 0$, $\chi = 0$, and $h/T_{c0} = 0.8$, 1 and 1.1. The plots demonstrate the paramagnetic effect on $T_c$ and are obtained on the basis of Eq.~(\ref{self_consistency_spin_triplet_paramagnetic}).}
\label{Fig:Fig7}
\end{figure}

The plots of the critical temperature as a function of the external magnetic field highlighting the joint effect of the orbital and paramagnetic mechanisms are presented in Fig.~\ref{Fig:joint_effect_spin_singlet}. We find that for $\beta \ll 1$ the orbital effect is rather small in comparison with the paramagnetic one, so, the results obtained for $\beta = 0.1$ in Figs.~\ref{Fig:joint_effect_spin_singlet}(a) and Figs.~\ref{Fig:joint_effect_spin_singlet}(b) mainly reveal the role of the paramagnetic effect. Corresponding $T_c(h)$ plots demonstrate that at first, the paramagnetic effect suppresses the interlayer spin-singlet superconductivity and leads to the appearance of the reentrant superconducting phase upon further increase in the Zeeman field. As it has been mentioned earlier, this effect occurs due to a compensation of the relative band shift by the Zeeman field. We find that the position of the reentrant phase, in turn, depends on the tunneling amplitude and the band shift. For band shifts $\chi$ noticeably exceeding the tunneling amplitude $t$ [see Fig.~\ref{Fig:joint_effect_spin_singlet}(a)] the reentrant phase is fully developed, while the decrease in $\chi$ leads to the shift of the center of the reentrant phase toward lower $h$ values [compare Fig.~\ref{Fig:joint_effect_spin_singlet}(b) and Fig.~\ref{Fig:joint_effect_spin_singlet}(a)].

\begin{figure}[htpb]
\centering
\includegraphics[scale = 1]{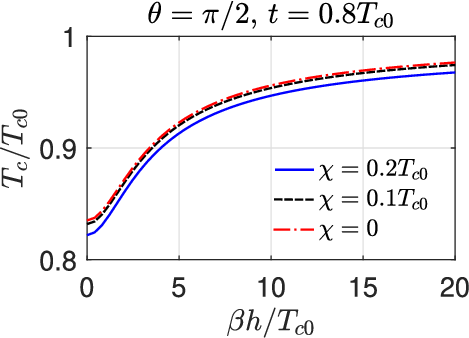}
\caption{Typical $T_c(H)$ behavior associated with the orbital effects on the spin-triplet interlayer superconductivity. The plots are obtained by numerical solution of the self-consistency equation~(\ref{self_consistency_orbital_paramagnetic_spin_triplet}). 
We choose $\theta = \pi/2$, $t = 0.8T_{c0}$, $\chi/T_{c0} = 0$, 0.1, and 0.2 to produce the plots.}
\label{Fig:Fig8}
\end{figure}

In accordance with our previous discussion of the orbital effect, we see that it manifests itself only for rather strong band shift in comparison with the tunneling amplitude [see, e.g., the plots obtained for $\beta = 1$ and 10 in Fig.~\ref{Fig:joint_effect_spin_singlet}(a)]. For this parameter range the orbital effect can lead to the increase in the rate of $T_c$ decrease for rather small magnetic fields. On the other hand, since the tunneling renormalization by the orbital mechanism can significantly affect the critical temperature [see, e.g., the results in Fig.~\ref{Fig:orbital_effect_illustration} for purely orbital effect], the orbital effect also influences the position and shape of the reentrant superconducting phases. It is interesting to note that the increase in the parameter $\beta$, first, results in the reduction of the parameter range corresponding to the reentrant phase. One can see that the reentrant phase is almost absent for $\beta = 1$ [see the black dashed-dotted line in Fig.~\ref{Fig:joint_effect_spin_singlet}(a)] and then reappears upon further increase in $\beta$.
The results shown in Fig.~\ref{Fig:joint_effect_spin_singlet}(b) indicate that in the opposite case (typical band shift is rather small in comparison with the tunneling amplitude), the orbital effect does not lead to noticeable modifications of the phase-transition line, which behavior is mainly determined by the paramagnetic mechanism of the superconductivity suppression.

\subsection{Magnetic field effects for the spin-triplet case}

\begin{figure*}[htpb]
\centering
\includegraphics[scale = 1.15]{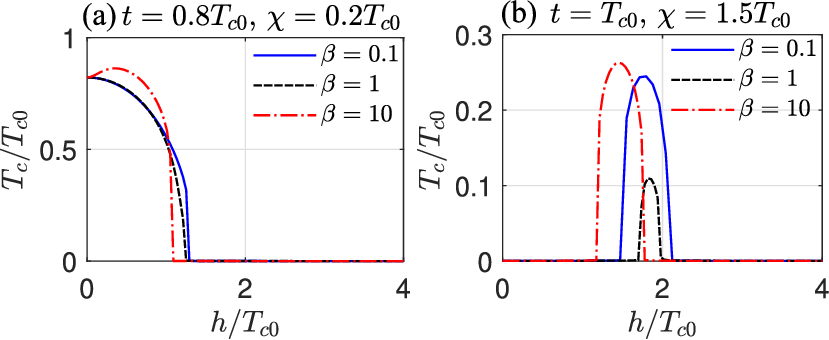}
\caption{Typical $T_c(h)$ dependencies demonstrating the joint influence of the orbital and paramagnetic effect on the spin-triplet interlayer pairing. The plots are obtained by direct numerical solution of the self-consistency equation~(\ref{self_consistency_orbital_paramagnetic_spin_triplet}).
We choose $\theta = 0$, $\beta = 0.1$, 1 and 10. The tunneling amplitude and the band shift are as follows: (a) $t = 0.8T_{c0}$, $\chi = 0.2T_{c0}$, (b) $t = T_{c0}$, $\chi = 1.5T_{c0}$.}
\label{Fig:paramagnetic_joint_spin_triplet}
\end{figure*} 

We proceed with study of the effects of the in-plane external magnetic field on the spin-triplet interlayer superconducting state. In general, the spin-triplet superconducting order parameter can be presented in the form $\hat{\Delta}_{\rm int} = \mathbf{d}\hat{\boldsymbol{\sigma}}(i\hat{\sigma}_y)$, where the unit vector $\mathbf{d}$ specifies the spin-structure of the interlayer Cooper pairs. In this section we focus on the case when $\mathbf{d}$ lies within the plane of the layers. As we show below, for such structure of the order parameter both the paramagnetic and the orbital effects affect the critical temperature. For definiteness, we choose $\hat{\Delta}_{\rm int} = \Delta_{\rm int}\hat{\sigma}_x(i\hat{\sigma}_y)$ in the linearized Eilenberger equations~(\ref{Eilenberger_magnetic_field_general}) and the superconducting gap function 
\begin{equation}\label{self_consistency_orbital_paramagnetic_spin_triplet}
\Delta_{\rm int} = -i\pi\lambda T\sum_{\omega_n} \int\frac{d\mathbf{n}}{2\pi} f_{12}^{(x)} \ ,
\end{equation}
is defined through the spin-triplet component $f_{12}^{(x)}$ of the anomalous interlayer Green's function $\hat{f}_{12} = (f_{12}^{(0)} + f_{12}^{(x)}\hat{\sigma}_x + f_{12}^{(y)}\hat{\sigma}_y)(i\hat{\sigma}_y)$. Note that substituting the anomalous Green's function $f_{ij} = (f_{ij}^{(0)} + f_{ij}^{(x)}\hat{\sigma}_x + f_{ij}^{(y)}\hat{\sigma}_y)(i\hat{\sigma}_y)$ into Eqs.~(\ref{Eilenberger_magnetic_field_general}) yields a 12$\times$12 linear system of algebraic equations for all the Green function components (the resulting equations are presented in Appendix~\ref{Eilenberger_equations_spin_triplet_magnetic_field}), which we solve for the relevant $f_{12}^{(x)}$ component.

Rather simple equation for the superconducting critical temperature follows from the general expressions if we disregard the orbital effect and quasiparticle tunneling. So, we put $H=0$ and $t=0$ in the expression for $f_{12}^{(x)}$
and arrive at the following equation  
\begin{eqnarray}\label{self_consistency_spin_triplet_paramagnetic}
\ln\left(\frac{T}{T_{c0}}\right) = \\
\nonumber
 2\pi T\sum_{\omega_n > 0}{\rm Re}\left[\frac{(\omega_n - i\chi)^2 + h_y^2}{(\omega_n - i\chi)[(\omega_n - i\chi)^2 + h^2]}-\frac{1}{\omega_n}\right] \ .
\end{eqnarray}
Typical behavior of the critical temperature is shown in Fig.~\ref{Fig:Fig7}. It follows from the above expression that the suppression of the spin-triplet interlayer superconductivity due to the paramagnetic mechanism is two-fold anisotropic. Indeed, for $\theta = \pi/2$ and $3\pi/2$ we get that $h_x = h\cos\theta = 0$ and the Zeeman field simply drops out of the equation for $T_c$. In this case possible decrease in the critical temperature can occur upon the increase in a band shift $\chi$. One can see from Fig.~\ref{Fig:Fig7} that maximal suppression of $T_c$ is realized when the Zeeman field is parallel or anti-parallel to the $x$ axis (the Zeeman field is directed parallel or anti-parallel to the unit vector, which determines the spin-structure of Cooper pairs).

The plots in Fig.~\ref{Fig:Fig8} obtained for $\theta = \pi/2$, $t = 0.8T_{c0}$, and $\chi/T_{c0} = 0$, 0.1 and $0.2$ highlight the effect of the orbital mechanism on the spin-triplet state. In comparison with the spin-singlet case we find that the orbital mechanism results in the increase of the critical temperature for the interlayer spin-triplet superconductivity. As it has been mentioned previously, such superconductivity enhancement arises due to the fact that the quasiparticle tunneling suppresses the spin-triplet superconductivity. One can see that $T_c(H)$ dependencies are saturated at rather large magnetic fields and the maximum $T_c$ is determined by the corresponding $T_c$ value at zero tunneling amplitude [see, e.g., Fig.~\ref{Fig:Tc_homogeneous_illustration}(b)].

Finally, let us discuss the joint influence of the paramagnetic and orbital mechanism on the spin-triplet interlayer superconductivity. Throughout the discussion we focus on the most interesting regime when both the orbital and paramagnetic effects can be rather strong, so we take $\theta = 0$ for both Figs.~\ref{Fig:paramagnetic_joint_spin_triplet}(a) and~\ref{Fig:paramagnetic_joint_spin_triplet}(b). The plots in Fig.~\ref{Fig:paramagnetic_joint_spin_triplet}(a) have been obtained for rather small band shift $\chi = 0.2T_{c0}$ and rather strong tunneling amplitude $t = 0.8T_{c0}$. One can clearly see that for $\beta = 0.1$ and 1 (the orbital effects are rather weak) the spin-triplet superconductivity is suppressed monotonically due to the paramagnetic mechanism. On the other hand, if both orbital and paramagnetic effects are rather strong (this corresponds to $\beta = 10$ in Fig.~\ref{Fig:paramagnetic_joint_spin_triplet}(a)) the critical temperature exhibits a nonmonotonic behavior: $T_c$ increases in weak fields and decreases upon further increase in the magnetic field. Such behavior naturally arises due to the competition of the orbital and paramagnetic effects discussed above in this section. The results in Fig.~\ref{Fig:paramagnetic_joint_spin_triplet}(b) are, in turn, obtained for rather large band shift $\chi = 1.5 T_{c0}$, so that the superconducting phase is absent for rather weak magnetic fields and reappears due to a compensation of the band shift by the Zeeman shift of the energy levels in strong magnetic fields. Similarly to the spin-singlet case, we find that the position of the reentrant phase and the shape of its phase-transition line are affected by the orbital effects due to renormalization of quasiparticle tunneling.

 \begin{table*}[htpb]
    \centering
    \caption{Summary of the results}
    \label{tab:summary_of_results}
    \begin{tabular}{|p{5cm}|p{5cm}|p{5cm}|}
    \hline
    \textbf{Property} & \textbf{Singlet pairing} & \textbf{Triplet pairing} \\
    \hline
    Effect of tunneling on homogeneous superconducting state & SC state is stabilized at larger band splittings ($T_c$ increases). & SC state is suppressed ($T_c$ decreases) \\
    \hline
    Effect of tunneling on
domain-wall superconductivity
 & $T_{cw} - T_c$ decreases & $T_{cw} - T_c$ increases (both $T_c$ are suppressed in the strong tunneling regime)  \\
    \hline 
    Orbital effect of the in-plane magnetic field & $T_c$ decreases (effect depends on $t$ and $\chi$) & $T_c$ increases (effect depends on $t$ and $\chi$) \\
    \hline
    Paramagnetic effect of the in-plane magnetic field & $T_c$ decreases, reentrant phases (sensitive to $t$). Anisotropy of the critical field can originate from the Fermi surface anisotropy & $T_c$ decreases, reentrant phases (sensitive to the magnetic field direction and $t$). Anisotropy of the critical field can originate from the spin structure of the Cooper pairs and the Fermi surface anisotropy \\
    \hline
    \end{tabular}
    \end{table*}

\section{Discussion}\label{discussion_section}

Our main results are summarized in Tab.~\ref{tab:summary_of_results}. Turning now to the discussion of our results in context of the experimental results in Ref.~\cite{JindalN2023}, we note that the increase in the critical temperature of the superconducting transition at the domain walls can naturally explain the observed interplay between the ferroelectric ordering and superconductivity. Indeed, assuming that the band offset for a completely polarized ferroelectric state fully suppresses the transition temperature $T_c$, we find that interlayer superconducting nuclei can exist only in the vicinity of the domain walls. The change in the externally applied electric field increases the concentration of domain walls being, thus, responsible for the resistivity drop. Finally, the resistivity should vanish only as soon as the concentration of the domain walls exceeds the critical one corresponding to the formation of an infinite superconducting cluster. The resulting temperature dependence of resistivity should be broadened and sensitive to the external magnetic field. This phenomenon should be also affected by the interlayer coupling, which can strongly reduce the difference between the critical temperatures of the homogeneous and localized nuclei both for the spin-singlet and spin-triplet states. For strongly coupled layers we find that the spin-singlet state always dominates comparing with the spin-triplet one. Provided that in the real experimental situation the interlayer coupling can be effectively tuned by pressure~\cite{JayaramanRMP1983,MaNL2021,NayakACSN2015,PeiACSN2022,PeiMD2022,JiaoRPP2023}, our results both for homogeneous and localized interlayer superconducting states imply that the critical temperature of the spin-singlet (spin-triplet) interlayer superconductivity should decrease (increase) upon the increase in pressure.

Regarding the discussion of our results in context of the experimental data on the effects of the in-plane magnetic field~\cite{LiPRL2024}, it should be noted that the existing experimental data don't provide unambiguous justification for the proposed interlayer pairing scenario. Although our results indicate the possibility of the in-plane anisotropy of the critical magnetic field 
both for the spin-singlet interlayer pairing (due to the anisotropy of the Fermi velocity) and the spin-triplet one (provided the unit vector $\mathbf{d}$ parametrizing the spin structure of the spin-triplet interlayer pairs $\hat{\Delta}_{\rm int} = \Delta_{\rm int}\mathbf{d}\hat{\boldsymbol{\sigma}}(i\hat{\sigma}_y)$ possesses the in-plane component), consistent account of the spin-orbit effects in bilayer T$_{\rm d}$-MoTe$_2$ structures can be important for uncovering the spin structure of the formed Cooper pairs in the system. 
Despite the fact that the nature of interlayer pairing is left beyond the scope of this work, it is natural to expect that the spin-orbit interaction can affect the distinctive features of the effective interlayer electronic attraction and, as a consequence, the resulting spin and spatial structure of the spin-triplet interlayer Cooper pairs. On a quantitative level, the effects of the in-plane magnetic field on the spin-triplet pairs, in turn, are known to depend on the spatial structure of the spin-triplet pairs (see, e.g., Ref.~\cite{BernatPRB2024}). In addition, these effects should be also sensitive to the possible momentum dependence of the $g$-factor~\cite{RothPR1959,RothPR1966,AndoRMP1982,SchoberPSS1986,SchoberPSS1987}.
The superconducting pairing of electrons from neighboring layers can be also present in multilayered or bulk systems. We expect the similar effect of the band offset in such systems, The renormalization of interlayer tunneling due to the orbital effect of the magnetic field as well as the anisotropy of the critical magnetic field governed by paramagnetic effect should also take place in multi-layers.
Nevertheless, the extensive analysis of the magnetic field effects and the domain-wall nucleation of the interlayer superconductivity performed in our work is of primary importance for revealing the possible role of this type of pairing in van der Waals systems with coexisting superconductivity and sliding ferroelectricity.

Finally, let us briefly compare the considered interlayer pairing scenario with recently proposed mechanism of intralayer superconductivity in sliding ferroelectrics based on the domain wall fluctuations and discussed in details for interlayer shear phonons~\cite{ChaudharyPRL2024}. In the latter case it was shown that the increase in the interlayer coupling results in the decrease of the transverse electric polarization, which, in turn, should be accompanied by the suppression of the $T_c$ enhancement for superconducting nuclei localized in the vicinity of the domain wall. An important point is that such mechanism can, in principle, coexist with another mechanisms of intralayer superconductivity, which can provide nonvanishing $T_c$ even within ferroelectric domains. Note that these mechanisms responsible for domain and domain-wall intralayer superconductivity may depend differently upon changes in the interlayer coupling. As we show in this work, irrespective to the particular type of the superconductivity nucleation (homogeneous or localized), one of the distinctive features of the interlayer pairing mechanism is its sensitivity to the interlayer coupling. The behavior of $T_c$ as a function of the interlayer coupling is, in turn, governed by the spin structure of interlayer pairs. For homogeneous interlayer superconducting states, we show that $T_c$ of the spin-singlet (spin-triplet) state increases (decreases) upon the increase in the interlayer coupling. The way changes in the interlayer coupling affect the $T_c$ for localized interlayer superconducting states is also determined by the spin structure of interlayer pairs.  We demonstrate that the increase in the interlayer coupling for the spin-singlet (spin-triplet) interlayer pairing leads to the decrease (increase) in the difference between the $T_c$ of the homogeneous state and the one localized in the vicinity of the domain wall. We believe that further experimental studies of the full superconducting phase diagram of the van der Waals bilayers under consideration for a wide range of magnetic fields, temperature, and interlayer coupling can provide new insights into the mechanism of the superconductivity in such heterostructures.

\section{Conclusion}\label{conclusion_section}

To sum up, we have suggested a theoretical model describing the interplay between the interlayer superconductivity and ferroelectricity in van der Waals bilayers. We have shown that the presence of ferroelectric domains can facilitate the emergence of interlayer superconductivity localized at the domain walls. Our analysis of the effects of the in-plane magnetic field on the interlayer superconductivity reveals that the joint effect of orbital and paramagnetic mechanisms are very sensitive to the spin structure of the interlayer Cooper pairs. In the spin-singlet case, we have demonstrated that both the orbital and paramagnetic effect result in the superconductivity suppression in weak magnetic fields, while the paramagnetic effect can be responsible for the appearance of reentrant superconducting phase in rather strong fields. For the spin-triplet state, we have shown that the paramagnetic mechanism can be highly anisotropic within the plane of the layers and the competition between the orbital and paramagnetic mechanisms can even lead to a nonmonotonic behavior of the superconducting critical temperature for rather weak magnetic fields. Theoretical results are discussed in the context of recent experimental data on the effect of electric polarization switching and applied magnetic field on the superconductivity in van der Waals bilayers.

The code for reproducing the data used to plot figures presented in the main text is available at Zenodo~\cite{Zenodo}.

\acknowledgements
The authors thank I.~V.~Bobkova for useful discussions. This work was supported by the Ministry of science and higher education of the Russian Federation (Grant No. 075-15-2025-010) in part of analytical calculations of the domain-wall superconductivity for the spin-singlet case, by the Russian Science Foundation (Grant No. 24-12-00152) in part of analytical calculations of the domain-wall superconductivity for the spin-triplet case, by MIPT Project No. FSMG-2023-0014 in part of analytical calculations of the magnetic field effects, and by the Federal Academic Leadership Program Priority 2030 (NUST MISIS Strategic Technology Project ``Quantum Internet'') in part of numerical calculations. A.A.K. acknowledges the financial support of the Foundation for the Advancement of Theoretical Physics and Mathematics BASIS (Grant No. 23-1-2-32).

\begin{widetext}
\appendix

\section{Derivation of Eqs.~(\ref{eilenberger_self_cons}) in the main text}\label{Eilenberger_derivation}

Here we present the detailed derivation of the Eilenberger equations~(\ref{eq:eilenberger_main}) and the self-consistency equation~(\ref{eq:self_cons_main}) in the main text. In the end of this section we briefly discuss the derivation of normalization condition for the quasiclassical matrix Green's functions in the generalized layer-Nambu-spin space as well as the derivation of the linearized Eilenberger equations. Following the standard approach~\cite{Svidzinski_book}, we introduce the Green's functions in the mixed representation
\begin{equation}\label{GF_mixed_representation}
\underline{G}\left(\mathbf{R}, \boldsymbol{\rho}\right)=\int\frac{d^{2}\mathbf{p}}{\left(2\pi\right)^{2}}e^{i\boldsymbol{\rho} \mathbf{p}}\underline{G}\left(\mathbf{R},\mathbf{p}\right) \ ,
\end{equation}
where $\mathbf{R}=(\mathbf{r}_1 + \mathbf{r}_2)/2$ and $\boldsymbol{\rho}=\mathbf{r}_1-\mathbf{r}_2$. The ``left'' system of the Gor'kov equations for the Green's function in the mixed representation~(\ref{GF_mixed_representation}) has the form of Eq.~(\ref{Gorkov_equations}) with the following substitutions: $\nabla_{\mathbf{r}}\rightarrow (1/2)\nabla_{\mathbf{R}}+i\mathbf{p}$ and $\mathbf{r}\rightarrow\mathbf{R}+(i/2)\nabla_{\mathbf{p}}$. Within the quasiclassical approximation the Green's functions
depend slowly on $\mathbf{R}$ (on a scale given by the superconducting coherence length) and $\mathbf{p}$ lies near the Fermi surface, so
\begin{equation}
\mathbf{p}=\mathbf{n}\left(p_{F}+\frac{\xi}{v_{F}}\right) \ , \ \ \xi=\frac{\mathbf{p}^{2}}{2m}-\frac{p_{F}^{2}}{2m} \ .
\end{equation}
Here the unit vector $\mathbf{n}$ parametrizes the momentum direction at the Fermi surface, $p_F$ is the Fermi momentum, $v_F = p_F/m$ is the Fermi velocity, and $\xi$ is of the order of the superconducting critical temperature. Taking into account these considerations, the operator of the intralayer energy relative to the chemical potential for the layer $j = 1,2$
can be written as
\begin{equation}
\hat{\xi}_{j}(\mathbf{r})\approx\xi-\frac{i}{2}v_{F}\mathbf{n}{\nabla}_{\mathbf{R}}-\frac{e}{c}v_{F}\mathbf{n}\mathbf{A}_{j}\left(\mathbf{R}\right)+U_{j}\left(\mathbf{R}\right) \ .
\end{equation}
Thus, the ``left'' equations for the Green's function in the mixed representation can be written in the form:
\begin{eqnarray}\label{left_Gorkov_mixed}
\begin{bmatrix}-i\omega_n + \check{\tau}_z\left(\xi + U_1 - \frac{i}{2}\mathbf{v}\nabla_{\mathbf{R}} - \check{\tau}_z\frac{e}{c}\mathbf{v}\mathbf{A}_1\right)&\check{t}\\ \check{t}^{\dagger}&-i\omega_n + \check{\tau}_z\left(\xi+U_2 - \frac{i}{2}\mathbf{v}\nabla_{\mathbf{R}} - \check{\tau}_z\frac{e}{c}\mathbf{v}\mathbf{A}_2\right)\end{bmatrix}\underline{G}(\mathbf{R},\mathbf{p}) = 1 \ .
\end{eqnarray}
Correspondingly, the set of the ``right'' equations takes the form:
\begin{eqnarray}\label{right_Gorkov_mixed}
\underline{G}(\mathbf{R},\mathbf{p}) 
\begin{bmatrix}-i\omega_n + \check{\tau}_z\left(\xi +U_1 + \frac{i}{2}\mathbf{v}\nabla_{\mathbf{R}} - \check{\tau}_z\frac{e}{c}\mathbf{v}\mathbf{A}_1\right)&\check{t}\\ \check{t}^{\dagger}&-i\omega_n + \check{\tau}_z\left(\xi+U_2 + \frac{i}{2}\mathbf{v}\nabla_{\mathbf{R}} - \check{\tau}_z\frac{e}{c}\mathbf{v}\mathbf{A}_2 \right]\end{bmatrix} = 1 \ .
\end{eqnarray}
As a next step, we multiply Eq.~(\ref{left_Gorkov_mixed}) by $\check{\tau}_z$ from the left and Eq.~(\ref{right_Gorkov_mixed}) from the right. Subtracting the resulting equations and introducing the quasiclassical Green function
\begin{equation}\label{quasiclassical_function_definition_appendix}
\underline{\tilde{g}}(\mathbf{R},\mathbf{n}) = \frac{i}{\pi}\int d\xi \  \underline{G}(\mathbf{R},\mathbf{n},\xi) \ ,
\end{equation}
we get
\begin{eqnarray}\label{Eilenberger_main}
-i\mathbf{v}\nabla_{\mathbf{R}}\underline{\tilde{g}} + [-i\omega_n\check{\tau}_z,\underline{\tilde{g}}] + \begin{bmatrix}0&U_{12} \check{\tilde{g}}_{12}\\-U_{12} \check{\tilde{g}}_{21}&0
\end{bmatrix} + \begin{Bmatrix}-\frac{e}{c}\mathbf{v}[\check{\tau}_z\mathbf{A}_1,\check{\tilde{g}}_{11}]& -\frac{e}{c}\mathbf{v}[\check{\tau}_z\mathbf{A}_1\check{\tilde{g}}_{12}-\check{\tilde{g}}_{12}\check{\tau}_z\mathbf{A}_2]\\ -\frac{e}{c}\mathbf{v}[\check{\tau}_z\mathbf{A}_2\check{\tilde{g}}_{21} - \check{\tilde{g}}_{21}\check{\tau}_z\mathbf{A}_1]&-\frac{e}{c}\mathbf{v}[\check{\tau}_z\mathbf{A}_2,\check{\tilde{g}}_{22}]\end{Bmatrix} + \\
\nonumber
+\begin{pmatrix}0&\check{\tau}_z\check{t}\\ \check{\tau}_z\check{t}^{\dagger}&0\end{pmatrix}\begin{pmatrix}\check{\tilde{g}}_{11}&\check{\tilde{g}}_{12}\\ \check{\tilde{g}}_{21}&\check{\tilde{g}}_{22}\end{pmatrix} - \begin{pmatrix}\check{\tilde{g}}_{11}&\check{\tilde{g}}_{12}\\ \check{\tilde{g}}_{21}&\check{\tilde{g}}_{22}\end{pmatrix}\begin{pmatrix}0&\check{t}\check{\tau}_z\\ \check{t}^{\dagger}\check{\tau}_z&0\end{pmatrix} = 0 \ .
\end{eqnarray}
Here $U_{12} = U_1(\mathbf{R})-U_2(\mathbf{R})$. Redefining the quasiclassical Green's function $\underline{\tilde{g}} = \check{\tau}_z\underline{g}$, we derive Eq.~(\ref{eq:eilenberger_main}) in the main text. Using the above mentioned definitions of the quasiclassical Green's functions, it is straightforward to see, that self-consistency equation~(\ref{self_cons_Gorkov}) takes the form presented by Eq.~(\ref{eq:self_cons_main}) in the main text.

Below we discuss the form of the normalization condition as well as the derivation of the linearized Eilenberger equations. Note that the form of the normalization condition can be inferred directly from the solution of the Gor'kov equations~(\ref{Gorkov_equations}) for a homogeneous state. For definiteness, we solve Eqs.~(\ref{Gorkov_equations}) for a spin-singlet interlayer pairing $\hat{\Delta}_{\rm int} = \Delta_{\rm int}(i\hat{\sigma}_y)$ and homogeneous relative band shift $U_1 - U_2 = -2\chi = {\rm const}$, integrate the solutions over the quasiparticle energy and then obtain the normal-state solutions by taking the limit $\Delta_{\rm int}\to 0$ in the end. Using Eq.~(\ref{quasiclassical_function_definition_appendix}), we find the following expressions for the normal quasiclassical Green's functions
\begin{subequations}
\begin{align}
\tilde{g}_{12} = \frac{i t}{2Q}\left[\frac{\omega_n^2 + |\Delta_{\rm int}|^2 - iQ}{\tilde{\xi}} + \frac{\omega_n^2 + |\Delta_{\rm int}|^2 + iQ}{\tilde{\xi}^*}\right] \ ,\\
\bar{\tilde{g}}_{12} = \frac{-it^*}{2Q}\left[\frac{\omega_n^2 + |\Delta_{\rm int}|^2 - iQ}{\tilde{\xi}} + \frac{\omega_n^2 + |\Delta_{\rm int}|^2 + iQ}{\tilde{\xi}^*}\right] \ ,\\
\tilde{g}_{21} = \frac{it^*}{2Q}\left[\frac{\omega_n^2 + |\Delta_{\rm int}|^2 - iQ}{\tilde{\xi}} + \frac{\omega_n^2 + |\Delta_{\rm int}|^2 + iQ}{\tilde{\xi}^*}\right] \ ,\\
 \bar{\tilde{g}}_{21} = \frac{-it}{2Q}\left[\frac{\omega_n^2 + |\Delta_{\rm int}|^2 - iQ}{\tilde{\xi}} + \frac{\omega_n^2 + |\Delta_{\rm int}|^2 + iQ}{\tilde{\xi}^*}\right] \ ,\\
\tilde{g}_{11} = \frac{i}{2Q}\left[\frac{i\omega_n(\chi^2 + |t|^2 + iQ)-\chi(\omega_n^2 - iQ)}{\tilde{\xi}} + \frac{i\omega_n(\chi^2 + |t|^2 - iQ)-\chi(\omega_n^2 + iQ)}{\tilde{\xi}^*}\right] \ ,\\
\bar{\tilde{g}}_{11} = \frac{i}{2Q}\left[\frac{i\omega_n(\chi^2 + |t|^2 + iQ) + \chi(\omega_n^2 - iQ)}{\tilde{\xi}} + \frac{i\omega_n(\chi^2 + |t|^2 - iQ)+\chi(\omega_n^2 + iQ)}{\tilde{\xi}^*}\right] \ ,\\
\tilde{g}_{22} = \frac{i}{2Q}\left[\frac{i\omega_n(\chi^2 + |t|^2 + iQ)+\chi(\omega_n^2 - iQ)}{\tilde{\xi}} + \frac{i\omega_n(\chi^2 + |t|^2 - iQ)+\chi(\omega_n^2 + iQ)}{\tilde{\xi}^*}\right] \ ,\\
\bar{\tilde{g}}_{22} = \frac{i}{2Q}\left[\frac{i\omega_n(\chi^2 + |t|^2 + iQ) - \chi(\omega_n^2 - iQ)}{\tilde{\xi}} + \frac{i\omega_n(\chi^2 + |t|^2 - iQ)-\chi(\omega_n^2 + iQ)}{\tilde{\xi}^*}\right] \ .
\end{align}
\end{subequations}
Here we introduced the notations
\begin{eqnarray}
\tilde{\xi} = \sqrt{-\omega_n^2 - |\Delta_{\rm int}|^2 + \chi^2 + |t|^2 + 2iQ} \ , \ \ \
Q = \sqrt{|t|^2(\omega_n^2 + |\Delta_{\rm int}|^2)+\chi^2\omega_n^2} \ .
\end{eqnarray}
It is easy to check that within the limit $\Delta_{\rm int}\to 0$ the above expressions satisfy the condition $(\underline{\eta}_z\check{\tau}_z\underline{\tilde{g}})^2 = (\underline{\eta}_z\underline{g})^2 = 1$, which is the normalization condition mentioned in the main text. Considering the above equations, we find the normal-state qusiclassical functions within the limit $t$, $\chi \to 0$, which are used in the derivation of the linearized Eilenberger equations. They have the following form: $g_{11} = g_{22} = \bar{g}_{11} = \bar{g}_{22} = -{\rm sgn}(\omega_n)$ and $g_{ij} = \bar{g}_{ij} = 0$ for $i \neq j$.

\section{Derivation of the linearized self-consistency equation for the spin-singlet domain wall state}\label{kernel_derivation_appendix}
Here we provide the derivation of the kernel~(\ref{self_cons_domain_wall_main_text}) in the main text. Our starting point is Eqs.~(\ref{eilenberger_domain_wall_without_spin_spin_singlet_case}). Seeking the solutions of the homogeneous system in the form $f_{ij}\propto e^{\lambda x}$, we get the characteristic equation
\begin{equation}
(v_x\lambda + 2\omega_n)^2[(v_x\lambda + 2\omega_n)^2 + 4(\chi^2 + t^2)] = 0 \ .
\end{equation}
Solving the above equation for the eigenvalues $\lambda$ and calculating the eigenvectors of the homogeneous system~(\ref{eilenberger_domain_wall_without_spin_spin_singlet_case}), we get that the solution of the homogeneous system can be written as
\begin{eqnarray}
\begin{pmatrix}f_{11}\\f_{12}\\f_{21}\\f_{22}\end{pmatrix} = e^{-2\omega_n x/v_x}\left[C_1\frac{1}{\sqrt{2}}\begin{pmatrix}1\\0\\0\\1\end{pmatrix} + C_2\frac{t}{\sqrt{2}\varkappa}\begin{pmatrix}-\chi/t\\ 1\\ 1\\ \chi/t\end{pmatrix}\right] +\\
\nonumber
+ e^{(-2\omega_n + 2i\varkappa)x/v_x}C_3\frac{t}{2\varkappa}\begin{bmatrix}1\\ t/(\varkappa-\chi)\\ - t/(\varkappa + \chi)\\ -1\end{bmatrix} + e^{(-2\omega_n - 2i\varkappa)x/v_x}C_4\frac{t}{2\varkappa}\begin{bmatrix}1\\ -t/(\varkappa+\chi)\\ t/(\varkappa-\chi)\\-1\end{bmatrix} \ ,
\end{eqnarray}
where $\varkappa = \sqrt{\chi^2 + t^2}$ and $C_i$ ($i = 1,2,3,4$) are arbitrary constants. Note that within our model description the intralayer Cooper pairing is disregarded, so we put $C_1 = 0$ in the following. Particular solutions are obtained using the variation of constant method, which leads us to the following set of equations
\begin{subequations}\label{variation_of_constant_system}
\begin{align}
-iv_x\frac{\partial C_2}{\partial x} = 4\Delta_{\rm int}(x){\rm sgn}(\omega_n)\frac{t}{\sqrt{2}\varkappa}e^{2\omega_n x/v_x} \ ,\\
-iv_x\frac{\partial C_3}{\partial x} = 2\Delta_{\rm int}(x){\rm sgn}(\omega_n)\frac{\chi }{\varkappa}e^{(2\omega_n - 2i\varkappa)x/v_x} \ ,\\
-iv_x\frac{\partial C_4}{\partial x} = 2\Delta_{\rm int}(x){\rm sgn}(\omega_n)\frac{\chi}{\varkappa}e^{(2\omega_n + 2i\varkappa)x/v_x} \ .
\end{align}
\end{subequations}
Performing the integration in Eqs.~(\ref{variation_of_constant_system}), we get the particular solutions
\begin{eqnarray}\label{particular_solutions}
\begin{pmatrix}f_{11}\\ f_{12}\\ f_{21}\\ f_{22}\end{pmatrix} =  \frac{2i{\rm sgn}(\omega_n)}{v_x}\frac{t^2}{\varkappa^2}\int_{c_2}^xds \ \Delta_{\rm int}(s)e^{-2\omega_n(x-s)/v_x}\begin{pmatrix}-\chi/t\\ 1\\ 1\\ \chi/t\end{pmatrix}  \\
\nonumber
+\frac{i{\rm sgn}(\omega_n)}{v_x}\frac{\chi t}{\varkappa^2}\int_{c_3}^xds \ \Delta_{\rm int}(s)e^{(-2\omega_n + 2i\varkappa)(x-s)/v_x}\begin{pmatrix}1\\ t/\varkappa_-\\ -t/\varkappa_+\\ -1\end{pmatrix}\\
\nonumber
+ \frac{i{\rm sgn}(\omega_n)}{v_x}\frac{\chi t}{\varkappa^2}\int_{c_4}^xds \ \Delta_{\rm int}(s)e^{(-2\omega_n - 2i\varkappa)(x-s)/v_x}\begin{pmatrix}1\\ -t/\varkappa_+\\ t/\varkappa_-\\ -1\end{pmatrix} ,
\end{eqnarray}
Here $\varkappa_{\pm} = \varkappa \pm \chi$ and $c_{i}$ ($i = 2,3,4$) are arbitrary constants. We have verified that particular solutions within $t\to 0$ limit and for a homogeneous band splitting, which can be easily obtained separately, coincide with expressions~(\ref{particular_solutions}) within the same limiting case. Our further strategy is to write down the above solutions in the regions of constant relative band shift (for $x<0$ and $x > 0$) and then match them continuously at the domain wall (at $x = 0$). The constants $c_2$, $c_3$, and $c_4$ in Eq.~(\ref{particular_solutions}) are chosen, so that the anomalous correlation functions $f_{ij}$ ($i,j = 1,2$) vanish within the limit $|x|\to \infty$.

The solution for $\omega_n > 0$, $v_x > 0$, and for $x < 0$ can be written as
\begin{eqnarray}\label{positive_wn_positive_vx_negative_x}
\begin{pmatrix}f_{11}\\f_{12}\\f_{21}\\f_{22}\end{pmatrix} = \frac{2i}{v_x}\frac{t^2}{\varkappa_L^2}\int_{-\infty}^0du \ \Delta_{\rm int}(x+u)e^{2\omega_n u/v_x}\begin{pmatrix}-\chi_L/t\\1\\1\\ \chi_L/t\end{pmatrix} \\
\nonumber
+\frac{i}{v_x}\frac{\chi_Lt}{\varkappa_L^2}\int_{-\infty}^0du \ \Delta_{\rm int}(x + u)e^{2(\omega_n-i\varkappa_L)u}\begin{pmatrix}1\\ t/\varkappa_{L-}\\ -t/\varkappa_{L+} \\ -1\end{pmatrix} 
+\frac{i}{v_x}\frac{\chi_Lt}{\varkappa_L^2}\int_{-\infty}^0du \ \Delta_{\rm int}(x + u)e^{2(\omega_n+i\varkappa_L)u}\begin{pmatrix}1\\ -t/\varkappa_{L+}\\ t/\varkappa_{L-} \\ -1\end{pmatrix} \ .
\end{eqnarray}
Here $\chi_L = \chi(x<0)$, $\varkappa_L = \varkappa(x<0)$, and $\varkappa_{L\pm} = \varkappa_L \pm \chi_L$. For $\omega_n > 0$, $v_x > 0$, and $x > 0$ the solution can be presented in the form:
\begin{eqnarray}\label{positive_wn_positive_vx_positive_x}
\begin{pmatrix}f_{11}\\f_{12}\\f_{21}\\f_{22}\end{pmatrix} = e^{-2\omega_n x/v_x}C_2\frac{t}{\sqrt{2}\varkappa_R}\begin{pmatrix}-\chi_R/t\\ 1\\ 1\\ \chi_R/t\end{pmatrix} \\
\nonumber
+ e^{(-2\omega_n + 2i\varkappa_R)x/v_x}C_3\frac{t}{2\varkappa_R}\begin{pmatrix}1\\ t/\varkappa_{R-}\\ - t/\varkappa_{R+}\\ -1\end{pmatrix} + e^{(-2\omega_n - 2i\varkappa_R)x/v_x}C_4\frac{t}{2\varkappa_R}\begin{pmatrix}1\\ -t/\varkappa_{R+}\\ t/\varkappa_{R-}\\-1\end{pmatrix}  \\
\nonumber
+ \frac{2i}{v_x}\frac{t^2}{\varkappa_R^2}\int_{0}^xds \ \Delta_{\rm int}(s)e^{-2\omega_n(x-s)/v_x}\begin{pmatrix}-\chi_R/t\\ 1\\ 1\\ \chi_R/t\end{pmatrix} \\
\nonumber
+\frac{i\chi_R t}{v_x\varkappa_R^2}\int_{0}^xds\Delta_{\rm int}(s)e^{-2(\omega_n - i\varkappa_R)(x-s)/v_x}\begin{pmatrix}1\\ t/\varkappa_{R-}\\ -t/\varkappa_{R+}\\ -1\end{pmatrix} 
+ \frac{i\chi_R t}{v_x\varkappa_R^2}\int_{0}^xds\Delta_{\rm int}(s)e^{-2(\omega_n + i\varkappa_R)(x-s)/v_x}\begin{pmatrix}1\\ -t/\varkappa_{R+}\\ t/\varkappa_{R-}\\ -1\end{pmatrix}.
\end{eqnarray}
Here $\chi_R = \chi(x>0)$, $\varkappa_R = \varkappa(x>0)$, and $\chi_{R\pm}=\chi_R \pm \chi_R$. Imposing the continuity condition on the solutions~(\ref{positive_wn_positive_vx_negative_x}) and (\ref{positive_wn_positive_vx_positive_x}) at the domain wall (at $x = 0$), we get a linear system for the coefficients
\begin{subequations}\label{linear_system_solution_matching_for_positive_wn_positive_vx}
\begin{align}
-C_2\frac{\chi_R}{\sqrt{2}\varkappa_R} + C_3\frac{t}{2\varkappa_R} + C_4\frac{t}{2\varkappa_R} = \frac{2i}{v_x}\frac{t}{\varkappa_L^2}\bar{d}_-\left(-\chi_L\right) + \frac{i}{v_x}\frac{\chi_Lt(\bar{d}_{--} + \bar{d}_{-+})}{\varkappa_L^2} \ ,\\
C_2\frac{t}{\sqrt{2}\varkappa_R} + C_3\frac{t^2}{2\varkappa_R\varkappa_{R-}} - C_4\frac{t^2}{2\varkappa_R\varkappa_{R+}} = \frac{2i}{v_x}\frac{t^2\bar{d}_-}{\varkappa_L^2} 
+\frac{i}{v_x}\frac{\chi_Lt^2}{\varkappa_L^2}\left(\frac{\bar{d}_{--}}{\varkappa_{L-}}-\frac{\bar{d}_{-+}}{\varkappa_{L+}}\right) \ ,\\
C_2\frac{t}{\sqrt{2}\varkappa_R} - C_3\frac{t^2}{2\varkappa_R\varkappa_{R+}} + C_4\frac{t^2}{2\varkappa_R\varkappa_{R-}} = \frac{2i}{v_x}\frac{t^2\bar{d}_-}{\varkappa_L^2} 
+\frac{i}{v_x}\frac{\chi_Lt^2}{\varkappa_L^2(\chi_L^2 + t^2)}\left(\frac{-\bar{d}_{--}}{\varkappa_{L+}}+\frac{\bar{d}_{-+}}{\varkappa_{L-}}\right) \ .
\end{align}
\end{subequations}
Here we introduced the following quantities:
\begin{subequations}
\begin{align}
\bar{d}_- = \int_{-\infty}^0du \ \Delta_{\rm int}(u)e^{e^{2\omega_n u/v_x}} \ , \ \bar{d}_{-\mp} = \int_{-\infty}^0du \ \Delta_{\rm int}(u)e^{e^{2(\omega_n\mp i\varkappa_L) u/v_x}} \ .
\end{align}
\end{subequations}
Solving the system~(\ref{linear_system_solution_matching_for_positive_wn_positive_vx}), we get 
\begin{subequations}\label{matching_coefficients_positive_wn_positive_vx}
\begin{align}
C_2 = \frac{2i}{v_x}\bar{d}_-\frac{2t(t^2+\chi_L\chi_R)}{\sqrt{2}\varkappa_R\varkappa_L^2}+\frac{2i}{v_x}(\bar{d}_{--}+\bar{d}_{-+})\frac{t\chi_L(\chi_L-\chi_R)}{\sqrt{2}\varkappa_R\varkappa_L^2} \ ,\\
C_3 = \frac{2i}{v_x}\frac{2t^2\bar{d}_-(\chi_R - \chi_L)}{2\varkappa_R\varkappa_L^2} + \frac{2i}{v_x}\frac{\chi_L(t^2 + \chi_L\chi_R)(\bar{d}_{-+}+\bar{d}_{--})+\chi_L\varkappa_R\varkappa_L(\bar{d}_{--}-\bar{d}_{-+})}{2\varkappa_R\varkappa_L^2} \ ,\\
C_4 = \frac{2i}{v_x}\frac{2t^2\bar{d}_-(\chi_R - \chi_L)}{2\varkappa_R\varkappa_L^2} + \frac{2i}{v_x}\frac{\chi_L(t^2 + \chi_L\chi_R)(\bar{d}_{--}+\bar{d}_{-+})-\chi_L\varkappa_R\varkappa_L(\bar{d}_{--}-\bar{d}_{-+})]}{2\varkappa_R\varkappa_L^2} \ .
\end{align}
\end{subequations}
To construct the solutions for $\omega_n < 0$ and $v_x<0$ it is convenient to use the symmetry of the Eilenberger equations~(\ref{eilenberger_domain_wall_without_spin_spin_singlet_case}) $(i,j= 1,2)$
\begin{equation}\label{eilenberger_symmetry_relation}
f_{ij}(x,v_x,\omega_n,t,\chi) = f_{ij}(x,-v_x,-\omega_n,-t,-\chi) \ .
\end{equation}

As a next step, we write down the solutions for negative Matsubara frequencies $\omega_n$ and positive velocity projections $v_x$. So, the solution for $\omega_n < 0$, $v_x > 0$, and $x<0$ can be written as
\begin{eqnarray}\label{negative_wn_positive_vx_negative_x}
\begin{pmatrix}f_{11}\\ f_{12}\\ f_{21}\\ f_{22}\end{pmatrix} = e^{2|\omega_n|x/v_x}C_2\frac{t}{\sqrt{2}\varkappa_L}\begin{pmatrix}-\chi_L/t\\1\\1\\ \chi_L/t\end{pmatrix} \\
\nonumber
+ e^{(2|\omega_n|+2i\varkappa_L)x/v_x}C_3\frac{t}{2\varkappa_L}\begin{pmatrix}1\\ t/\varkappa_{L-}\\ -t/\varkappa_{L+}\\ -1\end{pmatrix} + e^{(2|\omega_n|-2i\varkappa_L)x/v_x}C_4\frac{t}{2\varkappa_L}\begin{pmatrix}1\\ -t/\varkappa_{L+} \\ t/\varkappa_{L-}\\ -1\end{pmatrix}  \\
\nonumber
-\frac{2i}{v_x}\frac{t^2}{\varkappa_L^2}\int_0^xds \ \Delta_{\rm int}(s)e^{2|\omega_n|(x-s)/v_x}\begin{pmatrix}-\chi_L/t\\1\\1\\ \chi_L/t\end{pmatrix} \\
\nonumber
- \frac{i\chi_Lt}{v_x\varkappa_L^2}\int_0^xds\Delta_{\rm int}(s)e^{2(|\omega_n| + i\varkappa_L)(x-s)/v_x}\begin{pmatrix}1\\ t/\varkappa_{L-}\\ -t/\varkappa_{L+}\\ -1\end{pmatrix} 
-\frac{i\chi_Lt}{v_x\varkappa_L^2}\int_0^xds \Delta_{\rm int}(s)e^{2(|\omega_n|-i\varkappa_L)(x-s)/v_x}\begin{pmatrix}1\\ -t/\varkappa_{L+}\\ t/\varkappa_{L-}\\ -1\end{pmatrix}.
\end{eqnarray}
The solution for $\omega_n<0$, $v_x > 0$, and $x>0$ can be presented in the form:
\begin{eqnarray}\label{negative_wn_positive_vx_positive_x}
\begin{pmatrix}f_{11}\\ f_{12}\\ f_{21}\\ f_{22}\end{pmatrix} = \frac{2i}{v_x}\frac{t^2}{\varkappa_R^2}\int_0^{+\infty} du \ \Delta_{\rm int}(x+u)e^{-2|\omega_n|u/v_x}\begin{pmatrix}-\chi_R/t\\1\\1\\ \chi_R/t\end{pmatrix} \\
\nonumber
+\frac{i}{v_x}\frac{\chi_Rt}{\varkappa_R^2}\int_0^{+\infty}du \ \Delta_{\rm int}(x+u)e^{-(2|\omega_n|+2i\varkappa_R)u/v_x}\begin{pmatrix}1\\ t/\varkappa_{R-} \\ -t/\varkappa_{R+} \\ -1\end{pmatrix} \\
\nonumber
+ \frac{i}{v_x}\frac{\chi_Rt}{\varkappa_R^2}\int_0^{+\infty}du \ \Delta_{\rm int}(x+u)e^{-(2|\omega_n|-2i\varkappa_R)u/v_x}\begin{pmatrix}1\\ -t/\varkappa_{R+} \\ t/\varkappa_{R-}\\ -1\end{pmatrix} \ .
\end{eqnarray}
Matching the solutions~(\ref{negative_wn_positive_vx_negative_x}) and (\ref{negative_wn_positive_vx_positive_x}) at the domain wall (at $x = 0$), we get the following linear system:
\begin{subequations}\label{linear_system_solution_matching_for_negative_wn_positive_vx}
\begin{align}
-C_2\frac{\chi_L}{\sqrt{2}\varkappa_L} + C_3\frac{t}{2\varkappa_L} + C_4\frac{t}{2\varkappa_L} = \frac{2i}{v_x}\frac{t}{\varkappa_R^2}\bar{d}_+\left(-\chi_R\right) + \frac{i}{v_x}\frac{\chi_Rt}{\varkappa_R^2}(\bar{d}_{+-}+\bar{d}_{++}) \ ,\\
C_2\frac{t}{\sqrt{2}q_L} + C_3\frac{t^2}{2\varkappa_L\varkappa_{L-}} - C_4\frac{t^2}{2\varkappa_L\varkappa_{L+}} = \frac{2i}{v_x}\frac{t^2}{\varkappa_R^2}\bar{d}_{+}+\frac{i}{v_x}\frac{\chi_Rt^2}{\varkappa_R^2}\left(\frac{\bar{d}_{+-}}{\varkappa_{R-}}-\frac{\bar{d}_{++}}{\varkappa_{R+}}\right) \ ,\\
C_2\frac{t}{\sqrt{2}\varkappa_L} - C_3\frac{t^2}{2\varkappa_L\varkappa_{L+}} + C_4\frac{t^2}{2\varkappa_L\varkappa_{L-}} = \frac{2i}{v_x}\frac{t^2}{\varkappa_R^2}\bar{d}_+ + \frac{i}{v_x}\frac{\chi_Rt^2}{\varkappa_R^2}\left(\frac{-\bar{d}_{+-}}{\varkappa_{R+}}+\frac{\bar{d}_{++}}{\varkappa_{R-}}\right) \ .
\end{align}
\end{subequations}
Here we introduced 
\begin{equation}
\bar{d}_+ = \int_0^{+\infty}du \ \Delta_{\rm int}(u)e^{-2|\omega_n|u/v_x} \ , \ \ \bar{d}_{+\pm} = \int_0^{+\infty}du \ \Delta_{\rm int}(u)e^{(-2|\omega_n|\pm 2i\varkappa_R)u/v_x} \ .
\end{equation}
Solving the system~(\ref{linear_system_solution_matching_for_negative_wn_positive_vx}), we get
\begin{subequations}\label{matching_coefficients_negative_wn_positive_vx}
\begin{align}
C_2 = \frac{2i}{v_x}\frac{2t\bar{d}_+(t^2 + \chi_L\chi_R)}{\sqrt{2}\varkappa_L\varkappa_R^2} + \frac{2i}{v_x}\frac{t\chi_R(\chi_R-\chi_L)(\bar{d}_{+-}+\bar{d}_{++})}{\sqrt{2}\varkappa_L\varkappa_R^2} \ ,\\
C_3 = \frac{2i}{v_x}\frac{2t^2\bar{d}_+(\chi_L - \chi_R)}{2\varkappa_L\varkappa_R^2}+\frac{2i}{v_x}\frac{\chi_R(t^2 + \chi_L\chi_R)(\bar{d}_{++}+\bar{d}_{+-})-\chi_R\varkappa_R\varkappa_L(\bar{d}_{++}-\bar{d}_{+-})]}{2\varkappa_L\varkappa_R^2} \ ,\\
C_4 = \frac{2i}{v_x}\frac{2t^2\bar{d}_+(\chi_L - \chi_R)}{2\varkappa_L\varkappa_R^2}+\frac{2i}{v_x}\frac{\chi_R(t^2 + \chi_L\chi_R)(\bar{d}_{+-}+\bar{d}_{++})+\chi_R\varkappa_R\varkappa_L(\bar{d}_{++}-\bar{d}_{+-})]}{2\varkappa_L\varkappa_R^2} \ .
\end{align}
\end{subequations}
The continuous solution of the Eilenberger equation for $\omega_n > 0$ and $v_x <0$ can be obtained using the above Eqs.~(\ref{negative_wn_positive_vx_negative_x}), (\ref{negative_wn_positive_vx_positive_x}), (\ref{matching_coefficients_negative_wn_positive_vx}) as well as the symmetry relation~(\ref{eilenberger_symmetry_relation}).

As a next step, we substitute the obtained solutions of the Eilenberger equations into the self-consistency equation~(\ref{eq:self_cons_main}) in the main text. It is convenient to recast the summation in the self-consistency equation to the one over positive Matsubara frequencies $\omega_n>0$ and positive velocity projections $v_x>0$. For this purpose, below we write down explicitly the sum of the obtained solutions.

For $x < 0$ we find
\begin{eqnarray}\label{sum_of_solutions_negative_x}
f_{12}^{-}(x) \equiv f_{12}(\omega_n > 0, v_x > 0, x<0) + f_{12}(\omega_n > 0, v_x < 0, x<0) \\
\nonumber
 + f_{12}(\omega_n < 0, v_x > 0, x<0) + f_{12}(\omega_n < 0, v_x < 0, x<0) = \\
\nonumber
 = \frac{2i}{|v_x|}\frac{2t^2}{\varkappa_L^2}\int_{-\infty}^0du \ \Delta_{\rm int}(x+u)e^{2|\omega_n| u /|v_x|} + \frac{2i}{|v_x|}\frac{\chi_L^2}{\varkappa_L^2}\int_{-\infty}^0du \ \Delta_{\rm int}(x+u)e^{2|\omega_n| u /|v_x|}2\cos\left(\frac{2\varkappa_L u}{|v_x|}\right) - \\
\nonumber
-\frac{2i}{v_x}\frac{2t^2}{\varkappa_L^2}\int_0^xds \ \Delta_{\rm int}(s)e^{2|\omega_n|(x-s)/|v_x|} - \frac{2i}{|v_x|}\frac{\chi_L^2}{\varkappa_L^2}\int_0^xds \ \Delta_{\rm int}(s)e^{2|\omega_n|(x-s)/|v_x|}2\cos\left[\frac{2\varkappa_L(x-s)}{|v_x|}\right] + \\
\nonumber
+ e^{2|\omega_n| x/|v_x|}\frac{t}{\sqrt{2}\varkappa_L}2C_2 + e^{(2|\omega_n| + 2i\varkappa_L)x/|v_x|}C_3(\chi_L/\varkappa_L) + e^{(2|\omega_n| - 2i\varkappa_L)x/|v_x|}C_4(\chi_L/\varkappa_L) \ .
\end{eqnarray}
In the above Eq.~(\ref{sum_of_solutions_negative_x}) the constants $C_2$, $C_3$, and $C_4$ are determined by Eqs.~(\ref{matching_coefficients_negative_wn_positive_vx}). Below we show the corresponding expressions for the above mentioned constants in the case of a stepwise relative band shift $\chi_L = \chi$ and $\chi_R = -\chi$
\begin{subequations}\label{matching_constants_symmetric_bandshift_negative_x}
\begin{align}
\frac{t}{\sqrt{2}\varkappa_L}2C_2 = \frac{2i}{v_x}\left[\frac{2t^2(t^2 - \chi^2)}{\varkappa^4}\bar{d}_+ + \frac{2\chi^2t^2}{\varkappa^4}(\bar{d}_{+-}+\bar{d}_{++})\right] \ ,\\
C_3\frac{\chi_L}{\varkappa_L} = \frac{2i}{v_x}\left\{\frac{2t^2\chi^2}{\varkappa^4}\bar{d}_+ - \frac{\chi^2t^2}{\varkappa^4}\bar{d}_{+-} + \frac{\chi^4}{\varkappa^4}\bar{d}_{++}\right\} \ ,\\
C_4\frac{\chi_L}{\varkappa_L} = \frac{2i}{v_x}\left\{\frac{2t^2\chi^2}{\varkappa^4}\bar{d}_+ - \frac{\chi^2t^2}{\varkappa^4}\bar{d}_{++}+\frac{\chi^4}{\varkappa^4}\bar{d}_{+-}\right\} \ .
\end{align}
\end{subequations}

For $x>0$ we obtain
\begin{eqnarray}\label{sum_of_solutions_positive_x}
f_{12}^{+}(x) \equiv f_{12}(\omega_n > 0, v_x > 0, x>0) + f_{12}(\omega_n > 0, v_x < 0, x>0) \\
\nonumber
 + f_{12}(\omega_n < 0, v_x > 0, x>0) + f_{12}(\omega_n < 0, v_x < 0, x>0) = \\
 \nonumber
= \frac{2i}{|v_x|}\frac{2t^2}{\varkappa_R^2}\int_0^{+\infty}du \ \Delta_{\rm int}(x+u)e^{-2|\omega_n| u/|v_x|} + \frac{2i}{|v_x|}\frac{\chi_R^2}{\varkappa_R^2}\int_0^{+\infty}du \ \Delta_{\rm int}(x+u)e^{-2|\omega_n| u/|v_x|}2\cos\left(\frac{2\varkappa_Ru}{|v_x|}\right) + \\
\nonumber
+ \frac{2i}{|v_x|}\frac{2t^2}{\varkappa_R^2}\int_0^xds \ \Delta_{\rm int}(s)e^{-2|\omega_n|(x-s)/|v_x|} + \frac{2i}{|v_x|}\frac{\chi_R^2}{\varkappa_R^2}\int_0^xds \ \Delta_{\rm int}(s)e^{-2|\omega_n|(x-s)/|v_x|}2\cos\left[\frac{2\varkappa_R(x-s)}{|v_x|}\right] + \\
\nonumber
+e^{-2|\omega_n|x/|v_x|}\frac{t}{\sqrt{2}\varkappa_R}2C_2 + e^{(-2|\omega_n| + 2i\varkappa_R)x/|v_x|}C_3\frac{\chi_R}{\varkappa_R} + e^{(-2|\omega_n| - 2i\varkappa_R)x/|v_x|}C_4\frac{\chi_R}{\varkappa_R} \ .
\end{eqnarray}
The constants $C_2$, $C_3$, and $C_4$ in Eq.~(\ref{sum_of_solutions_positive_x}) are determined by Eqs.~(\ref{matching_coefficients_positive_wn_positive_vx}). Below we show the appropriate expressions for the constants in the case of a stepwise band shift $\chi_L = \chi$ and $\chi_R = -\chi$
\begin{subequations}\label{matching_constants_symmetric_bandshift_positive_x}
\begin{align}
\frac{2t}{\sqrt{2}\varkappa_R}C_2 = \frac{2i}{v_x}\left[\frac{2t^2(t^2 - \chi^2)}{\varkappa^4}\bar{d}_- + \frac{2t^2\chi^2}{\varkappa^4}(\bar{d}_{--} + \bar{d}_{-+})\right] \ ,\\
C_3\frac{\chi_R}{\varkappa_R} = \frac{2i}{v_x}\left\{\frac{2t^2\chi^2}{\varkappa^4}\bar{d}_- - \frac{\chi^2t^2}{\varkappa^4}\bar{d}_{--}+\frac{\chi^4}{\varkappa^4}\bar{d}_{-+}\right\} \ ,\\
C_4\frac{\chi_R}{\varkappa_R} = \frac{2i}{v_x}\left\{\frac{2t^2\chi^2}{\varkappa^4}\bar{d}_- - \frac{\chi^2t^2}{\varkappa^4}\bar{d}_{-+}+\frac{\chi^4}{\varkappa^4}\bar{d}_{--}\right\} \ .
\end{align}
\end{subequations}

The functions $f_{12}^{\pm}(x)$ enter the self-consistency equation in the following fashion
\begin{subequations}\label{self_cons_real_space}
\begin{align}
\Delta_{\rm{int}}(x>0) = -i\pi\lambda T\sum_{\omega_n > 0}\int_{v_x>0}\frac{d\mathbf{n}}{2\pi}f_{12}^+(x,\mathbf{n}) \ ,\\
\Delta_{\rm{int}}(x<0) = -i\pi\lambda T\sum_{\omega_n > 0}\int_{v_x>0}\frac{d\mathbf{n}}{2\pi}f_{12}^-(x,\mathbf{n}) \ .
\end{align}
\end{subequations}
Substitution of Eqs.~(\ref{sum_of_solutions_negative_x}), (\ref{sum_of_solutions_positive_x}) into Eqs.~(\ref{self_cons_real_space}) yields that for a symmetric stepwise band shift both the anomalous function $f_{12}(x)$  and the superconducting gap function $\Delta_{\rm int}(x)$ are of even spatial parity. Thus, in this case one has the following relations $\bar{d}_- = \bar{d}_+$ and $\bar{d}_{-\mp} = \bar{d}_{+\pm}$. Substituting Eqs.~(\ref{sum_of_solutions_negative_x}), (\ref{matching_constants_symmetric_bandshift_negative_x}), (\ref{sum_of_solutions_positive_x}), and (\ref{matching_constants_symmetric_bandshift_positive_x}) into Eqs.~(\ref{self_cons_real_space}), and introducing the Fourier transformed functions 
\begin{equation}
\Delta_{\rm int}(x) = \int\frac{dk}{2\pi}\Delta_{\rm int}(k)e^{ikx} \ ,
\end{equation}
we derive the following self-consistency equation in the momentum representation
\begin{eqnarray}\label{self_cons_domain_wall_appendix}
\Delta_{\rm int}(k) = -i\pi\lambda T \sum_{\omega_n > 0}\int_{v_x>0}\frac{d\mathbf{n}}{2\pi} \biggl\{ \ \int_{-\infty}^{+\infty}\frac{dk'}{2\pi} \Delta_{\rm int}(k') 2\pi \delta(k-k')\biggl\{\frac{4it^2}{\varkappa^2}\frac{\omega_n}{(\omega_n^2 + k^2v_x^2/4)} \\
\nonumber
+ \frac{4i\chi^2}{\varkappa^2}{\rm Re}\left[\frac{(\omega_n - i\varkappa)}{(\omega_n - i\varkappa)^2 + k^2v_x^2/4}\right]\biggl\} \\
\nonumber
+\int_{-\infty}^{+\infty}\frac{dk'}{2\pi}\Delta_{\rm int}(k')\biggl\{\frac{-2it^2v_x}{\varkappa^2}{\rm Re}\left[\frac{1}{(\omega_n - ik'v_x/2)(\omega_n - ikv_x/2)}\right] \\
\nonumber
-\frac{i\chi^2v_x}{\varkappa^2}{\rm Re}\left[\frac{1}{(\omega_n + i\varkappa - ik'v_x/2)(\omega_n + i\varkappa-ikv_x/2)}+\frac{1}{(\omega_n - i\varkappa - ik'v_x/2)(\omega_n - i\varkappa - ikv_x/2)}\right]
\biggl\} \\
\nonumber
+iv_x\int_{-\infty}^{+\infty}\frac{dk'}{2\pi}\Delta_{\rm int}(k'){\rm Re}\biggl\{
\frac{2t^2(t^2-\chi^2)}{\varkappa^4}\frac{1}{(\omega_n - ik'v_x/2)(\omega_n-ikv_x/2)} \\
\nonumber
+\frac{4t^2\chi^2}{\varkappa^4}\frac{(\omega_n - ik'v_x/2)}{(\omega_n - ikv_x/2)[(\omega_n - ik'v_x/2)^2 + \varkappa^2]}+\frac{4t^2\chi^2}{\varkappa^4}\frac{(\omega_n - ikv_x/2)}{(\omega_n - ik'v_x/2)[(\omega_n - ikv_x/2)^2 + \varkappa^2]} \\
\nonumber
-\frac{\chi^2t^2}{\varkappa^4}\left[\frac{1}{(\omega_n + i\varkappa - ik'v_x/2)(\omega_n + i\varkappa - ikv_x/2)}+\frac{1}{(\omega_n - i\varkappa-ik'v_x/2)(\omega_n - i\varkappa - ikv_x/2)}\right] \\
\nonumber
+\frac{\chi^4}{\varkappa^4}\left[\frac{1}{(\omega_n - i\varkappa - ik'v_x/2)(\omega_n + i\varkappa-ikv_x/2)}+\frac{1}{(\omega_n + i\varkappa - ik'v_x/2)(\omega_n-i\varkappa-ikv_x/2)}\right]
\biggl\}\biggl\} \ .
\end{eqnarray}
The resulting self-consistency equation~(\ref{self_cons_domain_wall_appendix}) can be cast to the form presented by Eqs.~(\ref{self_cons_domain_wall_main_text}) in the main text with the kernel:
\begin{subequations}\label{self_cons_kernel_singlet_appendix}
    \begin{align}
        K(k,k')=2\pi\delta\left(k-k'\right)K_h(k)+K_{\rm inh}(k,k'), \\
        K_h(k)=4\pi \lambda T \sum_{\omega_n > 0}\int_{-\pi/2}^{\pi/2}\frac{d\varphi}{2\pi}\text{Re}\left[\frac{4t^{2}\omega_{n}}{\varkappa^{2}\left(\omega_{n}^{2}+k^{2}v_x^{2}/4\right)}+\frac{4\chi^{2}}{\varkappa^{2}}\frac{(\omega_{n}-i\varkappa)}{(\left(\omega_{n}-i\varkappa\right)^{2}+k^{2}v_x^{2}/4)}\right], \\
        K_{\rm inh}(k,k')=
        \pi\lambda T\sum_{\omega_{n}>0}\int_{v_x>0}\frac{d\mathbf{n}}{2\pi}\text{Re}\biggl\{ v_x\frac{\chi^{4}}{\varkappa^{4}}\frac{4\varkappa^{2}}{\left(\varkappa^{2}+\left(\omega_{n}-ik'v_x/2\right)^{2}\right)\left(\varkappa^{2}+\left(\omega_{n}-ikv_x/2\right)^{2}\right)} + \\ \nonumber
        +v_x\frac{4t^{2}\chi^{2}}{\varkappa^{4}}\frac{\left(\varkappa^{2}+2\left(\omega_{n}-ik'v_x/2\right)\left(\omega_{n}-ikv_x/2\right)\right)\left(\varkappa^{2}-\left(\omega_{n}-ik'v_x/2\right)\left(\omega_{n}-ikv_x/2\right)\right)}{\left(\varkappa^{2}+\left(\omega_{n}-ik'v_x/2\right)^{2}\right)\left(\varkappa^{2}+\left(\omega_{n}-ikv_x/2\right)^{2}\right)\left(\omega_{n}-ik'v_x/2\right)\left(\omega_{n}-ikv_x/2\right)}\biggl\}.
    \end{align}
\end{subequations}
It is easy to see from the above equation that the inhomogeneous part of the kernel vanishes within the limit $\chi \to 0$ for arbitrary tunneling amplitude $t$. We have also verified that the resulting expressions~(\ref{self_cons_kernel_singlet_appendix}) are consistent with the corresponding results in the absence of tunnel coupling, which we derived separately from the Eilenberger equations~(\ref{eilenberger_domain_wall_without_spin_spin_singlet_case}) along the lines mentioned in this section. So, the self-consistency equation for $t = 0$ and the stepwise relative band shift $\chi_L = \chi$, $\chi_R = -\chi$ has the form similar to Eq.~(\ref{selfcons_singlet_form}) but with the following kernel 
\begin{subequations}
\begin{align}
K(k,k') = 2\pi \delta(k-k')K_h(k) + \\
\nonumber
+ 4\pi\lambda T {\rm Re}\sum_{\omega_n > 0}\int_{-\pi/2}^{\pi/2}\frac{d\varphi}{2\pi}\frac{\chi^2v_F\cos\varphi}{\{[\omega_n + ikv_F\cos(\varphi)/2]^2 + \chi^2\}\{[\omega_n + ik'v_F\cos(\varphi)/2]^2 + \chi^2\}} \ ,\\
K_h(k) = 4\pi \lambda T {\rm Re}\sum_{\omega_n > 0}\int_{-\pi/2}^{\pi/2}\frac{d\varphi}{2\pi}\frac{\omega_n + i\chi}{(\omega_n + i\chi)^2 + (k^2v_F^2\cos^2\varphi)/4} \ .
\end{align}
\end{subequations}

\section{Derivation of Eqs.~(\ref{delta_potential_approximation_equation})-(\ref{Tc_inhomogeneous_main_result}) in the main text (linearized Ginzburg-Landau-type equation for the spin-singlet domain wall state)}\label{Ginzburg_Landau_appendix}

Here we present the derivation of Eqs.~(\ref{delta_potential_approximation_equation})-(\ref{Tc_inhomogeneous_main_result}) in the main text. First, we provide the derivation details for the Ginzburg-Landau equation ~(\ref{delta_potential_approximation_equation}) and various coefficients in the Ginzburg-Landau theory given in Eqs.~(\ref{temperature_dependent_coefficients}). Second, we derive the equation for the critical temperature of the localized state~(\ref{Tc_localized_equation}) and its expansion in the vicinity of the critical temperature $T_{c0}$ for vanishing relative band offset and tunneling amplitude~(\ref{Tc_inhomogeneous_main_result}).

Our starting point is the expression for the kernel~(\ref{self_cons_kernel_singlet_appendix}) entering the self-consistency equation~(\ref{selfcons_singlet_form}). As a first step, we perform the gradient expansion of the homogeneous part of the kernel $K_h(k)$ over $k$ up to the term $\propto k^2$
\begin{eqnarray}\label{homogeneous_kernel_expansion_appendix}
K_{h}\left(k\right)  = 4\pi\lambda T {\rm Re}\sum_{\omega_n > 0}\int_{-\pi/2}^{\pi/2}\frac{d\varphi}{2\pi}\frac{1}{\varkappa^2}\left[\frac{t^{2}\omega_{n}}{\left(\omega_{n}^{2}+k^{2}v_x^{2}/4\right)}+\frac{\chi^{2}\left(\omega_{n}-i\varkappa\right)}{\left(\omega_{n}-i\varkappa\right)^{2}+k^{2}v_x^{2}/4}\right] \\
\nonumber
 \approx 4\pi\lambda T\sum_{\omega_n > 0}\int_{-\pi/2}^{\pi/2}\frac{d\varphi}{2\pi}\frac{1}{\varkappa^2}\left[\frac{t^{2}}{\omega_{n}}+\frac{\chi^{2}\omega_{n}}{\omega_{n}^{2}+\varkappa^{2}}-\frac{k^{2}v_x^{2}}{4}\left(\frac{t^{2}}{\omega_{n}^{3}}+\frac{\chi^{2}\omega_{n}\left(\omega_{n}^{2}-3\varkappa^{2}\right)}{\left(\omega_{n}^{2}+\varkappa^{2}\right)^{3}}\right)\right] \ .
\end{eqnarray}
Here $\varkappa = \sqrt{\chi^2 + |t|^2}$ and $v = v_F\cos(\varphi)$. The summation over Matsubara frequencies in Eq.~(\ref{homogeneous_kernel_expansion_appendix}) can be performed, using the following relations
\begin{subequations}\label{sums_over_Matsubara_frequencies_appendix}
\begin{align}
\sum_{\omega_{n}>0}\frac{1}{\omega_{n}}\approx\frac{\ln\left(2\gamma\omega_{D}/\pi T\right)}{2\pi T} \ , \ \ \ \sum_{\omega_{n}>0}\frac{1}{\omega_{n}^{3}}=\frac{7\zeta\left(3\right)}{8\pi^{3}T^{3}} \ ,\\
\sum_{\omega_{n}>0}\frac{\omega_{n}}{\omega_{n}^{2}+\varkappa^{2}}\approx \frac{\ln\left(2\gamma\omega_{D}/\pi T\right)}{2\pi T}+\frac{1}{2\pi T}\text{Re}\left[\psi\left(\frac{1}{2}\right)-\psi\left(\frac{1}{2}-i\frac{\varkappa}{2\pi T}\right)\right] \ ,\\
\,\sum_{\omega_{n}>0}\frac{\omega_{n}\left(\omega_{n}^{2}-3\varkappa^{2}\right)}{\left(\omega_{n}^{2}+\varkappa^{2}\right)^{3}}=-\frac{1}{16\pi^{3}T^{3}}\text{Re}\left[\psi_2\left(\frac{1}{2}-i\frac{\varkappa}{2\pi T}\right)\right] \ ,
\end{align}
\end{subequations}
where $\gamma \approx 1.78$, $\omega_D$ is the cut-off frequency, $\zeta(x)$ is the Riemann zeta function, $\psi(x)$ is the digamma function, and $\psi_2(x) = d^2\psi/dx^2$. Substituting Eq.~(\ref{sums_over_Matsubara_frequencies_appendix}) into Eq.~(\ref{homogeneous_kernel_expansion_appendix}), performing the trivial integration over the momentum directions, we obtain the relevant expansion of the homogeneous part of the kernel. Thus, in the absence of the domain wall (for a homogeneous relative band shift), the above expansion gives us the linearized self-consistency equation
\begin{eqnarray}
\biggl[-\frac{v_{F}^{2}k^{2}}{16}\left(\frac{7\zeta\left(3\right)t^{2}}{8\pi^{3}T^{3}}-\frac{\chi^{2}}{16\pi^{3}T^{3}}\text{Re}\left[\psi_2\left(\frac{1}{2}-i\frac{\varkappa}{2\pi T}\right)\right]\right) \\
\nonumber
-\frac{\varkappa^{2}}{4\pi T}\ln\left(\frac{T}{T_{c0}}\right)+\frac{\chi^{2}}{4\pi T}\text{Re}\left[\psi\left(\frac{1}{2}\right)-\psi\left(\frac{1}{2}-i\frac{\varkappa}{2\pi T}\right)\right]\biggl]\Delta\left(k\right)=0 \ .
\end{eqnarray}
As a second step, we consider the inhomogeneous part of the kernel $K_{\rm inh}(k,k')$ in Eq.~(\ref{self_cons_kernel_singlet_appendix}) and then put $k = k' = 0$. We find
\begin{eqnarray}\label{kernel_inhomogeneous_appendix}
K_{\rm inh}\left(0,0\right) = 4\pi\lambda T\sum_{\omega_n>0}\int_{-\pi/2}^{\pi/2}\frac{d\varphi}{2\pi}\frac{v_x\chi^{2}\left(\omega_{n}^{2}-t^{2}\right)}{\omega_{n}^{2}\left(\omega_{n}^{2}+\varkappa^{2}\right)^{2}} \ .
\end{eqnarray}
Performing the summation over the Matsubara frequencies in Eq.~(\ref{kernel_inhomogeneous_appendix}) 
\begin{equation}\label{inhomogeneous_frequency_sum}
\sum_{\omega_{n}}\frac{\left(\omega_{n}^{2}-t^{2}\right)}{\omega_{n}^{2}\left(\omega_{n}^{2}+\varkappa^{2}\right)^{2}}  =-\frac{\varkappa^{3}+\varkappa t^{2}\left(2+\cosh\left(\frac{\varkappa}{T}\right)\right)-\left(\varkappa^{2}+3t^{2}\right)T\sinh\left(\frac{\varkappa}{T}\right)}{16\varkappa^{5}T^{2}\cosh^{2}\left(\frac{\varkappa}{2T}\right)} \ ,
\end{equation}
the integration over the momentum directions, we get the additional contribution to the linearized self-consistency equation describing the effect of the Dirac delta well. Combining Eqs.~(\ref{selfcons_singlet_form}) and (\ref{homogeneous_kernel_expansion_appendix})-(\ref{inhomogeneous_frequency_sum}), we obtain the linearized self-consistency equation~(\ref{delta_potential_approximation_equation}) in the main text with the coefficients given by Eqs.~(\ref{temperature_dependent_coefficients}).

We get the following equation governing the critical temperature of the localized state
\begin{equation}
\frac{\varkappa^2}{4\pi T}\ln\left(\frac{T}{T_{c0}}\right)-\frac{\chi^2}{4\pi T}{\rm Re}\left[\psi\left(\frac{1}{2}\right)-\psi\left(\frac{1}{2}-i\frac{\varkappa}{2\pi T}\right)\right] = \frac{M \varkappa^4\alpha^2}{32\pi^2T^2} \ .
\end{equation}
Let us expand the right-hand side:
\begin{equation}
2M\alpha^2 = \frac{\chi^4}{T^4}\frac{\left[(\varkappa/T)^3+(\varkappa/T)(t/T)^2(2+\cosh(\varkappa/T))-((\varkappa/T)^2 + 3(t/T)^2\sinh(\varkappa/T))\right]^2}{\left[\frac{14\zeta(3)t^2}{16\pi^3T^3}-\frac{\chi^2}{16\pi^3T^3}{\rm Re}\left(\psi_2\left(\frac{1}{2}-i\frac{\varkappa}{2\pi T}\right)\right)\right](\varkappa/T)^{10}\cosh^4(\varkappa/2T)} \ .
\end{equation}
We have the following expansion of the denominator
\begin{eqnarray}
\frac{14\zeta(3)t^2}{16\pi^3T^3}-\frac{\chi^2}{16\pi^3T^3}{\rm Re}\left(\psi_2\left(\frac{1}{2}-i\frac{\varkappa}{2\pi T}\right)\right) \approx\\
\nonumber
\approx - \frac{\psi_2}{4\pi T}\left(\frac{\varkappa}{2\pi T}\right)^2\left[1 - \left(\frac{\chi}{2\pi T}\right)^2\left(\frac{\psi_4}{2\psi_2}-\frac{1}{4!}\frac{\psi_6}{\psi_2}\left(\frac{\varkappa}{2\pi T}\right)^2 + \frac{1}{6!}\frac{\psi_8}{\psi_2}\left(\frac{\varkappa}{2\pi T}\right)^4\right) + ...\right] \ .
\end{eqnarray}
Expansion of the right-hand-side yields
\begin{eqnarray}
\frac{2M \varkappa^2\alpha^2\pi}{16\pi T} \approx  -\frac{\pi^2}{\psi_2}\left(\frac{\chi}{T}\right)^4\left[\frac{1}{36}-\frac{1}{60}\left(\frac{t}{T}\right)^2 - \frac{1}{90}\left(\frac{\chi}{T}\right)^2\right]\\
\nonumber
\left\{1 + \tilde{\chi}^2\left[\frac{\psi_4}{2\psi_2}-\frac{1}{4!}\frac{\psi_6}{\psi_2}\tilde{\varkappa}^2 + \frac{1}{6!}\frac{\psi_8}{\psi_2}\tilde{\varkappa}^4\right] + \tilde{\chi}^4\left[\left(\frac{\psi_4}{2\psi_2}\right)^2 - \frac{\psi_4}{\psi_2}\frac{1}{4!}\frac{\psi_6}{\psi_2}\tilde{\varkappa}^2\right]+\tilde{\chi}^6\left(\frac{\psi_4}{2\psi_2}\right)^3\right\} \ .
\end{eqnarray}
Here $\tilde{\chi} = \chi/2\pi T$ and $\tilde{\varkappa} = \varkappa/2\pi T$. As a next step, we substitute the above expression into the self-consistency equation. We will seek the solution for the shift in the critical temperature in the form
\begin{equation}
\frac{T}{T_{c0}} = 1 + \epsilon \ ; \ \ \ \epsilon = \epsilon_2 + \epsilon_4 + \epsilon_6 \ ,
\end{equation}
where $\epsilon_{2,4,6}$ contain the terms $\propto (\chi/T_{c0})^n(t/T_{c0})^m$ with $m+n = 2$, 4, and $6$, respectively. Substituting the following expansion
\begin{eqnarray}
\label{expansion_1}
\ln\left(1 + \epsilon\right)\approx (\epsilon_2 + \epsilon_4 + \epsilon_6)-\frac{1}{2}(\epsilon_2^2 + 2\epsilon_2\epsilon_4)+\frac{1}{3}\epsilon_2^3 \ ,\\ \label{expansion_2} 
\frac{1}{(1+\epsilon)^2}\approx 1 - 2(\epsilon_2 + \epsilon_4)+3\epsilon_2^2 \ , \ \ \ \frac{1}{(1+\epsilon)^4} \approx 1 - 4\epsilon_2 \ ,
\end{eqnarray}
into the equation for the critical temperature, we obtain 
\begin{eqnarray}
\epsilon_2 + \epsilon_4 + \epsilon_6 - \frac{1}{2}(\epsilon_2^2 + 2\epsilon_2\epsilon_4)+\frac{1}{3}\epsilon_2^3 \\ \nonumber 
- \frac{\chi^2}{\varkappa^2}\left[\frac{\psi_2}{2}\left(\frac{\varkappa}{2\pi T_{c0}}\right)^2\left[1 - 2(\epsilon_2 + \epsilon_4)+3\epsilon_2^2\right]-\frac{\psi_4}{4!}\left(\frac{\varkappa}{2\pi T_{c0}}\right)^4(1-4\epsilon_2)+\frac{\psi_6}{6!}\left(\frac{\varkappa}{2\pi T_{c0}}\right)^6\right] =\\
\nonumber
=-\frac{4\pi^4}{\psi_2}\left(\frac{\chi}{T_{c0}}\right)^4\left(1 - 4\epsilon_2\right)\left[\frac{1}{36}-\frac{1}{60}\left(\frac{t}{T_{c0}}\right)^2  - \frac{1}{90}\left(\frac{\chi}{T_{c0}}\right)^2+\left(\frac{\chi}{2\pi T_{c0}}\right)^2\frac{\psi_4}{2\psi_2}\frac{1}{36}\right] \ .
\end{eqnarray}
Equating the terms with same powers of small parameters, we find the following expressions for $\epsilon_{2,4,6}$
\begin{subequations}
    \begin{align}
        \epsilon_2=\frac{\psi_2}{2}\left(\frac{\chi}{2\pi T_{c0}}\right)^2, \\
        \epsilon_4 = -\frac{3}{2}\epsilon_2^2-\frac{\psi_4}{4!}\left(\frac{\chi}{2\pi T_{c0}}\right)^2\left(\frac{\varkappa}{2\pi T_{c0}}\right)^2-\frac{\pi^2}{36\psi_2}\left(\frac{\chi}{T_{c0}}\right)^4, \\
        \epsilon_6=-\epsilon_2\epsilon_4+\frac{8}{3}\epsilon_2^3+4\epsilon_2\frac{\psi_4}{4!}\left(\frac{\chi}{2\pi T_{c0}}\right)^2\left(\frac{\varkappa}{2\pi T_{c0}}\right)^2+\frac{\psi_6}{6!}\left(\frac{\chi}{2\pi T_{c0}}\right)^2\left(\frac{\varkappa}{2\pi T_{c0}}\right)^4+\\ \nonumber
        +4\epsilon_2\frac{\pi^2}{36\psi_2}\left(\frac{\chi}{T_{c0}}\right)^4+\frac{\pi^2}{60\psi_2}\left(\frac{\chi}{T_{c0}}\right)^4\left(\frac{t}{T_{c0}}\right)^2+\frac{\pi^2}{90\psi_2}\left(\frac{\chi}{T_{c0}}\right)^6-\frac{\pi^2\psi_4}{72\psi_2^2\left(2\pi\right)^2}\left(\frac{\chi}{T_{c0}}\right)^6,
    \end{align}
\end{subequations}
and the shift of the critical temperature for a localized superconducting state
\begin{eqnarray}
\frac{T_{cw}(\chi,t)-T_c(\chi,t)}{T_{c0}} \approx -\frac{\pi^2}{36\psi_2 }\left(\frac{\chi}{T_{c0}}\right)^4 + \frac{\pi^2}{60\psi_2}\left(\frac{\chi}{T_{c0}}\right)^4\left(\frac{t}{T_{c0}}\right)^2 + \\
\nonumber
+\frac{\pi^2}{\psi_2}\left(\frac{1}{90}-\frac{\psi_4}{72\psi_2 (2\pi)^2} + \frac{5}{72}\frac{\psi_2}{(2\pi)^2}\right)\left(\frac{\chi}{T_{c0}}\right)^6 \ .
\end{eqnarray}
Evaluating the constant prefactors in the above expression, we get Eq.~(\ref{Tc_inhomogeneous_main_result}) in the main text.

\section{Derivation of the linearized self-consistency equation for the spin-triplet domain wall state}
\label{kernel_derivation_triplet_appendix}
In the case of a triplet spin structure, particularly, $\hat{\Delta}\left(x\right)=\Delta_{\rm int}\left(x\right)\hat{\sigma}_{x}\left(i\hat{\sigma}_{y}\right)$
the relevant system of equations for the anomalous Green functions is presented by Eqs.~(\ref{eilenberger_domain_wall_without_spin_spin_triplet_case}) in the main text.
The solution of the homogeneous system can be written as
\begin{align}
\left(\begin{array}{c}
f_{11}\\
f_{12}\\
f_{21}\\
f_{22}
\end{array}\right) & =e^{-2\omega_{n}x/v_{x}}\left[\frac{C_{1}}{\sqrt{2}}\left(\begin{array}{c}
1\\
0\\
0\\
1
\end{array}\right)+C_{2}\frac{t}{\sqrt{2}\varkappa}\left(\begin{array}{c}
-\chi/t\\
1\\
1\\
\chi/t
\end{array}\right)\right]+\\
 & +e^{\left(-2\omega_{n}+2i\varkappa\right)x/v_{x}}C_{3}\frac{t}{2\varkappa}\left(\begin{array}{c}
1\\
t/\left(\varkappa-\chi\right)\\
-t/\left(\varkappa+\chi\right)\\
-1
\end{array}\right)+e^{\left(-2\omega_{n}-2i\varkappa\right)x/v_{x}}C_{4}\frac{t}{2\varkappa}\left(\begin{array}{c}
1\\
-t/\left(\varkappa+\chi\right)\\
t/\left(\varkappa-\chi\right)\\
-1
\end{array}\right),\nonumber 
\end{align}
where $\varkappa=\sqrt{\chi^{2}+t^{2}}$ and $C_{i}$ ($i=$ 1, 2, 3, 4) are
arbitrary constants. Disregarding the intralayer Cooper pairing, we
put $C_{1}=0$ in the following. To find particular solution we use
the variation of constants method, which leads us to the set of equations:
\begin{subequations}
\begin{align}
iv_{x}\frac{\partial C_{2}}{\partial x}=0, \\
iv_{x}\frac{\partial C_{3}}{\partial x}=-2\Delta_{\rm int}\left(x\right)\text{sgn}\omega_{n}e^{\left(2\omega_{n}-2i\varkappa\right)x/v_{x}}, \\
iv_{x}\frac{\partial C_{4}}{\partial x}=2\Delta_{\rm int}\left(x\right)\text{sgn}\omega_{n}e^{\left(2\omega_{n}+2i\varkappa\right)x/v_{x}}.
\end{align}    
\end{subequations}
Performing the integration in equations above, we get the particular
solutions
\begin{eqnarray}
\left(\begin{array}{c}
f_{11}\\
f_{12}\\
f_{21}\\
f_{22}
\end{array}\right)=\frac{2i\text{sgn}\left(\omega_{n}\right)}{v_{x}}\frac{t}{2\varkappa}\int_{c_{3}}^{x}ds\Delta_{\rm int}\left(s\right)e^{\left(-2\omega_{n}+2i\varkappa\right)\left(x-s\right)/v_{x}}\left(\begin{array}{c}
1\\
t/\varkappa_{-}\\
-t/\varkappa_{+}\\
-1
\end{array}\right)\\
-\frac{2i\text{sgn}\left(\omega_{n}\right)}{v_{x}}\frac{t}{2\varkappa}\int_{c_{4}}^{x}ds\Delta_{\rm int}\left(s\right)e^{\left(-2\omega_{n}-2i\varkappa\right)\left(x-s\right)/v_{x}}\left(\begin{array}{c}
1\\
-t/\varkappa_{+}\\
t/\varkappa_{-}\\
-1
\end{array}\right)+c_{2}\frac{t}{\sqrt{2}\varkappa}e^{-2\omega_{n}x/v_{x}}\left(\begin{array}{c}
-\chi/t\\
1\\
1\\
\chi/t
\end{array}\right),\nonumber 
\end{eqnarray}
where we defined $\varkappa_{\pm}=\varkappa\pm\chi$ and $c_{i}$ ($i=$ 2, 3, 4)
are arbitrary constants. Now we proceed with writing down the above
solutions in the regions of constant relative band shift (for $x>0$
and $x<0$) and then match them continuously at $x=0$. The constants
$c_{i}$ are chosen, so that the anomalous functions $f_{ij}$
vanish at $\left|x\right|\rightarrow\infty$ .

The solution for positive $\omega_{n},\,v_{x}>0$ and for $x<0$ is
\begin{eqnarray}
\left(\begin{array}{c}
f_{11}\\
f_{12}\\
f_{21}\\
f_{22}
\end{array}\right)=\frac{2i}{v_{x}}\frac{t}{2\varkappa_{L}}\int_{-\infty}^{0}du \Delta_{\rm int}\left(x+u\right)e^{2\left(\omega_{n}-i\varkappa_{L}\right)u/v_{x}}\left(\begin{array}{c}
1\\
t/\varkappa_{L-}\\
-t/\varkappa_{L+}\\
-1
\end{array}\right)\label{eq:triplet_positive_wn_positive_vx_negative_x}\\
-\frac{2i}{v_{x}}\frac{t}{2\varkappa_{L}}\int_{-\infty}^{0}du\Delta_{\rm int}\left(x+u\right)e^{2\left(\omega_{n}+i\varkappa_{L}\right)u/v_{x}}\left(\begin{array}{c}
1\\
-t/\varkappa_{L+}\\
t/\varkappa_{L-}\\
-1
\end{array}\right) \nonumber
\end{eqnarray}
For $\omega_{n},\,v_{x}>0$ and $x>0$ it has the form:
\begin{eqnarray}
\left(\begin{array}{c}
f_{11}\\
f_{12}\\
f_{21}\\
f_{22}
\end{array}\right)=C_{2}\frac{t}{\sqrt{2}\varkappa_{R}}e^{-2\omega_{n}x/v_{x}}\left(\begin{array}{c}
-\chi_{R}/t\\
1\\
1\\
\chi_{R}/t
\end{array}\right)+e^{-2\left(\omega_{n}-i\varkappa_{R}\right)x/v_{x}}C_{3}\frac{t}{2\varkappa_{R}}\left(\begin{array}{c}
1\\
t/\varkappa_{R-}\\
-t/\varkappa_{R+}\\
-1
\end{array}\right)\label{eq:triplet_positive_wn_positive_vx_positive_x}\\
+e^{-2\left(\omega_{n}+i\varkappa_{R}\right)x/v_{x}}C_{4}\frac{t}{2\varkappa_{R}}\left(\begin{array}{c}
1\\
-t/\varkappa_{R+}\\
t/\varkappa_{R-}\\
-1
\end{array}\right)+\frac{2i}{v_{x}}\frac{t}{2\varkappa_{R}}\int_{0}^{x}ds\Delta_{\rm int}\left(s\right)e^{\left(-2\omega_{n}+2i\varkappa_{R}\right)\left(x-s\right)/v_{x}}\left(\begin{array}{c}
1\\
t/\varkappa_{R-}\\
-t/\varkappa_{R+}\\
-1
\end{array}\right)\nonumber \\
-\frac{2i}{v_{x}}\frac{t}{2\varkappa_{R}}\int_{0}^{x}ds\Delta_{\rm int}\left(s\right)e^{\left(-2\omega_{n}-2i\varkappa_{R}\right)\left(x-s\right)/v_{x}}\left(\begin{array}{c}
1\\
-t/\varkappa_{R+}\\
t/\varkappa_{R-}\\
-1
\end{array}\right)\nonumber 
\end{eqnarray}
Imposing the continuity condition on the solutions (\ref{eq:triplet_positive_wn_positive_vx_negative_x})
and (\ref{eq:triplet_positive_wn_positive_vx_positive_x}) at the
domain wall (at $x=0$), we get a linear system for the coefficients
\begin{subequations}
\begin{align}
-\frac{\chi_{R}}{\sqrt{2}\varkappa_{R}}C_{2}+\frac{t}{2\varkappa_{R}}C_{3}+\frac{t}{2\varkappa_{R}}C_{4}&=\frac{2i}{v_{x}}\frac{t}{2\varkappa_{L}}\overline{d}_{--}-\frac{2i}{v_{x}}\frac{t}{2\varkappa_{L}}\overline{d}_{-+}, \\ 
\frac{t}{\sqrt{2}\varkappa_{R}}C_{2}+\frac{t^{2}}{2\varkappa_{R}\varkappa_{R-}}C_{3}-\frac{t^{2}}{2\varkappa_{R}\varkappa_{R+}}C_{4} & =\frac{2i}{v_{x}}\frac{t^{2}}{2\varkappa_{L}\varkappa_{L-}}\overline{d}_{--}+\frac{2i}{v_{x}}\frac{t^{2}}{2\varkappa_{L}\varkappa_{L+}}\overline{d}_{-+}, \\
\frac{t}{\sqrt{2}\varkappa_{R}}C_{2}-\frac{t^{2}}{2\varkappa_{R}\varkappa_{R+}}C_{3}+\frac{t^{2}}{2\varkappa_{R}\varkappa_{R-}}C_{4}&=-\frac{2i}{v_{x}}\frac{t^{2}}{2\varkappa_{L}\varkappa_{L+}}\overline{d}_{--}-\frac{2i}{v_{x}}\frac{t^{2}}{\varkappa_{L}\varkappa_{L-}}\overline{d}_{-+}.
\end{align}
\end{subequations}

Here we introduced the following quantities:
\begin{equation}
\overline{d}_{-\mp}=\int_{-\infty}^{0}du\Delta_{\rm int}\left(u\right)e^{2\left(\omega_{n}\mp i\varkappa_{L}\right)u/v_{x}}.
\end{equation}

Solving the system, we get
\begin{subequations}\label{eq:matching_coefficients_positive_wn_positive_vx}
\begin{align}
C_{2}=\frac{2i}{v_{x}}\left(\overline{d}_{--}-\overline{d}_{-+}\right)\frac{t\left(\varkappa_{R+}^{2}\chi_{L}-\varkappa_{L+}\left(\varkappa_{L+}-2\chi_{L}\right)\chi_{R}-2\varkappa_{R+}\chi_{L}\chi_{R}\right)}{\sqrt{2}\varkappa_{L}\varkappa_{R+}\left(\varkappa_{L}^{2}-\chi_{L}^{2}\right)}, \\
C_{3}=\frac{2i}{v_{x}}\frac{\varkappa_{R+}\left(\varkappa_{R+}-2\chi_{R}\right)\left[\left(\overline{d}_{--}-\overline{d}_{-+}\right)\varkappa_{L+}\left(\varkappa_{L}-\chi_{L}\right)+\left(\overline{d}_{--}+\overline{d}_{-+}\right)\left(\varkappa_{L}\varkappa_{R+}-\varkappa_{L+}\chi_{R}\right)+2\overline{d}_{--}\chi_{L}\chi_{R}\right]}{2\varkappa_{R}\varkappa_{L}\left(\varkappa_{L}^{2}-\chi_{L}^{2}\right)}, \\
C_{4}=-\frac{2i}{v_{x}}\frac{\varkappa_{R+}\left(\varkappa_{R+}-2\chi_{R}\right)\left[\left(\overline{d}_{-+}-\overline{d}_{--}\right)\varkappa_{L+}\left(\varkappa_{L}-\chi_{L}\right)+\left(\overline{d}_{--}+\overline{d}_{-+}\right)\left(\varkappa_{L}\varkappa_{R+}-\varkappa_{L+}\chi_{R}\right)+2\overline{d}_{-+}\chi_{L}\chi_{R}\right]}{2\varkappa_{R}\varkappa_{L}\left(\varkappa_{L}^{2}-\chi_{L}^{2}\right)}.
\end{align}    
\end{subequations}
To construct the solutions for $\omega_{n}<0$ and $v_{x}<0$ one
can make use of the symmetry of the Eilenberger equations (\ref{eilenberger_domain_wall_without_spin_spin_triplet_case})
($i,j=1,\,2$)
\begin{equation}
f_{ij}\left(x,v_{x},\omega_{n},t,\chi\right)=f_{ij}\left(x,-v_{x},-\omega_{n},-t,-\chi\right).\label{eq:triplet_symmetry_relation}
\end{equation}

As a next step, we write down the solutions for negative Matsubara
frequencies $\omega_{n}$ and positive velocity projections $v_{x}$,
for $x<0$ they are of the form
\begin{eqnarray}
\left(\begin{array}{c}
f_{11}\\
f_{12}\\
f_{21}\\
f_{22}
\end{array}\right)=C_{2}\frac{t}{\sqrt{2}\varkappa_{L}}e^{2\left|\omega_{n}\right|x/v_{x}}\left(\begin{array}{c}
-\chi_{L}/t\\
1\\
1\\
\chi_{L}/t
\end{array}\right)+e^{2\left(\left|\omega_{n}\right|+i\varkappa_{L}\right)x/v_{x}}C_{3}\frac{t}{2\varkappa_{L}}\left(\begin{array}{c}
1\\
t/\varkappa_{L-}\\
-t/\varkappa_{L+}\\
-1
\end{array}\right)\label{eq:triplet_negative_wn_positive_vx_negative_x}\\
+e^{2\left(\left|\omega_{n}\right|-i\varkappa_{L}\right)x/v_{x}}C_{4}\frac{t}{2\varkappa_{L}}\left(\begin{array}{c}
1\\
-t/\varkappa_{L+}\\
t/\varkappa_{L-}\\
-1
\end{array}\right)+\frac{2i}{v_{x}}\frac{t}{2\varkappa_{L}}\int_{0}^{x}ds\Delta_{\rm int}\left(s\right)e^{2\left(\left|\omega_{n}\right|+i\varkappa_{L}\right)\left(x-s\right)/v_{x}}\left(\begin{array}{c}
1\\
t/\varkappa_{L-}\\
-t/\varkappa_{L+}\\
-1
\end{array}\right)\nonumber \\
-\frac{2i}{v_{x}}\frac{t}{2\varkappa_{L}}\int_{0}^{x}ds\Delta_{\rm int}\left(s\right)e^{2\left(\left|\omega_{n}\right|-i\varkappa_{L}\right)\left(x-s\right)/v_{x}}\left(\begin{array}{c}
1\\
-t/\varkappa_{L+}\\
t/\varkappa_{L-}\\
-1
\end{array}\right)\nonumber 
\end{eqnarray}

The solution for $\omega_{n}<0$, $v_{x}>0$ and $x>0$ can be written
in the form:
\begin{eqnarray}
\left(\begin{array}{c}
f_{11}\\
f_{12}\\
f_{21}\\
f_{22}
\end{array}\right)=\frac{2i}{v_{x}}\frac{t}{2\varkappa_{R}}\int_{0}^{+\infty}du\Delta_{\rm int}\left(x+u\right)e^{-2\left(\left|\omega_{n}\right|+i\varkappa_{R}\right)u/v_{x}}\left(\begin{array}{c}
1\\
t/\varkappa_{R-}\\
-t/\varkappa_{R+}\\
-1
\end{array}\right)\label{eq:triplet_negative_wn_positive_vx_positive_x}\\
-\frac{2i}{v_{x}}\frac{t}{2\varkappa_{R}}\int_{0}^{+\infty}du\Delta_{\rm int}\left(x+u\right)e^{-2\left(\left|\omega_{n}\right|-i\varkappa_{R}\right)u/v_{x}}\left(\begin{array}{c}
1\\
-t/\varkappa_{R+}\\
t/\varkappa_{R-}\\
-1
\end{array}\right).\nonumber 
\end{eqnarray}
 Matching the solutions (\ref{eq:triplet_negative_wn_positive_vx_negative_x})
and (\ref{eq:triplet_negative_wn_positive_vx_positive_x}) at the
domain wall ($x=0$), we get the following linear system:
\begin{subequations}
\begin{align}
-\frac{\chi_{L}}{\sqrt{2}\varkappa_{L}}C_{2}+\frac{t}{2\varkappa_{L}}C_{3}+\frac{t}{2\varkappa_{L}}C_{4}=\frac{2i}{v_{x}}\frac{t}{2\varkappa_{R}}\overline{d}_{+-}-\frac{2i}{v_{x}}\frac{t}{2\varkappa_{R}}\overline{d}_{++}, \\
\frac{t}{\sqrt{2}\varkappa_{L}}C_{2}+\frac{t^{2}}{2\varkappa_{L}\varkappa_{L-}}C_{3}-\frac{t^{2}}{2\varkappa_{L}\varkappa_{L+}}C_{4}=\frac{2i}{v_{x}}\frac{t^{2}}{2\varkappa_{R}\varkappa_{R-}}\overline{d}_{+-}+\frac{2i}{v_{x}}\frac{t^{2}}{2\varkappa_{R}\varkappa_{R+}}\overline{d}_{++}, \\
\frac{t}{\sqrt{2}\varkappa_{L}}C_{2}-\frac{t^{2}}{2\varkappa_{L}\varkappa_{L+}}C_{3}+\frac{t^{2}}{2\varkappa_{L}\varkappa_{L-}}C_{4}=-\frac{2i}{v_{x}}\frac{t^{2}}{2\varkappa_{R}\varkappa_{R+}}\overline{d}_{+-}-\frac{2i}{v_{x}}\frac{t^{2}}{2\varkappa_{R}\varkappa_{R-}}\overline{d}_{++}.
\end{align}    
\end{subequations}
Here we introduced
\begin{equation}
\overline{d}_{+\pm}=\int_{0}^{+\infty}du\Delta_{\rm int}\left(u\right)e^{\left(-2\left|\omega_{n}\right|\pm2i\varkappa_{R}\right)u/v_{x}}.
\end{equation}
 Solving the system above, we get
\begin{subequations}\label{eq:matching_coefficients_negative_wn_positive_vx}
\begin{align}
C_{2}=\frac{2i}{v_{x}}\left(\overline{d}_{+-}-\overline{d}_{++}\right)\frac{t\left(-\varkappa_{R+}^{2}\chi_{L}+\varkappa_{L+}\left(\varkappa_{L+}-2\chi_{L}\right)\chi_{R}+2\varkappa_{R+}\chi_{L}\chi_{R}\right)}{\sqrt{2}\varkappa_{R}\varkappa_{L}\left(\varkappa_{R}^{2}-\chi_{R}^{2}\right)}, \\
C_{3}=\frac{2i}{v_{x}}\frac{\varkappa_{L+}\left(\varkappa_{L+}-2\chi_{L}\right)\left[\left(\overline{d}_{+-}+\overline{d}_{++}\right)\varkappa_{L+}\varkappa_{R}+\overline{d}_{+-}\left(\varkappa_{R+}-\chi_{L}\right)\left(\varkappa_{R}-\chi_{R}\right)-\overline{d}_{++}\varkappa_{R+}\left(\varkappa_{R}+\chi_{L}-\chi_{R}\right)\right]}{2\varkappa_{L}\varkappa_{R}\left(\varkappa_{R}^{2}-\chi_{R}^{2}\right)}, \\
C_{4}=-\frac{2i}{v_{x}}\frac{\varkappa_{L+}\left(\varkappa_{L+}-2\chi_{L}\right)\left[\left(\overline{d}_{+-}+\overline{d}_{++}\right)\varkappa_{L+}\varkappa_{R}+\overline{d}_{++}\left(\varkappa_{R+}-\chi_{L}\right)\left(\varkappa_{R}-\chi_{R}\right)-\overline{d}_{+-}\varkappa_{R+}\left(\varkappa_{R}+\chi_{L}-\chi_{R}\right)\right]}{2\varkappa_{L}\varkappa_{R}\left(\varkappa_{R}^{2}-\chi_{R}^{2}\right)}.
\end{align}
\end{subequations}
The continuous solution of the Eilenberger equation for $\omega_{n}>0$
and $v_{x}<0$ can be obtained using the above Eqs.~ (\ref{eq:triplet_negative_wn_positive_vx_negative_x}), (\ref{eq:triplet_negative_wn_positive_vx_positive_x}), (\ref{eq:matching_coefficients_negative_wn_positive_vx}) and
the symmetry relation (\ref{eq:triplet_symmetry_relation}). Now we need to substitute the obtained solutions into the self-consistency
equation (\ref{eq:self_cons_main}) in the main text. It is convenient to recast the summation
in the self-consistency equation to the one over positive Matsubara
frequencies $\omega_{n}>0$ and positive velocity projections $v_{x}>0$.
For this purpose, in the following we write explicitly the sum of the obtained
solutions.

For $x<0$ we find
\begin{eqnarray}
f_{12}^{-}\left(x\right)\equiv f_{12}\left(\omega_{n}>0,v_{x}>0,x<0\right)+f_{12}\left(\omega_{n}>0,v_{x}<0,x<0\right)\\
+f_{12}\left(\omega_{n}<0,v_{x}>0,x<0\right)+f_{12}\left(\omega_{n}<0,v_{x}<0,x<0\right)=\nonumber \\
=\frac{4i}{\left|v_{x}\right|}\int_{-\infty}^{0}du\Delta_{\rm int}\left(x+u\right)e^{2\left|\omega_{n}\right|u/\left|v_{x}\right|}\cos\left(\frac{2\varkappa_{L}u}{\left|v_{x}\right|}\right)+C_{3}e^{2\left(\left|\omega_{n}\right|+i\varkappa_{L}\right)x/\left|v_{x}\right|}\nonumber \\
-C_{4}e^{2\left(\left|\omega_{n}\right|-i\varkappa_{L}\right)x/\left|v_{x}\right|}+\frac{4i}{v_{x}}\int_{0}^{x}ds\Delta_{\rm int}\left(s\right)e^{2\left|\omega_{n}\right|\left(x-s\right)/\left|v_{x}\right|}\cos\left(\frac{2\varkappa_{L}\left(x-s\right)}{\left|v_{x}\right|}\right).\nonumber 
\end{eqnarray}
In the above equation the constants $C_{3}$ and $C_{4}$ are defined
by Eqs.~(\ref{eq:matching_coefficients_negative_wn_positive_vx}). Below we provide the corresponding expressions for these
constants in the case of a stepwise relative band shift $\chi_{L}=-\chi_{R}=\chi$
\begin{subequations}
\begin{align}
C_{3}=\frac{2i}{\left|v_{x}\right|\varkappa^2}\left[t^{2}\overline{d}_{+-}+\chi^{2}\overline{d}_{++}\right], \\
C_{4}=-\frac{2i}{\left|v_{x}\right|\varkappa^2}\left[t^{2}\overline{d}_{++}+\chi^{2}\overline{d}_{+-}\right].
\end{align}
\end{subequations}
For $x>0$ we obtain
\begin{eqnarray}
f_{12}^{+}\left(x\right)\equiv f_{12}\left(\omega_{n}>0,v_{x}>0,x>0\right)+f_{12}\left(\omega_{n}>0,v_{x}<0,x>0\right)\\
+f_{12}\left(\omega_{n}<0,v_{x}>0,x>0\right)+f_{12}\left(\omega_{n}<0,v_{x}<0,x>0\right)=\nonumber \\
=\frac{4i}{\left|v_{x}\right|}\int_{0}^{+\infty}du\Delta_{\rm int}\left(x+u\right)e^{-2\left|\omega_{n}\right|u/\left|v_{x}\right|}\cos\left(\frac{2\varkappa_{R}u}{\left|v_{x}\right|}\right)+C_{3}e^{-2\left(\left|\omega_{n}\right|-i\varkappa_{R}\right)x/\left|v_{x}\right|}\nonumber \\
-C_{4}e^{-2\left(\left|\omega_{n}\right|+i\varkappa_{R}\right)x/\left|v_{x}\right|}+\frac{4i}{\left|v_{x}\right|}\int_{0}^{x}ds\Delta_{\rm int}\left(s\right)e^{-2\left|\omega_{n}\right|\left(x-s\right)/\left|v_{x}\right|}\cos\left(\frac{2\varkappa_{R}\left(x-s\right)}{\left|v_{x}\right|}\right)\nonumber 
\end{eqnarray}
Constants $C_{3}$ and $C_{4}$ are determined by Eqs.~(\ref{eq:matching_coefficients_positive_wn_positive_vx}).
Below we provide explicit expressions for these constants in the case
of a stepwise band shift $\chi_{L}=-\chi_{R}=\chi$
\begin{subequations}
\begin{align}
C_{3}=\frac{2i}{\left|v_{x}\right|\varkappa^2}\left[t^{2}\overline{d}_{--}+\chi^{2}\overline{d}_{-+}\right], \\
C_{4}=-\frac{2i}{\left|v_{x}\right|\varkappa^2}\left[\chi^{2}\overline{d}_{--}+t^{2}\overline{d}_{-+}\right].
\end{align}
\end{subequations}
So resulting self-consistency equation reads as
\begin{equation}
\Delta_{\text{int}}\left(x>0\right)=-i\pi\lambda T\sum_{\omega_{n}>0}\int_{v_{x}>0}\frac{d\mathbf{n}}{2\pi}f_{12}^{+}\left(x,\mathbf{n}\right),
\end{equation}
\begin{equation}
\Delta_{\text{int}}\left(x<0\right)=-i\pi\lambda T\sum_{\omega_{n}>0}\int_{v_{x}>0}\frac{d\mathbf{n}}{2\pi}f_{12}^{-}\left(x,\mathbf{n}\right).
\end{equation}
For a symmetric stepwise band shift both the anomalous function $f_{12}\left(x\right)$
and superconducting gap function $\Delta_{\text{int}}\left(x\right)$
are of even spatial parity. Thus, in this case one has the following
relations $\overline{d}_{-\mp}=\overline{d}_{+\pm}$.
Introducing
the Fourier-transformed superconducting gap
\begin{equation}
\Delta_{\text{int}}\left(x\right)=\int\frac{dk}{2\pi}\Delta_{\text{int}}\left(k\right)e^{ikx},
\end{equation}
we derive the self-consistency equation in momentum representation 

\begin{subequations}\label{self_cons_domain_wall_triplet_main_text}
\begin{align}
\label{self_cons_triplet}
\Delta_{\text{int}}\left(k\right)=\int_{-\infty}^{+\infty}\frac{dk'}{2\pi}K\left(k,k'\right)\Delta_{\text{int}}\left(k'\right) \ ,\\
\label{self_cons_kernel_triplet}
K\left(k,k'\right) = \pi\lambda T \sum_{\omega_n > 0}\int_{v_x>0}\frac{d\mathbf{n}}{2\pi} \biggl\{ 2\pi \delta(k-k')
{\rm Re}\left[\frac{4(\omega_n - i\varkappa)}{(\omega_n - i\varkappa)^2 + k^2v_x^2/4}\right]\\
\nonumber
-v_x{\rm Re}\biggl\{\frac{1}{(\omega_n + i\varkappa - ik'v_x/2)(\omega_n + i\varkappa-ikv_x/2)} +\frac{1}{(\omega_n - i\varkappa - ik'v_x/2)(\omega_n - i\varkappa - ikv_x/2)}
\\
\nonumber
-\frac{t^2}{\varkappa^2}\left[\frac{1}{(\omega_n + i\varkappa - ik'v_x/2)(\omega_n + i\varkappa - ikv_x/2)}+\frac{1}{(\omega_n - i\varkappa-ik'v_x/2)(\omega_n - i\varkappa - ikv_x/2)}\right]\\
\nonumber
-\frac{\chi^2}{\varkappa^2}\left[\frac{1}{(\omega_n - i\varkappa - ik'v_x/2)(\omega_n + i\varkappa - ikv_x/2)}+\frac{1}{(\omega_n + i\varkappa-ik'v_x/2)(\omega_n - i\varkappa - ikv_x/2)}\right]\biggl\}\biggl\}.
\end{align}
\end{subequations}

\section{Derivation of Eqs.~(\ref{GL_spin_triplet_main_text})-(\ref{Tc_DW_expansion_spin_triplet}) in the main text (linearized Ginzburg-Landau-type equation for the spin-triplet domain wall state)}\label{GL_triplet_appendix}
Here we present the derivation of Eqs.~(\ref{GL_spin_triplet_main_text})-(\ref{Tc_DW_expansion_spin_triplet}) in the main text. First, we provide the derivation details for the Ginzburg-Landau equation ~(\ref{GL_spin_triplet_main_text}) and various coefficients in the Ginzburg-Landau theory given in Eqs.~(\ref{GL_coefficients_triplet}). Second, we derive the equation for the critical temperature of the localized state~(\ref{Tc_localized_triplet_equation}) and its expansion in the vicinity of the critical temperature $T_{c0}$ for vanishing relative band offset and tunneling amplitude~(\ref{Tc_DW_expansion_spin_triplet}).

Our starting point is the expression for the kernel~(\ref{self_cons_kernel_triplet}) entering the self-consistency equation~(\ref{self_cons_triplet}).
The first term in~(\ref{self_cons_kernel_triplet}) is the homogeneous part of the kernel $K_h\left(k,k'\right)\propto\delta\left(k-k'\right)$, for convenience we introduce the function $K_h\left(k\right)$, defined as
\begin{equation}
K_{h}\left(k,k'\right)=2\pi\delta\left(k-k'\right)K_{h}\left(k\right).
\end{equation}
The expansion of the above function up to the second order by momentum $k$ has the following form
\begin{eqnarray}\label{hom_kernel_triplet_expansion}
    K_{h}\left(k\right)=4\pi\lambda T\sum_{\omega_{n}>0}\int_{-\pi/2}^{\pi/2}\frac{d\varphi}{2\pi}\text{Re}\left[\frac{\left(\omega_{n}-i\varkappa\right)}{\left(\omega_{n}-i\varkappa\right)^{2}+k^{2}v_{x}^{2}/4}\right] \\ \nonumber
    \approx4\pi\lambda T\sum_{\omega_{n}>0}\int_{-\pi/2}^{\pi/2}\frac{d\varphi}{2\pi}\left[\frac{\omega_{n}}{\omega_{n}^{2}+\varkappa^{2}}-\frac{k^{2}v_{F}^{2}\cos^{2}\varphi}{4}\cdot\frac{\omega_{n}\left(\omega_{n}^{2}-3\varkappa^{2}\right)}{\left(\omega_{n}^{2}+\varkappa^{2}\right)^{3}}\right].
\end{eqnarray}
Here $\varkappa=\sqrt{\chi^2+\left|t\right|^2}$ and $v_x=v_F\cos\left(\varphi\right)$.
Performing summation over Matsubara frequencies (using the Eqs.~(\ref{sums_over_Matsubara_frequencies_appendix})) and angle integration, we obtain the homogeneous part of the linearized self-consistency equation
\begin{eqnarray}\label{self_cons_triplet_hom}
        \left(\frac{1}{4\pi T}\text{Re}\left[\psi\left(\frac{1}{2}\right)-\psi\left(\frac{1}{2}-i\frac{\varkappa}{2\pi T}\right)\right]+\frac{k^{2}v_{F}^{2}}{16}\cdot\frac{1}{16\pi^{3}T^{3}}\text{Re}\left[\psi_2\left(\frac{1}{2}-i\frac{\varkappa}{2\pi T}\right)\right]\right)\Delta_{{\rm int}}\left(k\right) = \\ \nonumber
        =\frac{1}{4\pi T}\ln\left(\frac{T}{T_{c0}}\right)\Delta_{{\rm int}}\left(k\right).
\end{eqnarray}
Evaluating the inhomogeneous part $K_{{\rm inh}}\left(k,k'\right)=K\left(k,k'\right)-K_{h}\left(k,k'\right)$ at $k=k'=0$, we get
\begin{eqnarray}\label{inhom_kernel_triplet_value}
K_{{\rm inh}}\left(0,0\right)	=4\pi\lambda T\sum_{\omega_{n}>0}\int_{-\pi/2}^{\pi/2}\frac{d\varphi}{2\pi}v_{F}\cos\varphi\frac{\chi^{2}}{\varkappa^{2}}\cdot\frac{\varkappa^{2}}{\left(\omega_{n}^{2}+\varkappa^{2}\right)^{2}}= \\ \nonumber
	=4\pi\lambda Tv_{F}\cdot\frac{\chi^{2}\left(\sinh\left(\frac{\varkappa}{T}\right)-\frac{\varkappa}{T}\right)}{16\pi \varkappa^{3}T\cosh^{2}\left(\frac{\varkappa}{2T}\right)}=4\pi\lambda T\frac{v_{F}}{T^{2}}\frac{\left(\frac{\chi}{T}\right)^{2}\left(\sinh\left(\frac{\varkappa}{T}\right)-\frac{\varkappa}{T}\right)}{16\pi\left(\frac{\varkappa}{T}\right)^{3}\cosh^{2}\left(\frac{\varkappa}{2T}\right)}.
\end{eqnarray}
Combining Eqs.~(\ref{self_cons_triplet}) and~(\ref{hom_kernel_triplet_expansion})~-~(\ref{inhom_kernel_triplet_value}), we obtain the linearized self-consistency equation~(\ref{GL_spin_triplet_main_text}) in the main text with the coefficients given by Eqs.~(\ref{GL_coefficients_triplet}). 
Solving the resulting equation~(\ref{GL_spin_triplet_main_text}) for the lowest eigenvalue, we get the equation~(\ref{Tc_localized_triplet_equation}), governing the critical temperature of the localized state formation $T_{cw}$. It's explicit form reads as
\begin{equation}\label{T_cw_triplet_explicit_appendix}
    \ln\left(\frac{T_{cw}}{T_{c0}}\right)-\text{Re}\left[\psi\left(\frac{1}{2}\right)-\psi\left(\frac{1}{2}-i\frac{\varkappa}{2\pi T_{cw}}\right)\right]=-\frac{\pi^{2}\left(\frac{\chi}{T_{cw}}\right)^{4}\left(\sinh\left(\frac{\varkappa}{T_{cw}}\right)-\frac{\varkappa}{T_{cw}}\right)^{2}}{\text{Re}\left[\psi_2\left(\frac{1}{2}-i\frac{\varkappa}{2\pi T_{cw}}\right)\right]\left(\frac{\varkappa}{T_{cw}}\right)^{6}\cosh^{4}\left(\frac{\varkappa}{2T_{cw}}\right)}.
\end{equation}
Now we proceed with the expansion of the above equation over the small $\chi/T_{c0},\,t/T_{c0},\,\varkappa/T_{c0}\ll1$, expanding as well the deviation of the critical temperature $T_{cw}$ from $T_{c0}$
\begin{equation}\label{dev_exp_triplet_appendix}
    \frac{T_{cw}}{T_{c0}}=1+\epsilon;\quad\epsilon = \epsilon_{2}+\epsilon_{4}+\epsilon_{6},
\end{equation}
where $\epsilon_{2,4,6}$ contain the terms $\propto\left(\chi/T_{c0}\right)^n\left(t/T_{c0}\right)^m$ with $m+n=2,4,\,\text{and}\,6,$ respectively.
Substituting expansions~(\ref{dev_exp_triplet_appendix}),~(\ref{expansion_1}), and~(\ref{expansion_2}) into the Eq.~(\ref{T_cw_triplet_explicit_appendix}), we get
\begin{subequations}
    \begin{align}
        \epsilon_{2}+\epsilon_{4}+\epsilon_{6}-\frac{1}{2}(\epsilon_{2}^{2}+2\epsilon_{2}\epsilon_{4})+\frac{1}{3}\epsilon_{2}^{3}\\ \nonumber
        -\left[\frac{\psi_{2}}{2}\left(\frac{\varkappa}{2\pi T_{c0}}\right)^{2}\left[1-2(\epsilon_{2}+\epsilon_{4})+3\epsilon_{2}^{2}\right]-\frac{\psi_{4}}{4!}\left(\frac{\varkappa}{2\pi T_{c0}}\right)^{4}(1-4\epsilon_{2})+\frac{\psi_6}{6!}\left(\frac{\varkappa}{2\pi T_{c0}}\right)^6\right] \\ \nonumber
    =-\frac{\pi^{2}}{\psi_{2}}\left(\frac{\chi}{T_{c0}}\right)^{4}\left(1-4\epsilon_{2}\right)\left[\frac{1}{36}-\frac{1}{90}\left(\frac{\varkappa}{T_{c0}}\right)^{2}+\frac{\psi_{4}}{2\psi_{2}}\frac{1}{36}\left(\frac{\varkappa}{2\pi T_{c0}}\right)^{2}\right].
    \end{align}
\end{subequations}
Equating the terms with same powers of small parameters, we find the solution for $\epsilon_{2,4,6}$
\begin{subequations}
    \begin{align}
    \epsilon_{2}=\frac{\psi_{2}}{2}\left(\frac{\varkappa}{2\pi T_{c0}}\right)^{2},\\
    \epsilon_{4}=-\frac{3\epsilon_{2}^{2}}{2}-\frac{\psi_{4}}{4!}\left(\frac{\varkappa}{2\pi T_{c0}}\right)^{4}-\frac{\pi^{2}}{\psi_{2}}\left(\frac{\chi}{T_{c0}}\right)^{4}\frac{1}{36},\\
    \epsilon_{6}=-\frac{1}{3}\epsilon_{2}^{3}-\epsilon_{2}\epsilon_{4}+3\epsilon_{2}^{3}-4\frac{\psi_{4}}{4!}\left(\frac{\varkappa}{2\pi T_{c0}}\right)^{4}\epsilon_{2}+\frac{4}{36}\frac{\pi^{2}}{\psi_{2}}\epsilon_{2}\left(\frac{\chi}{T_{c0}}\right)^{4} - \\ \nonumber+
    \frac{\psi_6}{6!}\left(\frac{\varkappa}{2\pi T_{c0}}\right)^6-\frac{\pi^{2}}{\psi_{2}}\left(\frac{\chi}{T_{c0}}\right)^{4}\left[-\frac{1}{90}\left(\frac{\varkappa}{T_{c0}}\right)^{2}+\frac{\psi_{4}}{2\psi_{2}}\frac{1}{36}\left(\frac{\varkappa}{2\pi T_{c0}}\right)^{2}\right],
    \end{align}
\end{subequations}
which leads to the following shift of the critical temperature for a localized superconducting state
\begin{subequations}
    \begin{align}
        \frac{T_{cw}\left(\chi,t\right)-T_{c}\left(\chi,t\right)}{T_{c0}}=-\frac{\pi^{2}}{36\psi_{2}}\left(\frac{\chi}{T_{c0}}\right)^{4}+\frac{\pi^{2}}{\psi_{2}}\left(\frac{\chi}{T_{c0}}\right)^{4}\left(\frac{t}{T_{c0}}\right)^{2}\left[\frac{5\psi_{2}}{72\left(2\pi\right)^{2}}+\frac{1}{90}-\frac{\psi_{4}}{72\psi_{2}\left(2\pi\right)^{2}}\right]+ \\ \nonumber
        +\frac{\pi^{2}}{\psi_{2}}\left(\frac{\chi}{T_{c0}}\right)^{6}\left[\frac{5\psi_{2}}{72\left(2\pi\right)^{2}}+\frac{1}{90}-\frac{\psi_{4}}{72\psi_{2}\left(2\pi\right)^{2}}\right]\approx \left(0.016+0.02\left(\frac{t}{T_{c0}}\right)^2\right)\left(\frac{\chi}{T_{c0}}\right)^4+0.02\left(\frac{\chi}{T_{c0}}\right)^6.
    \end{align}
\end{subequations}
Thus, reproducing the Eq.~(\ref{Tc_DW_expansion_spin_triplet}) in the main text.

\section{Solution of the linearized Eilenberger equations in the presence of the magnetic field} \label{Eilenberger_equations_spin_triplet_magnetic_field}

Here we present the solutions of the linearized Eilenberger equations that we use for the analysis of the effects of the in-plane magnetic field on the spin-triplet and spin-singlet interlayer superconductivity. In the singlet case we make substitution $\hat{f}_{ij} = \left[f_{ij}^{(0)} + \mathbf{n}_h\hat{\boldsymbol{\sigma}}f_{ij}^{(1)}\right](i\hat{\sigma}_y)$ in Eqs.~(\ref{Eilenberger_magnetic_field_general}) and obtain the self-consistency equation
\begin{subequations}
    \begin{align}
        1 = 2\pi\lambda T\sum_{\omega_n > 0}\int\frac{d\mathbf{n}}{2\pi}\omega_nP_0(\mathbf{n},\omega)/Q_0(\mathbf{n},\omega) \ , \\
        P_0(\mathbf{n},\omega) = \left[(h^2 - q^2)^2 + 2(h^2 + q^2)\omega_n^2+\omega_n^4\right]\left(h^2 + \omega_n^2 + \chi^2\right) + (h^2 + \omega_n^2)|t|^4 \\
\nonumber
+|t|^2\left[(2\omega_n^2 + \chi^2)(q^2 + \omega_n^2)+h^2(2q^2 + \chi^2)-2h^4\right] \ ,\\
Q_0(\mathbf{n},\omega) = \left[h^4 + 2h^2(\omega_n^2 - q^2)+(q^2+\omega_n^2)^2\right]\left[h^4 + 2h^2(\omega_n^2-\chi^2)+(\omega_n^2 + \chi^2)^2\right] + (h^2 + \omega_n^2)^2|t|^4 + \\
\nonumber
+2|t|^2\left[\chi^2(h^4-h^2(q^2-2\omega_n^2)+\omega_n^2(\omega_n^2 + q^2))-(h^2 + \omega_n^2)^2(h^2 - q^2 - \omega_n^2)\right] \ .
    \end{align}
\end{subequations}
Here $q = e\mathbf{v}\mathbf{A}_1/c = ev_FdH\sin(\theta - \varphi)/2c$, and $\theta$ ($\varphi$) is the angle between the magnetic field direction (trajectory direction) and $x$-axis. In the limit $h=0$ the above expressions reproduce the Eq.~(\ref{self_cons_orbital_main_text}) in the main text.

Substituting the anomalous Green's function $f_{ij} = (f_{ij}^{(0)} + f_{ij}^{(x)}\hat{\sigma}_x + f_{ij}^{(y)}\hat{\sigma}_y)(i\hat{\sigma}_y)$ and the superconducting gap function $\hat{\Delta}_{\rm int} = \Delta_{\rm int}\hat{\sigma}_x(i\hat{\sigma}_y)$ into Eqs.~(\ref{Eilenberger_magnetic_field_general}), we get the following system

\begin{equation}
\left[
\begin{array}{*{12}c}
 a_{1,1} & a_{1,2} & a_{1,3} & a_{1,4} & 0 & 0 & a_{1,7} & 0 & 0 & 0 & 0 & 0 \\
a_{2,1} & a_{2,2} &0& 0 & a_{2,5} & 0 & 0 & a_{2,8} & 0 & 0 & 0 & 0 \\
a_{3,1}&0& a_{3,3}& 0& 0& a_{3,6}& 0& 0& a_{3,9}& 0& 0& 0 \\
a_{4,1}&0&0& a_{4,4}& a_{4,5}& a_{4,6}& 0& 0& 0& a_{4,10}& 0& 0\\
0& a_{5,2}& 0& a_{5,4}& a_{5,5}& 0& 0& 0& 0& 0& a_{5,11}& 0\\
0& 0& a_{6,3}& a_{6,4}&0& a_{6,6}& 0& 0& 0& 0& 0& a_{6,12}\\
a_{7,1}& 0& 0& 0&0&0& a_{7,7}& a_{7,8}& a_{7,9}& a_{7,10}& 0& 0\\
0& a_{8,2}& 0& 0& 0& 0& a_{8,7}&a_{8,8}&0&0&a_{8,11}&0\\
0& 0& a_{9,3}& 0& 0& 0& a_{9,7}&0& a_{9,9}& 0& 0& a_{9,12}\\
0& 0& 0& a_{10,4}& 0& 0& a_{10,7}&0&0& a_{10,10}& a_{10,11}& a_{10,12}\\
0& 0& 0& 0& a_{11,5}& 0& 0& a_{11,8}& 0&a_{11,10}&a_{11,11}&0\\
0& 0& 0& 0& 0& a_{12,6}& 0& 0& a_{12,9}&a_{12,10}&0&a_{12,12} 
\end{array}
\right]
\begin{bmatrix}
f_{11}^{(0)}\\
f_{11}^{(x)}\\
f_{11}^{(y)}\\
f_{12}^{(0)}\\
f_{12}^{(x)}\\
f_{12}^{(y)}\\
f_{21}^{(0)}\\
f_{21}^{(x)}\\
f_{21}^{(y)}\\
f_{22}^{(0)}\\
f_{22}^{(x)}\\
f_{22}^{(y)} 
\end{bmatrix} = 
\begin{bmatrix}
0\\0\\0\\0\\-2\Delta_{\rm int}\\0\\0\\2\Delta_{\rm int}\\0\\0\\0\\0
\end{bmatrix} \ .
\end{equation}
Here 
\begin{eqnarray}
a_{1,1} = a_{2,2} = a_{3,3} = 2(i\omega_n - q) \ , \ \ a_{4,4} = a_{5,5} = a_{6,6} = 2(i\omega_n - \chi) \ ,\\
a_{7,7} = a_{8,8} = a_{9,9} = 2(i\omega_n + \chi) \ , \ \ a_{10,10} = a_{11,11} = a_{12,12} = 2(i\omega_n + q) \ ,\\
a_{1,2} = a_{2,1} = a_{4,5} = a_{5,4} = a_{7,8} = a_{8,7} = a_{10,11} = a_{11,10} = 2h_x \ ,\\
a_{1,3} = a_{3,1} = a_{4,6} = a_{6,4} = a_{7,9} = a_{9,7} = a_{10,12} = a_{12,10} = 2h_y \ ,\\
a_{1,4} = a_{2,5} = a_{3,6} = a_{7,10} = a_{8,11} = a_{9,12} = t \ ,\\
a_{1,7} = a_{2,8} = a_{3,9} = a_{4,10} = a_{5,11} = a_{6,12} = -t \ ,\\
a_{4,1} = a_{5,2} = a_{6,3} = a_{10,7} = a_{11,8} = a_{12,10} = t^* \ ,\\
a_{7,1} = a_{8,2} = a_{9,3} = a_{10,4} = a_{11,5} = a_{12,6} = -t^* \ .
\end{eqnarray}
Solving the above system of equations for $f_{12}^{(x)}$, we obtain the self-consistency equation in the form
\begin{subequations}
    \begin{align}
        1 = 2\pi\lambda T\sum_{\omega_n > 0}\int\frac{d\mathbf{n}}{2\pi}\omega_n P_x/Q_{x1}Q_{x2} \ , \\
        P_x = (q^2+\omega_n^2)\left[h^4-2h^2(q^2-\omega_n^2)+(q^2+\omega_n^2)^2\right]\left[h_y^4-2h_y^2(\chi^2-\omega_n^2)+(\chi^2+\omega_n^2)^2+h_x^2(\chi^2+h_y^2+\omega_n^2)\right] \\
\nonumber
+|t|^2\biggl\{h_x^6\omega_n^2+h_x^4\left[q^2\chi^2-2q^2h_y^2-(q^2-2\chi^2)\omega_n^2+4\omega_n^4\right] \\
\nonumber
+2(q^2+\omega_n^2)\left[-h_y^6+h_y^4(q^2+\chi^2-\omega_n^2)+\omega_n^2(q^2+\omega_n^2)(\chi^2+\omega_n^2)+h_y^2(-q^2\chi^2+2(q^2+\chi^2)\omega_n^2 + \omega_n^4)\right] \\
\nonumber
+h_x^2[q^4\chi^2+q^2(2q^2+\chi^2)\omega_n^2 + (3q^2+4\chi^2)\omega_n^4+5\omega_n^6 \\
\nonumber
-h_y^4(4q^2+3\omega_n^2)+h_y^2(2q^4+3q^2\chi^2-(q^2-4\chi^2)\omega_n^2+2\omega_n^4)]
\biggl\} + \\
\nonumber
+|t|^4\biggl\{(h^2+\omega_n^2)[h_y^2q^2+\omega_n^2(q^2+\omega_n^2+h^2)]\biggl\} \ ,\\
Q_{x1} = (q^2+\omega_n^2)(\omega_n^2+\chi^2)+\omega_n^2|t|^2 \ ,\\
Q_{x2} = [(h^2-q^2)^2+2(h^2+q^2)\omega_n^2+\omega_n^4][(h^2+\omega_n^2)^2-2(h^2-\omega_n^2)\chi^2+\chi^4]+(h^2+\omega_n^2)^2|t|^4 - \\
\nonumber
-2|t|^2(h^2-q^2-\omega_n^2)(h^2+\omega_n^2)^2+2|t|^2\chi^2h^2(h^2-q^2)+2|t|^2\omega_n^2\chi^2(2h^2+q^2)+2|t|^2\omega_n^4\chi^2 \ .
    \end{align}
\end{subequations}

At $q=0$ and $t=0$ we obtain the Eq.~(\ref{self_consistency_spin_triplet_paramagnetic}) in the main text.
\end{widetext}

\end{document}